\documentclass[twocolumn,trackchanges]{aastex631}

\hypersetup{linkcolor=red,citecolor=green,filecolor=cyan,urlcolor=magenta}
\usepackage{comment}
\usepackage{hyperref}
\usepackage{amsmath}
\usepackage{longtable}
\usepackage{longfigure}

\usepackage{supertabular}

\begin{document}

\title{Probing X-ray Timing and Spectral Variability in the Blazar PKS 2155-304 Over a Decade of \textit{{\it XMM-Newton}} Observations}

\author[0000-0002-0705-6619]{Gopal Bhatta}
\affiliation{Janusz Gil Institute of Astronomy, University of Zielona Góra, ul. Szafrana 2, 65-516 Zielona Góra, Poland}
\email{g.bhatta@ia.uz.zgora.pl}

\author[0000-0002-9126-1817]{Suvas Chandra Chaudhary}
\affiliation{Inter-University Centre for Astronomy and Astrophysics, Pune, Maharashtra 411007, India}
\email{suvas0101phy@gmail.com}

\author{Niraj Dhital}
\affiliation{Central Department of Physics, Tribhuvan University, Kirtipur 44613, Nepal}

\author[0000-0003-4586-0744]{Tek P. Adhikari}
\affiliation{CAS Key Laboratory for Research in Galaxies and Cosmology, Department of Astronomy, University of Science and Technology of China, Hefei, Anhui 230026, China}
\affiliation{School of Astronomy and Space Science, University of Science and Technology of China, Hefei, Anhui 230026, China}
      \email{tek@ustc.edu.cn }

\author{Maksym Mohorian}
\affiliation{School of Mathematical and Physical Sciences, Macquarie University, Sydney, NSW 2109, Australia}
\affiliation{Astronomy, Astrophysics and Astrophotonics Research Centre, Macquarie University, Sydney, NSW 2109, Australia}

\author{Adithiya Dinesh}
\affiliation{IPARCOS Institute and EMFTEL Department, Universidad Complutense de Madrid, E-28040 Madrid, Spain}

\author{Radim P{\'a}nis}
\affiliation{Research Centre for Theoretical Physics and Astrophysics, Institute of Physics,\\
Silesian University in Opava, Bezru{\v c}ovo n{\'a}m.13, CZ-74601 Opava, Czech Republic}

\author{Raghav Neupane}
 \affiliation{Central Department of Physics, Tribhuvan University, Kirtipur 44613, Nepal}

 \author{Yogesh Singh Maharjan}
 \affiliation{Department of Physics, Amrit Campus, Tribhuvan University, Kathmandu, Nepal}

\begin{abstract}
Blazars, a class of active galactic nuclei (AGN) powered by supermassive black holes, are known for their remarkable variability across multiple timescales and wavelengths. Despite significant advancements in our understanding of AGN central engines, 
thanks to both ground- and space-based telescopes, the details of the mechanisms driving this variability remain elusive.
The primary objective of this study is to constrain the X-ray variability properties of the TeV blazar PKS 2155-304. We conducted a comprehensive X-ray spectral and timing analysis, focusing on both long-term and intra-day variability (IDV), using data from 22 epochs of {\it XMM-Newton} observations collected over 15 years (2000 to 2014).
For the timing analysis, we estimated the fractional variability, variability amplitude, minimum variability timescales, flux distribution, and power spectral density. In the spectral analysis, we fitted the X-ray spectra using power-law, log-parabola, and broken power-law models to determine the best-fitting parameters.
We observed moderate IDV in the majority of the light curves. Seven out of the 22 observations showed a clear bimodal flux distribution, indicating the presence of two distinct flux states. Our analysis revealed a variable power spectral slope. Most hardness ratio plots did not show significant variation with flux, except for two observations,  where the hardness ratio changed considerably with flux. The fitted X-ray spectra favored the broken power law model for the majority of observations, indicating break in the spectral profiles.
The findings of this work shed light on the IDV of blazars, providing insights into the non-thermal jet processes that drive the observed flux variations.

\end{abstract}


\section{Introduction} \label{sec:intro}
Blazars belong to a subclass of active galactic nuclei (AGN) characterized by the presence of relativistic jets directed towards our line of sight, with an angle between the jet axis and the line of sight of $\leq 10$ degrees \citep{urry1995}. These sources are distinguished by their high luminosity and flux and polarization variability across diverse timescales. Additionally, they exhibit Doppler-boosted broad-band continuum emission spanning from radio to TeV wavebands. The broadband spectral energy distribution (SED) of blazars features
 two distinct humps in the $\nu$ vs.
 $\nu f_{\nu}$ plane. The low-energy hump peaks in the optical/X-ray range, whereas the high-energy hump peaks in the GeV/TeV energy range. The lower energy feature is mostly attributed to synchrotron emission from relativistic particles in the magnetic fields present in the jets. However, the origin of the hump at higher energy is still under wide discussion \citep[see][and references therein]{2019Galax...7...20B}. In leptonic models of blazar emission, where the broadband radiative output is considered to be primarily generated by leptons (such as electrons and possibly positrons), the origin of the high-energy feature is most plausibly explained by Inverse-Compton scattering of low-energy seed photons by relativistic particles within the jets. 
In such a scenario, the seed photons can originate from synchrotron photons within the jets, as proposed in the synchrotron self-Compton model (e.g., \citealt{1992ApJ...397L...5M}), and/or from low-energy photons emitted by the accretion disk \citep{1993ApJ...416..458D}, broad-line region \citep{1994ApJ...421..153S}, and dusty torus \citep{2000ApJ...545..107B}, collectively referred to as external-Compton (EC) models.
For certain blazar sources, the SED suggests a better explanation within hadronic models. In these models, high-energy emission is associated with synchrotron radiation from protons and/or secondary leptons generated through proton-photon interactions \cite[e. g., see][]{1993A&A...269...67M,2003APh....18..593M}.
One possible way to categorize the blazars is based on whether the rest-frame equivalent width (EW) of their broad optical emission line is less than or greater than 5 Å \citep{1991ApJ...374..431S,1991ApJS...76..813S}. This categorization introduces two classes of blazars: flat-spectrum radio quasars (FSRQs) and BL Lacertae (BL Lacs).

FSRQs are powerful sources dominated by Compton emission, with the synchrotron peak occurring at a lower frequency. In contrast, BL Lac objects, though relatively less powerful, are mostly found to be TeV sources, with their inverse Compton peak located in the highest $\gamma$-ray bands. The dichotomy between BL Lacs and FSRQs is based on intrinsic differences in the nature of the accretion disc and the physical origins of high-energy emissions in these sources. This includes the larger gamma-ray luminosity of FSRQs and the harder gamma-ray spectral characteristics of BL Lacs \citep[see][and references therein]{2024MNRAS.528..976B}. Furthermore, in the case of high-energy emission via external inverse-Compton \citep[e. g.,][]{1994ApJ...421..153S,1993ApJ...416..458D}, the seed photon field in FSRQs is contributed by different populations of photons external to the jet component, whereas in BL Lacs it is primarily contributed through synchrotron self-Compton \citep[e. g.,][]{1985A&A...146..204G}. Blazars can also be classified based on their location of the synchrotron peak-frequencies ($\nu_{\rm peak}$): low ($\nu_{\rm peak} < 10^{14} {\rm \ Hz}$), intermediate ($10^{14} {\rm \ Hz } < \nu_{\rm peak} < 10^{15} {\rm \ Hz}$), and high ($\nu_{\rm peak} > 10^{15}{\rm \ Hz}$) \citep{2010ApJ...716...30A}. Moreover, high-synchrotron peaked blazars with synchrotron peak frequencies above 10$^{17}$ Hz are specifically classified as extreme high-energy synchrotron peaked blazars \citep[see][and references therein]{2019A&A...632A..77C}.

Blazars exhibit high-amplitude rapid flux variability across diverse timescales throughout the entire electromagnetic spectrum \citep[see e. g,][]{2023MNRAS.520.2633B,2021ApJ...923....7B,2020ApJ...891..120B}. This makes multifrequency variability studies particularly effective in exploring the central engine of AGN. The temporal variability can be broadly classified into long-term, short-term, and intra-day variability (IDV). Long-term variability, occurring over several months to several years, can lead to flux changes by up to an order of magnitude \citep[e. g.,][]{2011A&A...531A.123K,2020A&A...634A..80R,2021ApJ...923....7B}. Short-term variability, spanning days to a few months, results in flux changes by a few factors \citep[e. g.,][]{2023MNRAS.518.1459P,2023MNRAS.520.2633B,2006A&A...455..871F}. Furthermore, IDV  or microvariability is characterized by rapid flux variations within a day \citep{2021Galax...9..114W}.

In particular, IDV offers insights into the dynamic nature of the relativistic jets and compact regions of blazars, providing strong motivation to explore multi-wavelength IDV in these sources. IDV in several blazars has been studied using observations in a wide range of energy/frequency bands \citep[see e.g.,][]{2007ApJ...664L..71A,2013A&A...558A..92B,2016ApJ...831...92B,2018Galax...6....2B,2018MNRAS.480.4873A,2023MNRAS.518.1459P}.

Similarly, X-ray variability properties have been extensively studied using observations made by different instruments like {\it  XMM-Newton} \citep{2023ApJ...955..121D,2022MNRAS.510.5280M,2022ApJS..262....4N}, Swift \citep[e. g.,][]{2014MNRAS.444.1077K}, Astrosat \citep{2024A&A...682A.134G,2022MNRAS.513.1662M,2020ApJ...897...25B}, NuSTAR \citep[e. g.,][]{2018A&A...619A..93B}, Chandra \citep[e. g.,][]{2018MNRAS.480.4873A} and RXTE \citep[e. g.,][]{2018ApJ...867...68W}. These studies offer crucial insights into blazar variability, including its structure, evolution, dynamics, and the emission mechanisms of the central engine. Blazar sources are found to exhibit significant flux variability in the X-ray band, with variability timescales ranging from a few minutes to a few decades. In BL Lac objects, the observed X-ray emission probes the relativistic particle population accelerated by internal shocks \citep[e. g.,][]{2010ApJ...711..445B}, turbulence, and/or magnetic reconnection \citep[e. g.,][]{2021Galax...9...27M,2019MNRAS.482...65C} prevalent within their jets. In contrast, FSRQs exhibit substantial thermal emission from their accretion disks, which predominantly appears as the characteristic `big blue bump' in the optical/UV bands \citep[e. g.,][]{1998A&A...340...47P}. As a result, the observed X-ray emission may partly arise from thermal instabilities within the accretion disk structures \citep[see][for discussion on disk-jet connection in blazars]{2009MNRAS.400.1521J}.
However, the details of the mechanism leading to IDV across diverse EM bands are still under discussion.

As part of a broader program investigating the MWL emission mechanisms in blazars, this work focuses on characterizing the X-ray variability of PKS 2155-304, a well-studied HSP blazar, through detailed multi-timescale analysis.
Our primary objective is to characterize both the intra-day and long-term variability properties of this source through detailed X-ray observations \citep[e. g., see our similar previous works on other sources][]{2022MNRAS.510.5280M,2018A&A...619A..93B,2023ApJ...955..121D}. By examining the temporal and spectral behavior across different timescales, we aim to better understand the physical mechanisms driving the MWL variability and high-energy emission in blazars. Blazar PKS 2155-304 serves as an excellent laboratory for this investigation due to its strong X-ray emission and documented variability across the electromagnetic spectrum, making it an ideal candidate for probing the underlying physics of blazar emission mechanisms and jet dynamics.

The paper is organized as follows: Section \ref{sec:2} provides a brief introduction to the source PKS~2155--304. Section \ref{sec:3} covers the details of the {\it XMM-Newton} data reduction methods. Analysis techniques and results are presented in Section \ref{sec:4}. Finally, Section \ref{sec:5} outlines the discussion and conclusion of this work.

\section{Blazar PKS~2155--304}\label{sec:2}
  The blazar PKS~2155--304 is one of the brightest sources in the Southern sky when observed in the UV and X-ray bands. Classified as a BL Lac object, its apparent magnitude in the V-band is 13.09. It is located in the constellation Piscis Austrinus, with right ascension 21h 58m 52.0s and declination -30° 13' 32''. It is located at a redshift of $z$ = 0.116 \citep{1993ApJ...411L..63F}.   The blazar PKS 2155--304 was initially detected as an X-ray source during observations conducted using the HEAO 1 satellite
\citep{1979ApJ...229L..53S,1979ApJ...234..810G}. Later the source was detected in the TeV range \citep{2005A&A...430..865A}.

Radio images from the Very Large Array reveal an extended radio jet located approximately 20 kpc from the nucleus of PKS 2155-304 \citep{2013AJ....145...73L}. The 43 GHz images of the source obtained by very-long-baseline interferometry reveal new morphological details, indicating a significant degree of jet bending within the inner milli-arcsecond of the source \citep{2010ApJ...723.1150P}.

The spectral power density of the source's long-term optical variability is found to be consistent with a broken power-law model, with a characteristic break timescale of approximately 2.7 years \citep{2011A&A...531A.123K}. Moreover, in a study involving long-term optical observations and gamma-ray observations from Fermi/LAT, it was found that the optical emission is highly correlated with the gamma-ray emission \citep{2021ApJ...923....7B,2021MNRAS.504.1772R}. Furthermore, time series analysis of long-term optical and gamma-ray observations of the source  revealed quasi-periodic oscillations (QPOs) with characteristic timescales of $\sim$ 600 days and $\sim$ 250 days, which were found to be significant over the red-noise inherent in blazars \citep{2021ApJ...923....7B,2020ApJ...896..134P,2019MNRAS.484..749C,2014ApJ...793L...1S}.

  During July/August 2006, the  source experienced an intense outburst in very high energy (VHE) flux, with peak fluxes approximately 7 times higher than the VHE flux level of the Crab Nebula. This event exhibited rapid fluctuations with a variability timescale of only 3 minutes \citep{ 2007ApJ...664L..71A}.

 Multiwavelength (MWL) analyses of the source have been conducted by modeling the broadband SEDs from different observation epochs using standard blazar models. A log-parabolic SED of PKS~2155--304 was found to be consistent across optical, UV, and X-ray wavelengths \citep{ 2014MNRAS.444.3647B}. During a MWL observation campaign spanning from optical to TeV energy ranges, it was discovered that in a low flux state, the object exhibited highly significant flux variability in the X-rays. Additionally, the broadband SED was found to be consistent with a one-zone synchrotron self-Compton model \citep{2020A&A...639A..42A}.

The flux and spectral variability of the BL Lac PKS 2155-304 have been extensively studied in the X-ray energy band using observations from several space telescopes, including Chandra, Astrosat, Swift-XRT, NuSTAR, and {\it XMM-Newton} \citep[see][for a recent review]{2020Galax...8...64G}. In a study using NuSTAR observations, \citet{2018A&A...619A..93B} reported the presence of steep spectra, with a photon index, $\Gamma \sim 3$, in the hard X-ray emission of the source. A study of the source using X-ray observations from the Suzaku satellite revealed large-amplitude flux and spectral intra-day variabilities in highly correlated soft and hard bands, with the blazar exhibiting harder spectra when brighter \citep{2021ApJ...909..103Z}.

In this work, we analysed 22 {\it XMM-Newton} {\it EPIC-PN} observations of blazar PKS~2155--304 observed between 2000 to 2014. Observation IDs along with their exposure IDs, modes and durations are presented in Table \ref{table1}.

\begin{figure*}
	\begin{minipage}{.35\textwidth} 
		\includegraphics[height=6cm]{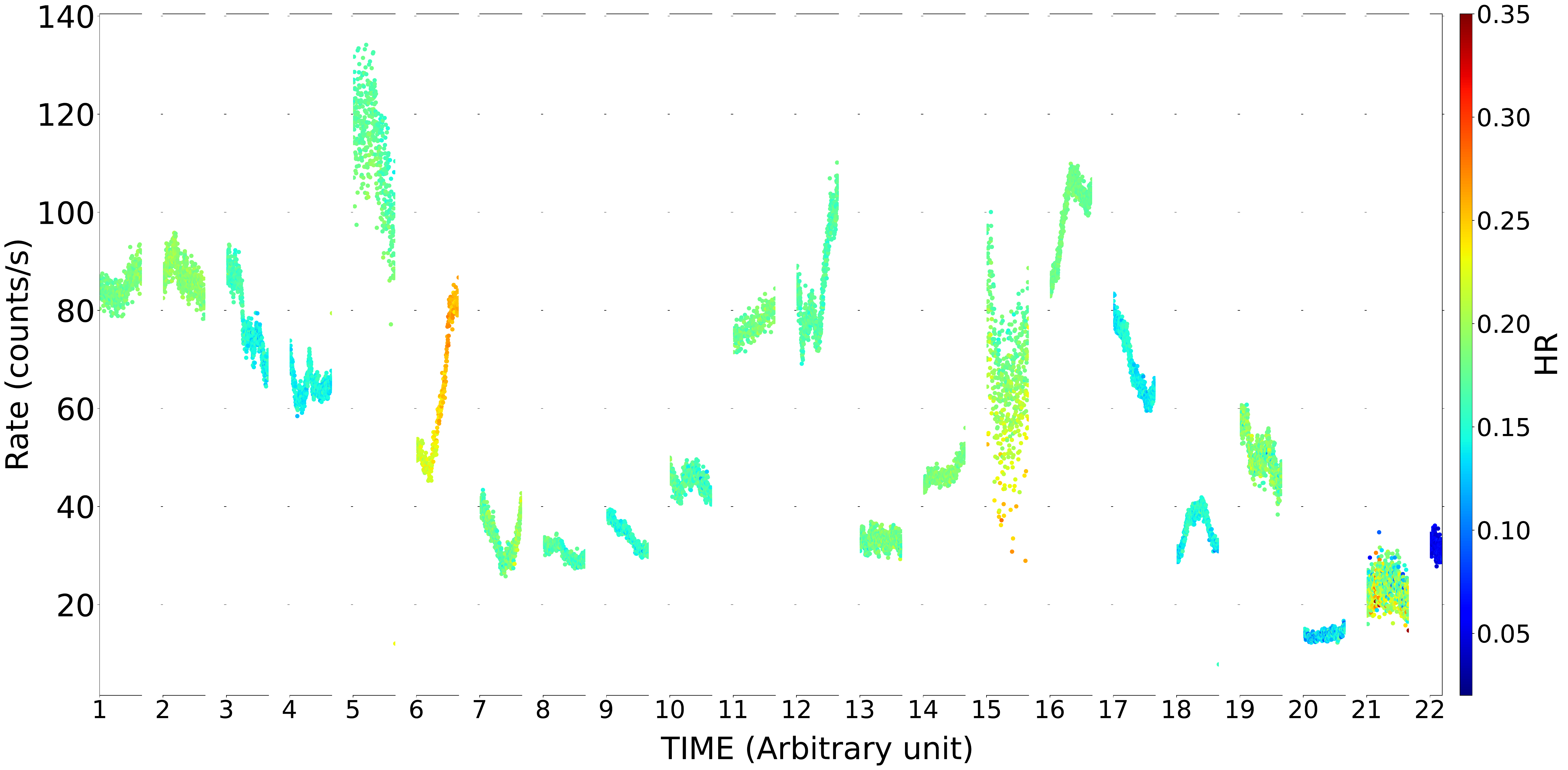}
   
	\end{minipage}
	\begin{minipage}{.2\textwidth}
	\  
	\end{minipage}%
	\begin{minipage}{.35\textwidth}  
		\raggedleft
		\includegraphics[height=7cm]{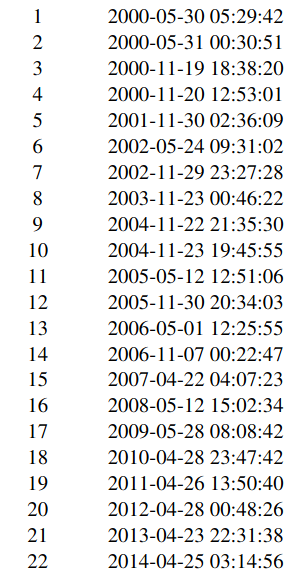}
	\end{minipage}
	\caption{{ Long-term X-ray light curves of PKS 2155-304 from \textit{{\it XMM-Newton}} observations spanning 2000 to 2014}. The flux points are color-coded based on their corresponding hardness ratio, with the starting observation dates and times indicated on the right side.}
	\label{figure1}
\end{figure*}

\begin{table*}
\caption{{\it {\it XMM-Newton}} EPIC PN observations of BL Lac PKS 2155-304 from 2000-2014. Col. 1: Observation ID; Col. 2: Instrument; Col. 3: Exposure ID ; Col. 4: Mode; Col. 5: Date; Col. 6: Length of observation (in ks).}
\label{table1}
\centering
\centering
\begin{tabular}{ccccccc}
  \hline
   & Obs .ID & Instrument & Exposure ID & Mode & Date & Duration (ks) \\ 
   & (1)& (2) & (3) & (4) & (5) & (6)  \\ 
  \hline

  1 & 0080940101 & PN & S003 & Imaging & 2000-11-19 18:38:20  & 57.2 \\
  2 & 0080940301 & PN & S003 & Imaging & 2000-11-20 12:53:01  & 58.1 \\
  3 & 0124930101 & PN  & S010  & Imaging & 2000-05-30 05:29:42  & 37.9 \\
  4 & 0124930201 & PN & S003 & Imaging & 2000-05-31 00:30:51  & 59.3 \\
  5 & 0124930301 & PN & S003 & Imaging & 2001-11-30 02:36:09  & 44.6 \\
  6 & 0124930501 & PN & S009 & Imaging & 2002-05-24 09:31:02  & 96.1 \\
  7 & 0124930601 & PN & S003 & Imaging & 2002-11-29 23:27:28  & 56.8 \\
  8 & 0158960101 & PN & S001 & Imaging & 2003-11-23 00:46:22  & 26.6 \\
  9 & 0158960901 & PN & S001 & Imaging & 2004-11-22 21:35:30  & 28.4 \\ 
  10 & 0158961001 & PN & S013 & Imaging & 2004-11-23 19:45:55  & 39.9 \\
  11 & 0158961101 & PN & S001 & Imaging & 2005-05-12 12:51:06  & 26.1 \\  
  12 & 0158961301 & PN & S001 & Imaging & 2005-11-30 20:34:03  & 59.9 \\
  13 & 0158961401 & PN & S001 & Imaging & 2006-05-01 12:25:55  & 64.3 \\
  14 & 0411780101 & PN & S001 & Imaging & 2006-11-07 00:22:47  & 29.9 \\
  15 & 0411780201 & PN & S001 & Imaging & 2007-04-22 04:07:23  & 58.5 \\
  16 & 0411780301 & PN & S001 & Imaging & 2008-05-12 15:02:34  & 60.7 \\
  17 & 0411780401 & PN & S001 & Imaging & 2009-05-28 08:08:42  & 64.3 \\
  18 & 0411780501 & PN & S001 & Imaging & 2010-04-28 23:47:42  & 69.1 \\ 
  19 & 0411780601 & PN & S001 & Imaging & 2011-04-26 13:50:40  & 63.3 \\
  20 & 0411780701 & PN & S001 & Imaging & 2012-04-28 00:48:26  & 53.6 \\
  21 & 0411782101 & PN & S001 & Imaging & 2013-04-23 22:31:38  & 76.0 \\
  22 & 0727770901 & PN & S001 & Imaging & 2014-04-25 03:14:56  & 65.0 \\
  \hline
\end{tabular}

\end{table*}

\section{\textit{{\it XMM-Newton}}: Observations and Data Analysis}\label{sec:3}
The blazar PKS~2155--304 has been the target of several MWL monitoring programs using various space and ground based telescopes. For our analysis, we specifically utilized X-ray archival data from {\it {\it XMM-Newton}} \citep{2001A&A...365L...1J}. The space observatory is equipped with three instruments: the European Photon Imaging Camera (EPIC), the Optical Monitor (OM), and the Reflection Grating Spectrometer (RGS) \citep{2000SPIE.4012..154B,2001A&A...365L..18S}. The EPIC instrument is particularly notable for its impressive features, including excellent angular resolution (Point Spread Function, PSF = 6 arcseconds Full Width at Half Maximum) and moderate energy resolution in the range of $E/\Delta E$ = 20-50 \citep{2001A&A...365L..18S}. In our investigation, we focused exclusively on the EPIC-PN data due to its greater collecting area, higher sensitivity compared to EPIC-MOS, and reduced susceptibility to pile-up effects that arise when multiple X-ray photons arrive at the detector within a short time interval, leading to a single, distorted detection event.

The observations included in this study are comprehensively listed in Table \ref{table1}. We conducted our analysis using the Science Analysis System (SAS) and followed the standard data analysis procedures outlined in the {\it {\it XMM-Newton}} ABC Guide. Notably, the EPIC instrument facilitates both imaging and spectroscopic studies within the energy range of 0.2 to 15 keV.

For each observation, we began the data processing by generating a summary of the Calibration Index File ({\it CIF}) and the Observation Data File ({\it ODF}) using the available calibrated data files. The {\textit{epproc}} command was used to reprocess the EPIC-PN data. We then created high-energy light curves for single events, identifying intervals with background flaring. These intervals are crucial for distinguishing high-energy events from hot pixels \citep{2004A&A...419..837D}. A standard rate cutoff criterion (RATE $\leq$ 0.4 for EPIC-PN) was applied to select periods with low and stable light curves.

To prepare for the final analysis, we produced an EPIC "clean" file using imaging event files, with all resulting filtered files stored under the identifier {\it EPICclean}. Additionally, for scientific purposes, good time intervals were stored with the prefix {\it EPICgti} for future reference. Given that many of the observations were affected by pile-up effects, we used the \textit{epatplot} tool for pile-up correction. This correction involved selecting annular regions within the image files, details of which are documented in Table \ref{table2} (Column 7).

We then created source and background light curves by applying a quality selection flag, specifically {\it FLAG == 0 $\&\&$ $PATTERN \leq 4$}, for all EPIC-PN observations. Circular source and background regions were defined using the {\it ds9} software. To obtain corrected source light curves that account for factors such as bad pixels, PSF variation, and quantum efficiency, we utilized the SAS task {\it epiclccorr}.

Light curves were extracted in three distinct energy bands: the soft band (0.3--2 keV), the hard band (2-10 keV), and the full {\bf {\it XMM-Newton}} band (0.3--10 keV), each with a constant time bin of 100 seconds\footnote{\url{https://www.cosmos.esa.int/web/xmm-newton/sas-threads}}. After extracting the light curves, we performed a spectral analysis of PKS~2155--304 for each observation. Source and background spectra were generated following standard procedures. The redistribution matrix was created using the SAS command \textit{RMFGEN}, and data pattern information was collected using the \textit{ARFGEN} command from the Data Sub Space (DSS). The \textit{specgroup} SAS command was used to produce grouped spectrum files, which were subsequently fitted with various {\it XSPEC} models \cite{1996ASPC..101...17A}. The spectrum was analyzed over the energy range of 0.3 to 10 keV, applying different {\it XSPEC} models to determine the most appropriate fitting parameters.

To account for absorption effects, we used a hydrogen column density value of $N_H=1.28\times10^{20} \rm cm^{-2}$, obtained from online $N_H$ estimators\footnote{\url{https://heasarc.gsfc.nasa.gov/cgi-bin/Tools/w3nh/w3nh.pl}}. We then generated the spectrum and performed spectral fitting using three {\it XSPEC} models to determine the most suitable parameters (see Section \ref{spec_fit}).

\begin{table*}
\caption{The table summarizes key properties of {\it XMM-Newton} EPIC PN observations of the BL Lac PKS 2155-304 from 2000 to 2014. The columns are defined as follows: Col. 1: observation ID; Col. 2: Mean Flux (counts/s); Col. 3: Fractional Variability in percentage; Col. 4: Variability Amplitude; Col. 5: Minimum Variability timescale; Col. 6: pile-up; Col. 7: pile-up region; Col. 8: Negative spectral power index.}
\label{table2}
\centering
\centering
\begin{tabular}{ccccccccc}
  \hline
    Obs .ID & $\langle$F$\rangle$ (counts/s) & $F_{\rm var}$($\%$) & VA & $\tau_{var}$ &  pile-up & Region & $-\beta_\mathrm{P}$ &\\ 
&(0.3--10 keV)&  (0.3--10 keV)&(0.3--10 keV)& ks & & & &\\
  \hline

 0080940101 & $78.46\pm7.42$ & $9.24\pm0.09$ & $1.05\pm0.06$ & $2.91\pm2.30$ & yes & $150\leq r\leq800$ & 2.46±0.14 &\\ 
 0080940301 & $64.56\pm2.65$ & $3.44\pm0.10$ & $0.92\pm0.59$ & $2.63\pm1.49$ & yes & $150\leq r\leq600$ & 2.70±0.16 &\\
 0124930101 & $84.96\pm3.26$ & $1.32\pm0.35$ & $1.14\pm0.88$ & $1.69\pm0.84$ &   yes & $150\leq r\leq800$ & 4.16±0.89 &\\ 
 0124930201 & $87.17\pm3.28$ & $3.16\pm0.09$ & $0.71\pm0.05$ & $1.83\pm0.97$ &  yes & $150\leq r\leq600$ & 4.14±0.08 &\\ 
 0124930301 & $111.82\pm11.50$ & $7.75\pm0.37$ & - & $1.08\pm0.96$ &yes & $150\leq r\leq800$ & 2.73±0.19 &\\ 
 0124930501 & $61.26\pm12.96$ & $21.06\pm0.14$ & $1.63\pm0.09$ & $1.42\pm0.76$ &yes & $150\leq r\leq800$ & 2.28±0.10 & \\ 
 0124930601 & $33.75\pm4.23$ & $12.11\pm0.14$ & $1.35\pm0.11$ & $2.32\pm2.26$ & yes & $150\leq r\leq800$ & 2.50±0.21 &\\ 
 0158960101 & $30.66\pm1.65$ & $4.84\pm0.15$ & $0.76\pm0.06$ & $2.31\pm1.75$ &  no & $150\leq r\leq800$ & 1.50±0.29 & \\ 
 0158960901 & $34.37\pm2.60$ & $7.25\pm0.13$ & $0.88\pm0.06$ & $2.43\pm1.75$ &  no & $ r\leq600$ & 1.85±0.14 &\\ 
 0158961001 & $45.01\pm2.11$ & $3.73\pm0.16$ & $0.73\pm0.07$ & $1.94\pm1.45$  &yes & $150\leq r\leq600$ & 2.11±0.07 &\\ 
 0158961101 & $77.22\pm2.55$ & $2.51\pm0.15$ & $0.63\pm0.05$ & $1.73\pm0.94$ & yes & $150\leq r\leq800$ & 2.97±0.82 &\\ 
 0158961301 & $85.12\pm10.02$ & $11.60\pm0.09$ & $1.24\pm0.07$ & $1.70\pm0.84$  &yes & $150\leq r\leq800$ & 2.86±0.15 &\\ 
 0158961401 & $33.10\pm1.31$ & $2.17\pm0.19$ & $0.77\pm0.08$ & $1.71\pm1.40$ & yes & $150\leq r\leq800$ & 0.95±0.17 &\\ 
 0411780101 & $46.89\pm2.07$ & $3.97\pm0.12$ & $0.84\pm0.08$ & $2.66\pm1.97$ &  no & $ r\leq800$ & 2.46±0.06 &\\ 
 0411780201 & $66.31\pm11.44$ & $9.52\pm0.78$ & $3.31\pm2.64$ & $0.36\pm0.22$ &  yes & $150\leq r\leq800$ & 4.49±0.13 &\\ 
 0411780301 & $99.58\pm7.26$ & $7.18\pm0.05$ & $0.87\pm0.03$ & $5.03\pm4.92$ & yes & $200\leq r\leq800$ & 2.39±0.06 &\\ 
 0411780401 & $69.04\pm6.30$ & $8.99\pm0.06$ & $0.98\pm0.04$ & $4.10\pm3.43$ & yes & $400\leq r\leq1000$ & 1.82±0.07 &\\ 
 0411780501 & $35.68\pm3.76$ & $10.32\pm0.08$ & - & $2.44\pm1.94$ &  yes & $150\leq r\leq800$ & 1.02±0.04 &\\ 
 0411780601 & $50.54\pm4.28$ & $7.54\pm0.16$ & $1.23\pm0.14$ & $1.37\pm1.00$ &  yes & $150\leq r\leq800$ & 2.09±0.06 &\\ 
 0411780701 & $13.91\pm0.95$ & $5.90\pm0.15$ & - & $1.83\pm1.61$ &  no & $ r\leq600$ & 0.92±0.08 &\\ 
 0411782101 & $23.29\pm3.05$ & $6.75\pm0.61$ & - & $0.37\pm0.23$ & yes &  $350\leq r\leq1000$ & 2.35±0.05 &\\ 
 0727770901 & $31.86\pm1.53$ & $3.45\pm0.16$ & $0.83\pm0.09$ & $1.51\pm1.07$ &yes & $150\leq r\leq600$ & 3.09±0.14 &\\ 
  \hline
\end{tabular}

\end{table*}


  
\section{ANALYSIS Methods} \label{sec:4}
A comprehensive summary of all the {\it XMM-Newton} observations used in our study is presented in Table \ref{table1}. The table lists the observation ID, instrument, exposure ID, observation mode, observation date, start and end times, and the total duration of each observation. To investigate flux variability, we employed several analytical techniques, each briefly described in this paper. The methodologies are broadly categorized into timing and spectral analyses. The timing analysis involved the estimation of several variability parameters, including excess variance, fractional variability, power spectral density, and flux distribution (see Table \ref{table2}). For our spectral analysis, we computed hardness ratios by obtaining count rates in two distinct energy bands and calculating their ratios. We then separately fitted the \textit{XMM-Newton} spectra with three commonly used non-thermal models to find which one is preferred over the 14 years of PKS 2155-304 observations.
\subsection{EXCESS VARIANCE: FLUX VARIABILITY \& VARIABILITY AMPLITUDE}
Figure \ref{figure1} presents the long-term X-ray light curve, constructed from {\it XMM-Newton} observations spanning 14 years. The light curve clearly demonstrates  significant flux variability on both intra-day and yearly timescales. To quantify the magnitude of the observed variability, three statistical measures are commonly employed: excess variance ($\sigma_{\rm XS}$), fractional variability ($F_{\rm var}$), and variability amplitude (VA).

Excess variance quantifies a source's intrinsic variability by subtracting the variance attributed to measurement errors from the total observed LC variance.
For a light curve comprising $N$ measured flux values $X_{i}$, each with associated finite uncertainties $\sigma_{\rm {err,i}}$  due to measurement errors, and given $S^{2}$ as the sample variance of the light curve, the excess variance is computed using the following relation:
\begin{equation}
\label{excess}
   \sigma^{2}_{XS} =  S^{2} - \Bar{\sigma}^{2}_{\rm err},
\end{equation}
\noindent where $\Bar{\sigma}^{2}_{\rm err}$ represents square of the mean measurement errors \citep{1997ApJ...476...70N,2002ApJ...568..610E}.   Fractional variability is calculated using the relation,


\begin{equation}
    F_{\rm var} = \sqrt{\frac{S^{2} - \bar{\sigma}^{2}_{\rm err}}{\bar{X}^2}}
\end{equation}
\citep{1990ApJ...359...86E,2003MNRAS.345.1271V}. 
The associated error in the fractional variability is obtained using,
\begin{equation}
   \sigma_{F_{\mathrm{var}}} =  \sqrt{\left(\frac{1}{\sqrt{2N}}\frac{\bar{\sigma}^{2}_{err}}{F_{var}}\frac{1}{\bar{X}^2}\right)^2 + \left(\sqrt{\frac{\bar{\sigma}^{2}_{err}}{N}}\frac{1}{\bar{X}^2}\right)^2} 
\end{equation}
\citep{2015A&A...576A.126A,2018Galax...6....2B}. 
Variability amplitude $VA$ gives the information about peak-to-peak flux variations, which can be defined as,
\begin{equation}
    VA = \dfrac{F_{\mathrm{max}}-F_{\mathrm{min}}}{F_{\mathrm{min}}}, \label{eqn:VA}
\end{equation}
where $F_{\mathrm{max}}$ and $F_{\mathrm{min}}$ are the maximum and minimum flux, respectively. 
The error in $VA$ is estimated using,
\begin{figure*}
	\centering
	\begin{minipage}{.4\textwidth} 
		\centering 
		\includegraphics[width=.99\linewidth]{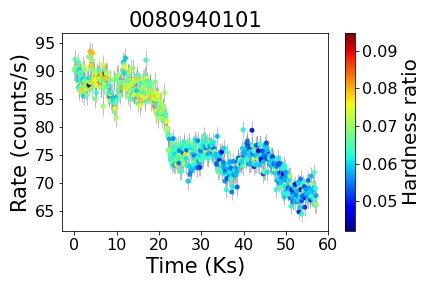}
	\end{minipage}
	\begin{minipage}{.1\textwidth} 
	\  
	\end{minipage}%
	\begin{minipage}{.4\textwidth} 
		\centering 
		\includegraphics[width=.99\linewidth]{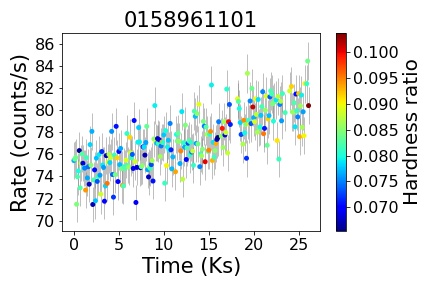}
	\end{minipage}
	\caption{{\it XMM-Newton} light curves in the 0.3--10 keV energy range are presented for two representative observations (IDs 0080940101 and 0158961101) of PKS 2155-304. Light curves for the remaining observations are provided in Appendix A.}
	\label{LCs}
	\end{figure*} 

\begin{equation}
    \sigma_{\mathrm{VA}} = (VA+1)\cdot\sqrt{\left(\dfrac{\sigma_{F_{\mathrm{max}}}}{F_{\mathrm{max}}}\right)^2+\left(\dfrac{\sigma_{F_{\mathrm{min}}}}{F_{\mathrm{min}}}\right)^2},
\end{equation}
which follows from the propagation of errors in flux in Equation \ref{eqn:VA} \citep{2018A&A...619A..93B,2022MNRAS.510.5280M}.
Since in the {\it VA} measurement we are only considering peak to peak fluxes, it may not provide the information about the overall variability. In such situation fractional variability, variability index and flux-histograms may provide deeper information about observed flux variability. Since variability of blazars are observed over diverse timescales, we also estimate variability timescale for the PKS~2155--304.   
The minimum timescale of such variability is \citep{1974ApJ...193...43B}

\begin{equation}
    \tau_{\mathrm{var}} = \left|\dfrac{dt}{d\ln F}\right|,
\end{equation}
where $dt$ is the time interval between flux measurements. Error or uncertainty in the variability timescale is given as:
\begin{equation}
    \sigma_{\tau_{\mathrm{var}}} \approx \sqrt{\dfrac{F_1^2\Delta F_2^2+F_2^2\Delta F_1^2}{F_1^2 F_2^2 (\ln[F_1/F_2])^4}}\cdot\Delta t,
\end{equation}
where $\Delta F_1$  and $\Delta F_2$ are the flux uncertainties used to estimate the minimum variability timescales for the fluxes $F_1$ and $F_2$, respectively \citep[see][]{, 2018A&A...619A..93B}.

Variability quantifying measures, including $F_{\rm var}$, {\it VA}, and $\tau_{\rm var}$, were computed for all 22 observations and are listed in Table \ref{table2}. We have also presented the light curves for two of the observations in Figure \ref{LCs}; the light curves for the remaining observations are included in the Appendix. All presented X-ray light curves of the source demonstrate significant fractional variability, as determined by considering $F_{\rm var} > 3\times \sigma_{F_{\mathrm{var}}}$ \cite[see e. g.,][]{2021MNRAS.506.1198D}. The fractional variability for the observations of the source ranges from 1.32$\pm$0.35\% (lowest for observation 0124930101) to 21.06$\pm$0.14\% (highest for observation 0124930501). Our results are also consistent with fractional variability results presented in \cite{2017ApJ...850..209G} for six {\it {\it XMM-Newton}} observations of PKS~2155--304 in the energy regime 0.6--10 keV. Furthermore, it can be seen from Table \ref{table2} that variability amplitude ranges from a minimum value of 0.71$\pm$0.05 to a maximum 3.31$\pm$2.64. 
Additionally, six light curves do not show any significant $VA$ in the full X-ray energy band (0.3--10 keV). We also noted that mean flux, mean fractional variability and mean variability amplitude throughout the {\it XMM-Newton} observation during 2000--2014 is about 57.48$\pm$4.83 counts/s, $(6.99\pm0.20)\%$ and 1.1$\%$, respectively. Furthermore, the fractional variability estimated from observations over the entire period is approximately 40\%, demonstrating the magnitude of the variability that blazars undergo over the timescale of more than a decade.

\begin{figure}
    \includegraphics[width=1.\linewidth]{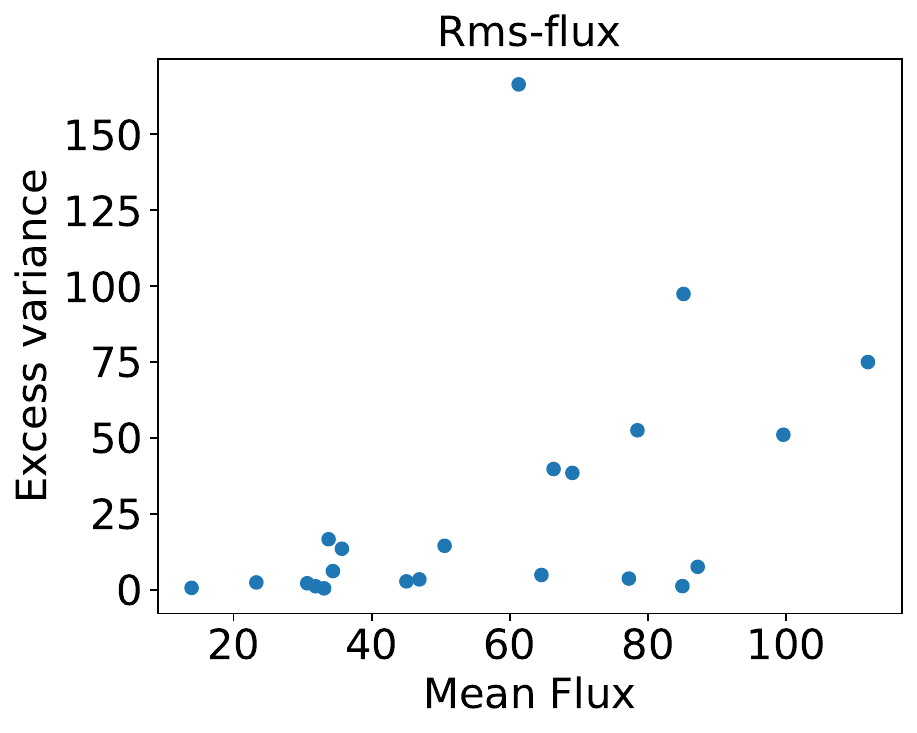}
    \caption{Relationship between the root mean square (RMS) variability and mean flux for the blazar PKS 2155-304.} 
    \label{rmsflux}
\end{figure}


\subsection{RMS--FLUX RELATION}


In time-series studies of high-energy astrophysical systems, examining the relationship between the root mean square (RMS) and mean flux offers insights into the nature of the system. In this method of analysis, the source light curve is divided into multiple sections, and the correlation between the RMS and mean flux for each section is analyzed. At first, a linear RMS-flux relationship was observed in the context of variability observed in X-ray binaries hosting accretion discs \citep[e.g.,][]{2001MNRAS.323L..26U,2004A&A...414.1091G,2012MNRAS.422.2620H}, and it was also reported later in the X-ray variability of Seyfert galaxies \citep{2004ApJ...612L..21G,2019MNRAS.482.2088A}. In recent years, a similar trend has been observed in various frequency bands of blazar observations, including optical, X-ray, and gamma-rays, from several sources \citep{wang2023comprehensive, 2022MNRAS.510.3688K,2021ApJ...923....7B,2020ApJ...891..120B,2020ApJ...897...25B,2013ApJ...766...16E}.

To further characterize the X-ray variability properties of the source, we analyzed the distribution of variability across different flux states using excess variance (as given by Equation \ref{excess}) as an estimator of intrinsic source variance.  Positive $\sigma^{2}_{\rm XS}$ values indicate intrinsic source variability, while negative values, potentially arising from observational noise, were excluded from the analysis.
We computed the root mean square (RMS) variability for each observation and examined its relationship with the mean flux, expressed as count rate. Figure \ref{rmsflux} presents resultant RMS-flux relationship. As seen in the figure, analysis of this relationship revealed no clear or simple correlation between RMS variability and mean flux, suggesting a complex underlying variability mechanism.


\begin{figure}[htb]
    \includegraphics[width=1.\linewidth]{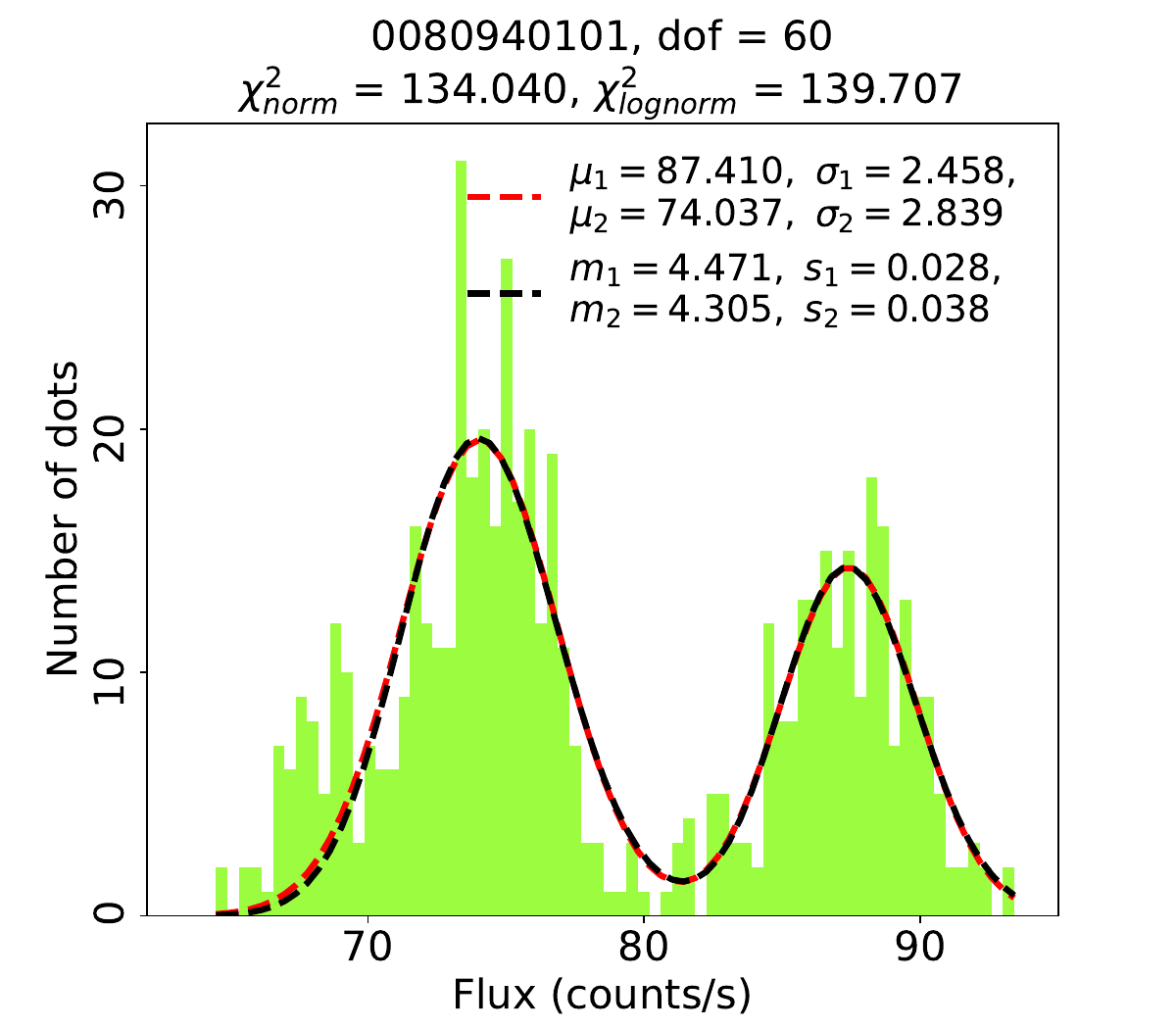}
    \caption{Histogram showing the flux distribution for one of the observation IDs of PKS 2155-304. Similar flux distribution histograms for all other observations are presented in Appendix}
    \label{histo}
\end{figure}

\begin{table*}[htb]

\caption{Normal and lognormal fit statistics for the X-ray flux distributions of each observation.
The mean ($\mu$) and the standard deviation ($\sigma$) of the normal fit are shown in the columns 2 and 3 while the mean location (m) and the scale parameters (s) of the lognormal fit are listed in the columns 5 and 6 respectively.}
\label{table3}
\centering
\centering
\begin{tabular}{llll|llll}
 \hline
 & &  Normal fit & &  & Lognormal fit & &\\
   Obs & $\mu \mid \mu_{1}$ & $\sigma \mid \sigma_{1}$ & $\chi^{2}/{\rm dof}$ & $m \mid m_{1}$ & $s \mid s_{1}$ & $\chi^{2}/{\rm dof}$ \\
       & $\mu_{2}$ & $\sigma_{2}$                &                      & $m_{2}$ &$s_{2}$ & & \\
  \hline
0080940101 & 74.04 & 2.84 & 134.04/60 & 4.31 & 0.04 & 139.71/60\\
           & 87.41 & 2.46 &  & 4.47 & 0.03 &\\
0080940301 & 76.42 & 2.10 & 64.23/33 & 4.16 & 0.03 &68.04/33\\
0124930101 & 84.43 & 2.55 & 30.16/15 & 4.44 & 0.03 &27.42/15\\
0124930201 & 86.90 & 3.40 & 79.63/42 & 4.47 & 0.04 &76.23/42\\
0124930301 & 113.91& 11.20 & 39.16/18 & 4.74 & 0.10 &50.64/18\\
0124930501 & 49.84 & 3.10 & 93.758/53 & 3.91 & 0.06 & 94.06/53\\
           & 80.60 & 2.24 &           & 4.39 & 0.03 &\\
0124930601 & 32.86 & 5.96 & 181.58/46 & 3.51 & 0.17 &165.95/46\\
0158960101 & 30.57 & 2.37 & 111.87/43 & 3.42 & 0.08 &111.45/43\\
0158960901 & 34.14 & 4.70 & 112.25/46 & 3.54 & 0.14 &113.35/46\\
0158961001 & 45.07 & 2.47 & 48.56/46 & 3.81 & 0.06 &52.95/46\\
0158961101 & 76.97 & 2.75 & 49.48/40 & 3.91 & 0.04 &48.44/40\\
0158961301 & 77.96 & 3.91 & 50.93/41 & 4.36 & 0.05 &47.56/41\\
           & 97.23 & 6.74 &          & 4.58 & 0.07 &\\
0158961401 & 33.10 & 1.33 & 34.93/43 & 3.50 & 0.04 &35.28/43\\

0411780101 & 46.12 & 1.17 & 38.88/53 & 3.83 & 0.03 &39.33/53\\
           & 50.18 & 1.06 &  & 3.92 & 0.02 & \\       
0411780201 & 66.21 & 10.13 & 50.29/44 & 4.20 & 0.15 &53.56/44\\
0411780301 & 87.84 & 3.11 & 86.68/57 & 4.48 & 0.04 &87.55/57\\
           & 104.12 & 2.64 &           & 4.65 & 0.03 &\\
           
0411780401 & 64.00 & 3.25 & 68.94/55 & 4.16 & 0.05 & 67.83/55\\
           & 76.37 & 2.33 &          & 4.34 & 0.03 &\\
           
0411780501 & 32.48 & 1.80 & 46.48/28 & 3.65 & 0.03 & 44.52/28\\
           & 38.60 & 1.36 &  & 3.48 & 0.06 &\\
           
0411780601 & 49.77 & 4.11 & 131.88/43 & 3.91 & 0.08 &113.33/43\\
0411780701 & 13.87 & 0.69 & 20.77/11& 2.63 & 0.05 & 15.12/11 \\
0411782101 & 23.27 & 2.98 & 24.49/23 & 3.15 & 0.13 &32.67/23\\
0727770901 & 31.75 & 1.67 & 65.53/41 & 3.46 & 0.05 &69.47/41\\\hline
\end{tabular}

\end{table*}

\subsection{Histograms and Log-normality}
The analysis of flux distribution of variable astrophysical sources offers valuable insights into their variability characteristics, including emission states and the underlying physical processes governing overall emission. Probability distribution functions (PDFs) that describe the shape of the flux distribution are believed to be indicative of the physical mechanisms involved. Typically, a normal flux distribution is expected if the overall emission results from additive processes, whereas a lognormal flux distribution often implies the presence of multiplicative processes. Studies of the flux distribution in both long- and short-term light curves have been conducted for numerous blazars across various energy and frequency bands \citep{2023ApJ...955..121D,2022MNRAS.510.5280M,2021ApJ...923....7B,2020ApJ...897...25B,2020ApJ...891..120B,2017ApJ...849..138K}.



We present histograms (see Figure \ref{histo} and Table \ref{table3}) for all selected observations of PKS~2155--304, fitting normal, lognormal, and in some cases double lognormal distributions. The expression for the normal distribution, where $\mu$ and $\sigma$ represent the mean and standard deviation, is given as:
\begin{equation}
        N(x) = \frac{1}{\sqrt{2\pi\sigma^2}} {\exp\left({-\ \frac{(x-\mu)^2}{2\sigma^2}}\right)}{\ \rm ,}
\end{equation}
Similarly the expression for binormal distribution composed of two sub-populations having means $\mu_1$ and $\mu_2$ and standard deviations $\sigma_1$ and $\sigma_2$ is 
\begin{multline}
    N_{\rm bimodal}(x) = \dfrac{1}{\sqrt{2\pi}\sigma_1}\exp\left(-\dfrac{(x-\mu_1)^2}{2\sigma_1^2}\right)\\
    +\dfrac{1}{\sqrt{2\pi}\sigma_2}\exp\left(-\dfrac{(x-\mu_2)^2}{2\sigma_2^2}\right),
\end{multline}

The lognormal distribution with location parameter $m$ and scale parameter $s$ is given as given as:
\begin{equation}
    Ln(x) =  \dfrac{1}{\sqrt{2\pi}sx}\exp\left(-\dfrac{(\ln x-m)^2}{2s^2}\right),
\end{equation}
The bimodal lognormal distribution is given by,
\begin{multline}
    Ln_{\mathrm{\rm bimodal}}(x) = \dfrac{1}{\sqrt{2\pi}s_1x}\exp\left(-\dfrac{(\ln x-m_1)^2}{2s_1^2}\right)\\
    +\dfrac{1}{\sqrt{2\pi}s_2x}\exp\left(-\dfrac{(\ln x-m_2)^2}{2s_2^2}\right),
\end{multline}
where $m_1$, $m_2$ and $s_1$, $s_2$ are location and scale parameters, respectively, for the two superposed lognormal distributions in the bimodal lognormal distribution. Also, the fit parameters obtained from fitting the normal and lognormal distributions are shown in Table \ref{table3}. 

\begin{figure}
    \includegraphics[width=1.\linewidth]{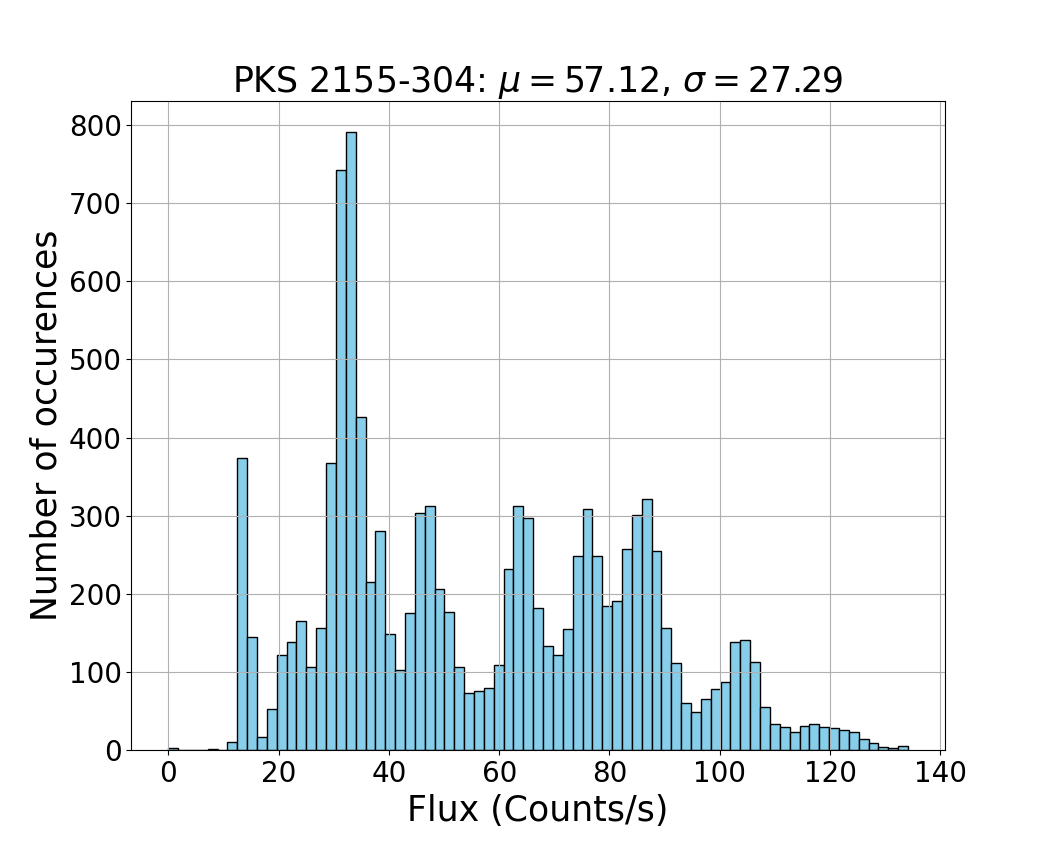}
    \caption{Histogram showing the overall flux distribution of the blazar PKS 2155-304 observed between 2000 and 2014. This distribution illustrates the long-term variability characteristics of the source over the 14-year observational period.} 
    \label{overall_histo}
\end{figure}

%
 Results from our flux histogram analysis show that while a majority of the observations follow a unimodal distribution, there are seven out of 22 observations which follow a bimodal distribution. As can be inferred from the values of $\chi^2/{\rm dof}$, lognormal PDF fits most of the unimodal flux histograms better than with normal PDF. The bimodal flux histograms also fit better with the bimodal lognormal distribution based on the values of reduced $\chi^2$, although marginally.
 
 We found that seven observations (with IDs 0080940101, 0124930501 0158961301, 0411780301 and 0411780301) out of 22 observations exhibit a bimodal distribution. In Figure \ref{histo}, we present the histogram for one of the observations, which shows a clear bimodal distribution. Blazar flux distributions showing bimodal feature are also reported in \cite{2010A&A...520A..83H,2022MNRAS.510.5280M, 2018RAA....18..141S}. The presence of the bimodal distribution suggests a clear distinction between low and high flux states. However, since the light curves often exhibit continuous flux variation, it is difficult to categorize specific flux points as belonging to either a quiescent state or a flaring state using a bimodal PDF.

 In addition, we present the X-ray flux (count rate) histograms for the entire set of observations in Figure \ref{overall_histo}. However, as seen from the figure, this distribution does not resemble either a normal or lognormal probability density function (PDF). The flux distribution is rather complex, with multiple peaks, implying that the overall flux could be a combination of fluxes from different emission states and/or zones.



\subsection{Power spectral density analysis}
Studying variability in AGN using methods based on both the time domain and frequency domain can help uncover its underlying physical processes. The Power Spectral Density (PSD) serves as an important analytical tool in the frequency domain, measuring the `variability power' at a given temporal frequency (or timescale). This enables us to understand the nature of variability and provides clues about its possible origin. The Discrete Fourier Transform (DFT) is widely used as a measure of PSD and is suitable for a discrete set of observations \citep[see][]{2003MNRAS.345.1271V,2012A&A...544A..80G, 2022MNRAS.510.5280M}. If events occur at discrete times
  $t_j$ with $j = 1, 2, ..., n$, and sampled at frequencies $\nu_{\rm min} = \frac{1}{T}$, $2 \nu_{\rm min}$ ... , $\nu_{\rm max} = \frac{1}{2\Delta t}$, DFT be expressed as,
\begin{equation}\label{eq8}
   |DFT(\nu)|^2 = P(\nu) = \dfrac{T}{\Bar{x}^2n^2}\left|\sum_{j=1}^n x(t_j)e^{-i2\pi\nu t_j}\right|^2,
\end{equation}
where $T$ is the length of the light curve, $\Delta t$ is the mean sampling step and $\Bar{x}$ is the mean flux (counts/s) of the lightcurve. Statistical fluctuations in the variability power associated with the detector are commonly referred to as Poisson noise and given by, 
\begin{equation}
P_{\rm Poisson} =
   \frac{T \sigma_{\rm stat}^{2}}{n \Bar{x}^2} \quad    \text{and} \quad
\sigma_{\rm statistical}^2 =
    \sum_{j=0}^{n-1} \frac{(\Delta x_{\rm j})^2 }{n} 
\end{equation}
where $\Delta x_{\rm j}$ represents error in the observed flux at a given time $t_j$. PSD study of the blazar PKS 2155-304 has been carried out assuming powerlaw variation of temporal frequency in both phenomenological as well as physical models using light curves in different energy regimes with various observations like {\it XMM-Newton}, Astrosat, Suzaku, Swift and Fermi-LAT \citep[see][]{2023ApJ...955..121D,2022ApJS..262....4N,2021ApJ...909..103Z, 2020ApJ...897...25B, 2020ApJ...891..120B}. The blazar PSD can be expressed using PLform of the temporal frequency
\begin{equation}
    P(\nu)=A\cdot \nu^{-\beta_{\rm{P}}}+C,
\end{equation}
where {\it A} is normalization constant, $\beta_{\rm{P}}$ is spectral power index and $C$ represents the Poisson noise level as discussed in \citep{2002MNRAS.332..231U, 2018A&A...620A.185N, 2022MNRAS.510.5280M}.
The slope of the powerlaw PSD can provide important clues about the physical processes driving the variability of the blazar.

\begin{figure}
    \includegraphics[width=1.\linewidth]{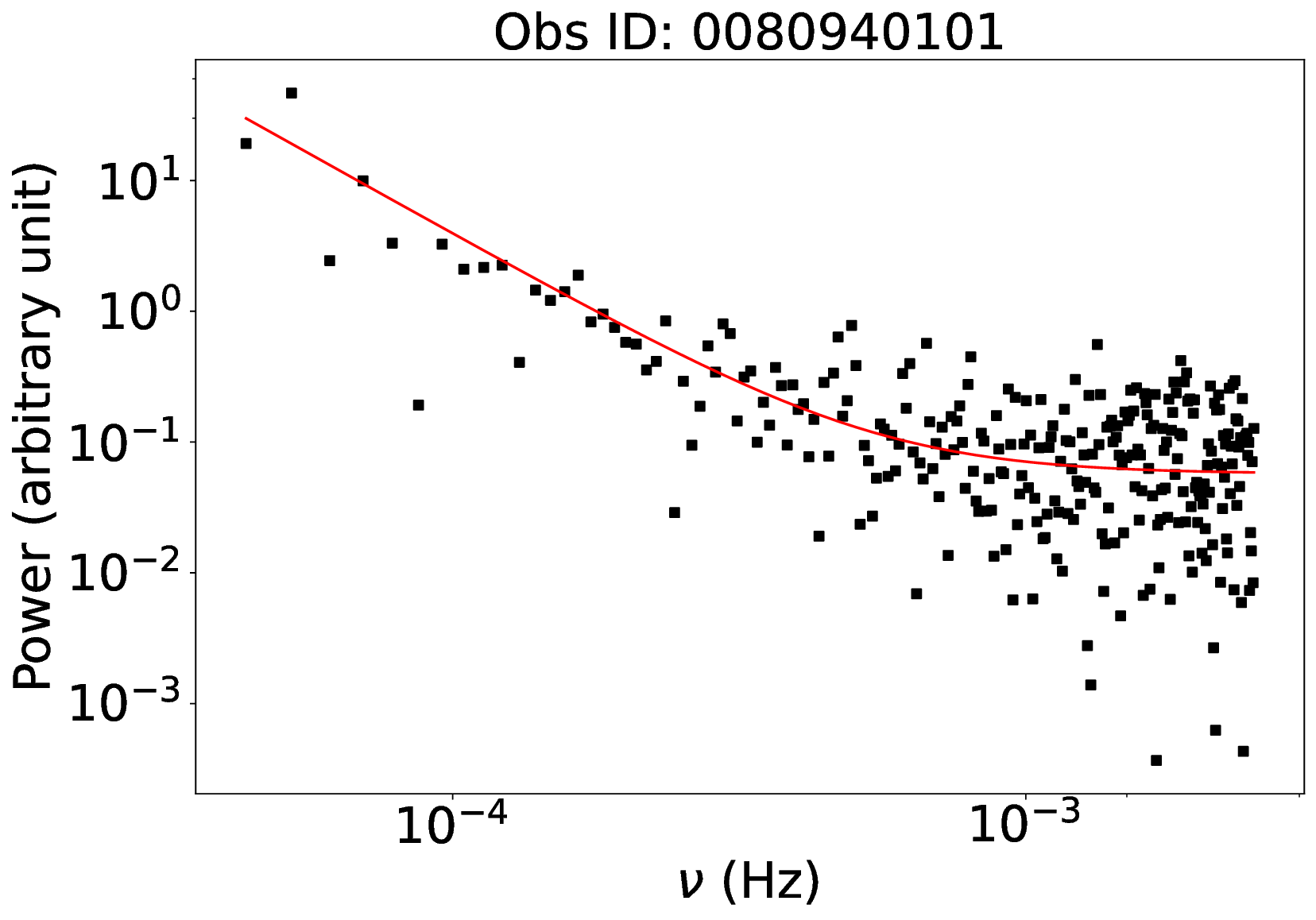}
    \caption{X-ray PSD of one of the observation IDs  on intra-day timescale. The PSD plots for other observations are presented in the Appendix.}
    \label{psd}
\end{figure}

In this study, we performed PSD fit using the optimization algorithm for curve fitting included in the SciPy python library \footnote{\url{https://docs.scipy.org/doc/scipy/reference/generated/scipy.optimize.curve_fit.html}}.
 Our analysis of the {\it XMM-Newton} X-ray light curves of PKS 2155-304 reveals a predominance of red noise, which is well characterized by a PL model. The PSD slopes ($\beta_\mathrm{P}$), as illustrated in Figure \ref{psd} and quantified in column 8 of Table \ref{table2}, exhibit significant variability across the 22 observations of PKS 2155-304. The mean $\beta_\mathrm{P}$ value is 2.49, with a range from 0.922 (observation 0411780701) to 4.491 (observation 0411780201). These variable PSD slopes correspond to different stochastic processes: flicker noise ($\beta_\mathrm{P} \sim 1$), red noise ($\beta_\mathrm{P} \sim 2$), and other noise processes with steeper PL index, that is, ~ $\beta_\mathrm{P}>2$ \citep{1978ComAp...7..103P}.

\subsection{SPECTRAL ANALYSIS}

\subsubsection{HARDNESS RATIO}
To study the spectral variability of the X-ray emission from the source light curves are produced in two energy bands: a soft band between 0.3--2 keV and a hard band between 2--10 keV. The blazar hardness ratio is often used to study the spectral properties of blazars and to search for spectral changes related to changes in the physical conditions in the jet or the accretion process onto the central supermassive black hole. It can also be used to classify blazars based on their spectral properties and to study the evolution of the emission from these objects over time. For this work, the hardness ratio is defined as
\begin{equation}
\label{hard}
    HR = \frac{H}{S},
\end{equation}
where $H$ and $S$ are the flux (counts/s) in the hard (2--10 keV) and soft (0.3--2 keV) bands, respectively. The hardness ratio is a commonly used model-independent method to study spectral variations over time and flux states. Also, the uncertainty in the $HR$ ($\sigma_{\rm HR}$) is estimated as,
\begin{equation}
    \sigma_{\rm HR} = \frac{2}{S^2}\sqrt{H^{2}\sigma^{2}_{\rm S} + S^{2}\sigma^{2}_{\rm H}}.
\end{equation}
where $\sigma_H$ and $\sigma_S$
are errors in hard and soft bands, respectively.

\begin{figure}
    \includegraphics[width=.9\linewidth]{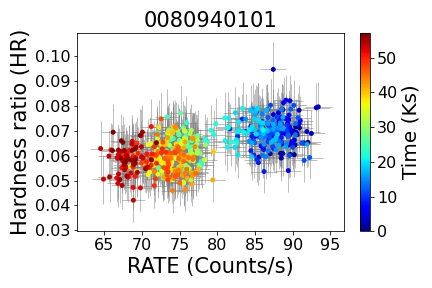}
    \caption{The source hardness ratio is plotted against flux for one of the observation IDs. To trace the time evolution, the symbols are color-coded according to the time of observation. Similar plots for the rest of the observations are presented in the Appendix.}
    \label{HR}
\end{figure}

\begin{figure}[!b]
    \centering
    \includegraphics[width=9.0cm]{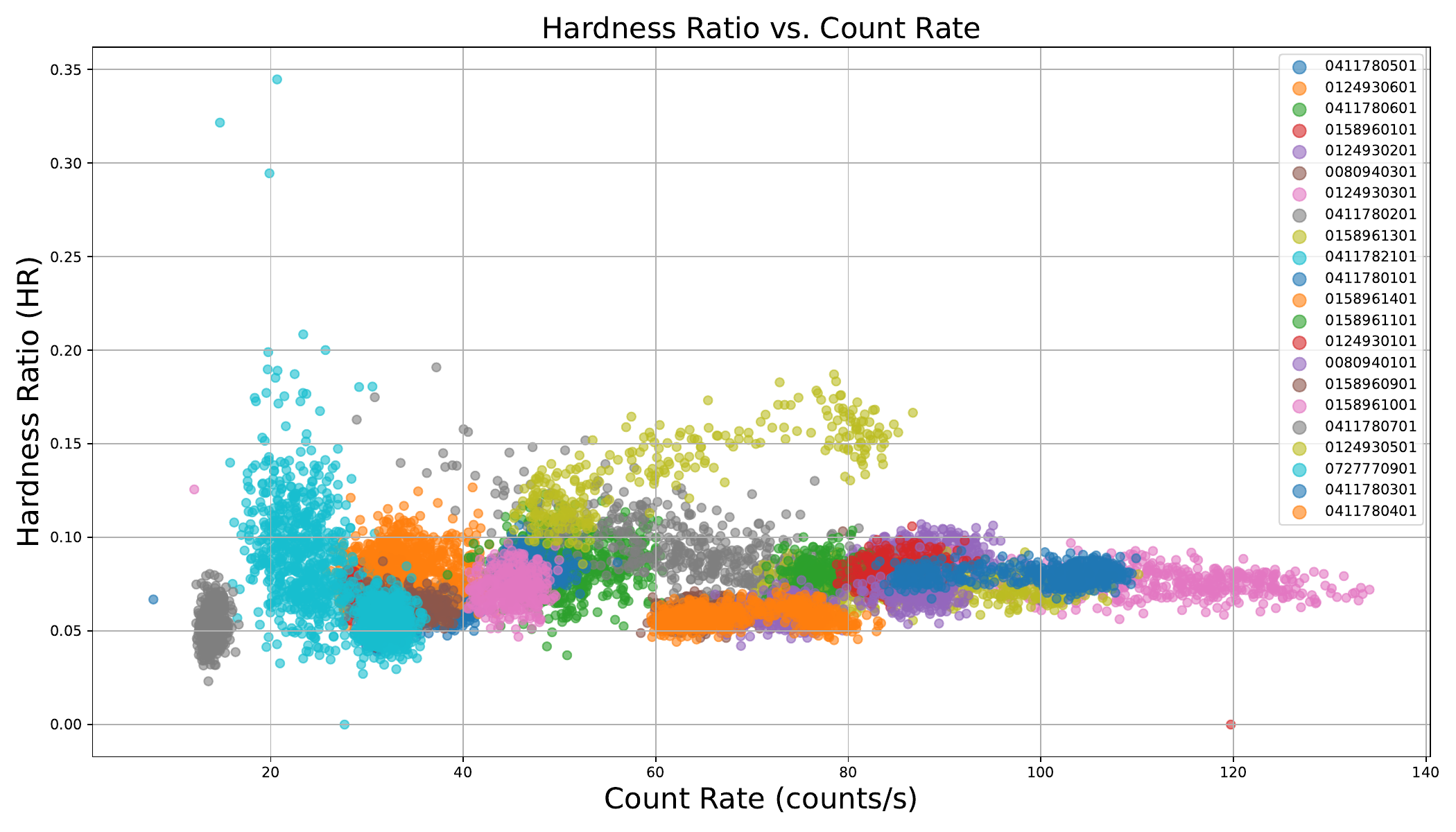}
    \caption{The figure shows the distribution of hardness ratios against count rate for the blazar PKS 2155-304, spanning more than a decade of observational data. The hardness ratio, as defined in Equation \ref{hard}, is plotted on the y-axis, with the corresponding count rate on the x-axis. Each data point represents a single observation, with colors corresponding to unique observation IDs. }
    \label{fig:FullHR}
\end{figure}

\begin{figure}[!t]
    \includegraphics[height=.99\linewidth, angle=0]{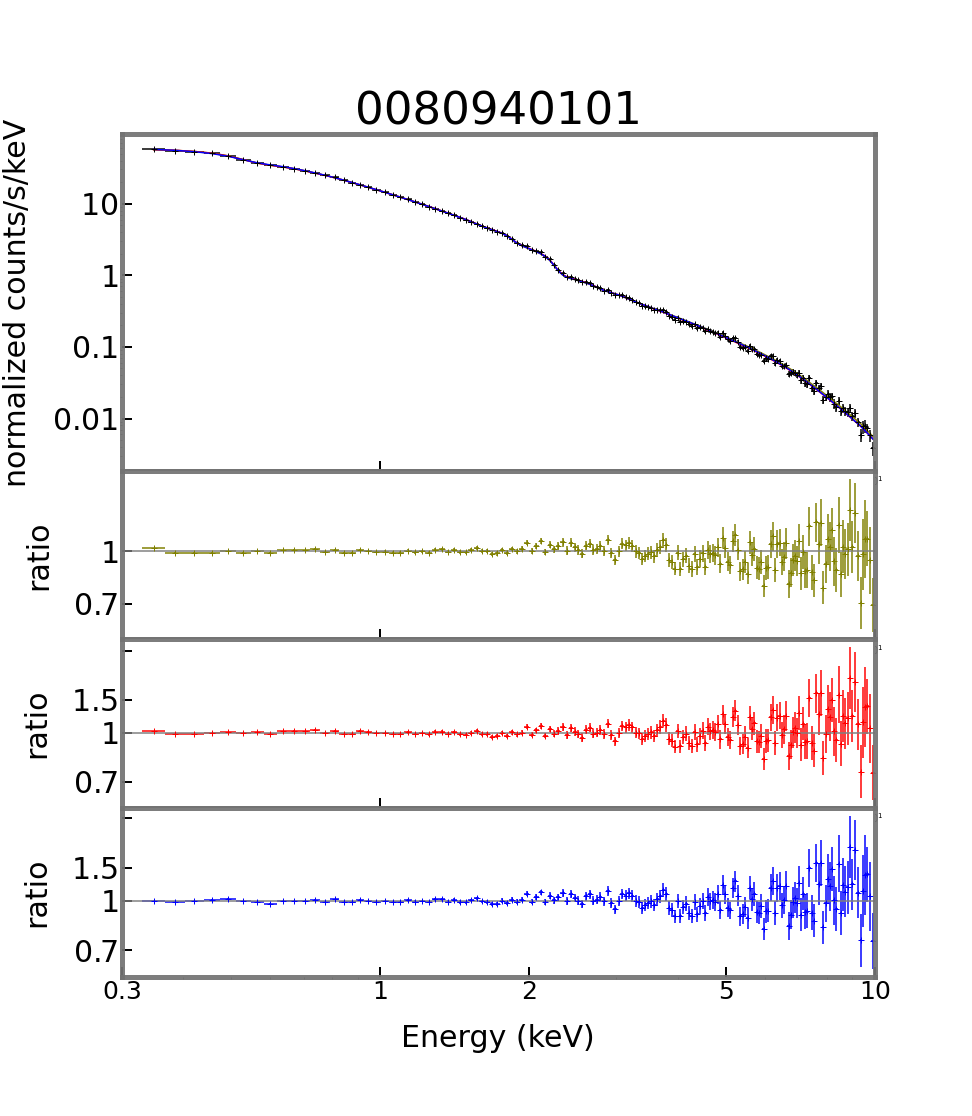}
    \caption{ The spectral ﬁtting of one of the observations of PKS 2155–304 in 0.3–10 keV. Each spectra is ﬁtted using the powerlaw, broken
PL and log parabolic model and the data-to-model ratio is shown in the three subpanels for each spectra in olive, red and blue
colors, respectively.} 
    \label{spectra}
\end{figure}

It is interesting to note that although we observed pronounced flux variability throughout the observations, spectral variations measured by HR exhibit a complex pattern, with no significant variation in the majority of observations.  Obs. IDs 0080940101,  0124930501, and 0411780501 reveal a trend (or a weak trend) of \emph{harder-when-brighter}, whereas, Obs. ID 0411780201 shows a trend of \emph{softer-when-brighter}  \citep[e. g., see][]{2018A&A...619A..93B}. In many observations, the error in HR at lower count rates is also higher, thus not providing conclusive HR variations with count rate. Additionally, Figure \ref{fig:FullHR} presents the distribution of HR across the entire set of observations as a function of count rate. Note that different observational epochs are plotted in distinct colors.
Although the data reveal no simple correlation between flux and HR, it is observed that long-term variability is characterized predominantly by achromatic variations, as indicated by the nearly flat behavior of HR distribution, suggesting that the spectral shape remains largely stable despite changes in overall flux. In contrast, chromatic variations, where the spectral shape undergoes considerable changes, are primarily observed on shorter timescales. Similar spectral behavior has been observed for a few sources in the optical band \citep[see e.g.,][]{2023MNRAS.522..102R,2024A&A...686A.228O}.

\subsubsection{SPECTRAL FITTING}
\label{spec_fit}
We also performed the spectral analysis of the X-ray emission from blazar PKS 2155―304 by extracting the spectra of the observations within the energy range 0.3--10 keV and, consequently, fitted them with  PL models from \emph{XSPEC}\footnote{\url{https://heasarc.gsfc.nasa.gov/xanadu/xspec/manual/XspecManual.html}}. To accurately characterize the underlying spectral shape of the emission, we employed the following three different PL models. \\
(i) {Power law (PL)}: 
This is a simple PL model in which the photon count rate distribution over the energy range is expressed solely by a single power index. The model can be written as
\begin{equation}
    \dfrac{dN}{dE} = K\cdot E^{-\Gamma},
\end{equation} 
where $K$ and $\Gamma$ represent normalization constant and photon index, respectively.\\
(ii) {Broken power law (BPL)}: In a BPL  model, a spectral shape is characterized by two PL indices that meet at a discontinuity at an energy point known as the break energy. The distribution can be expressed as:
\begin{equation}
    \dfrac{dN}{dE} = 
    \begin{cases}
      K\cdot E^{-\Gamma_1} & \text{if $E\geq E_{b}$},\\
      K\cdot E^{-\Gamma_2} & \text{otherwise},
    \end{cases}
\end{equation}
Here $K$, $\Gamma_{1}$, $\Gamma_{2}$ and $E_b$ are normalization constant, low- and high-energy photon indexes and break energy, respectively.\\
(iii) Log parabola (LP): LP is one of the models widely used in the spectrum analysis of blazars \citep{2004A&A...413..489M,2006A&A...448..861M}. The distribution function for LP is
\begin{equation}
\label{eq:eplp}
\dfrac{dN}{dE} = K\cdot10^{-\beta\left(\log\ (E/E_{\rm p})\right)^2}/E^2,
\end{equation}
where $K$, $E_p$ and $\beta$ are the normalization constant, peak energy and curvature parameter.

In this study, we fit the X-ray spectra obtained from \textit{{\it XMM-Newton}} observations listed in Table \ref{table1} using three models available in the X-ray fitting package, \textit{xspec}. To minimize instrumental artifacts, we restricted our analysis to the 0.3--10 keV energy range. We initially applied the simple PL model, which failed to yield a significant reduced chi-square ($\chi^2_r$). To improve the fit, we then applied more complex models: a BPL and LP. Both these models resulted in significantly improved $\chi^2_r$ values. Furthermore, we noted spectral curvature at lower energies. To accurately capture this feature, we utilized the LP model, a variant of the standard log-parabola model described in Equation \ref{eq:eplp}. The results from our spectral fits are presented in Table \ref{table4}, and a representative fitted spectrum from one observation is shown in Figure \ref{spectra}.
The spectral fits for the remaining observations are presented in the Appendix.
Our results indicate that the BPL and LP models each fit distinct subsets of observations better than the PL model, achieving better overall fits for the complete dataset and suggesting the presence of spectral curvature.
A similar investigation of the spectral curvature of the PKS 2155-304 source was conducted by \citet{2017ApJ...850..209G}, utilizing {\it XMM-Newton} data from 2009 to 2014 and employing log-parabola model fitting techniques.

\begin{table*}
\centering
\caption{X-ray spectral properties of the blazar PKS 2155-304. Col. 2: observation ID; Col. 3: Spectral models, {\it power law} (PL), broken {\it power law} (BPL); Col 4: photon index (PL), high-energy photon index (BPL);  Col. 5:  curvature parameter (LP), low-energy photon index (BPL); Col. 6: break/peak energy in keV; Col. 7: $\chi^2$/degrees of freedom;  .}
\label{table4}
\begin{tabular}{lllllllll}
  \hline
  & Obs .ID & Model &  $ \alpha | \Gamma1 $ & $\beta | \Gamma2$ & Break energy/Peak Energy & \\ 
 && &   & & $E_{b}/E_{p}$ (keV) & $\chi^2_{r}$ &\\
  \hline
  1 & 0080940101 & PL  & 2.759±0.004 & -- & -- & 1.618 (266.9/165) &  \\
      && \textbf{BPL}  &  2.729±0.006 & 2.830±0.016 & 2.4668±0.224 & 1.275 (214.2/168) &   \\
      && LP & - & 0.104±0.010 & 0.0006±0.0004 & 1.368 (224.4/164) &\\
      
  2 & 0080940301  & PL  &  2.827±0.003 & -- & -- & 2.678 (450/168) & \\
      && BPL  &  2.791±0.004 & 2.933±0.013 & 2.6257±0.137 & 1.774 (294.5/166) &  \\
      && \textbf{LP} & - & 0.127$\pm$0.009 & 0.0012±0.0006 & 1.567 (261.7/167) &\\
  
  3 & 0124930101  & PL  & 2.597±0.003 & -- & -- & 4.212 (720.3/171) &  \\
      && \textbf{BPL}  & 2.526±0.005 & 2.733±0.01 & 2.381±0.073 & 1.455 (245.9/169) &  \\
      && LP & - & 0.162±0.008 & 0.0322±0.0070 & 2.291 (389.6/170)  &\\
      
  4 & 0124930201  & PL  &  2.546±0.002 & -- & -- & 4.837 (841.6/174) & \\
      && \textbf{BPL}  &  2.487±0.003 & 2.671±0.008 & 2.502±0.07 & 1.502 (258.4/172) & \\
      && LP & - & 0.134±0.006 & 0.0211±0.0047 & 2.747 (475.3/173) &\\

  5 & 0124930301  & PL   & 2.750±0.002 & -- & -- & 18.531 (3168.8/171) &   \\
      && \textbf{BPL}  & 2.624±0.03 & 3.007±0.008 & 2.340±0.030 & 2.275 (384.5/169) &  \\
      && LP & - & 0.312±0.006 & 0.1341±0.0075 & 4.048 (688.1/170) &\\
      
 6 & 0124930501 & PL   & 2.387±0.003 & -- & -- & 2.438 (416.9/171) &  \\
      && BPL   & 2.295±0.003 & 3.007±12560.5 &  10.127±192340 & 11.50 (1944.6/169) & \\
      && \textbf{LP} & - & 0.141±0.011 & 0.1077±0.0269 & 1.566 (266.3/170) &\\
      
  7 & 0124930601  & PL  & 2.826±0.004 & -- & -- & 3.116 (517.3/166) &  \\
      && BPL  & 2.747±0.007 & 2.956±0.013 & 2.309±0.087 & 1.465 (240.3/164) &   \\
      && \textbf{LP} & - & 0.213±0.012  & 0.0265±0.0068 & 1.379 (227.5/165) &\\
      
  8 & 0158960101  & PL  & 2.867±0.006 & -- & -- &   1.251 (192.7/154) & \\
      && BPL &  2.825±0.013 &  2.921±0.018 & 2.119±0.263 & 1.126 (171.1/152) &  \\
      && \textbf{LP} &  & 0.100±0.022 & 0.0001±0.0002 & 1.108 (169.6/153) &\\
      
  9 & 0158960901 & PL  & 2.998±0.006 & -- & -- & 1.451 (220.6/152) &  \\
      && \textbf{BPL}  &  2.928±0.013 & 3.065±0.015 & 1.885±0.142 & 1.105 (165.8/150) &  \\
      && LP & - & -0.305±0.020 & 99.9926±24.7221 & 5.007 (756.2/151) &\\
      
  10 & 0158961001 & PL & 2.885±0.004 & -- & -- & 2.516 (412.7/164) &  \\
      && BPL  & 2.812±0.008 & 2.975±0.013 &   2.115±0.107 & 1.653 (267.8/162) &  \\
      && \textbf{LP}  & --- & 0.188±0.015 & 0.0105±0.0045 & 1.546 (252/163) &  \\
      
  11 & 0158961101 & PL  & 2.578±0.003 & -- & -- & 1.791 (306.2/171) &  \\
      && BPL  & 2.880±0.004 & 2.601±0.004 & 0.488±0.014 & 1.442 (246.7/171) &  \\
      && \textbf{LP} & --- & 0.091±0.010 & 0.0014±0.0010 & 1.352 (231.3/171) &  \\
      
  12 & 0158961301 & PL   & 2.621±0.002 & -- & -- & 4.065 (703.2/173) &  \\
      && \textbf{BPL} &  2.601±0.003 & 2.762±0.008 & 2.414±0.077 & 1.508 (265.5/176) &  \\
      && LP & --- & 0.125±0.006 & 0.0053±0.0016 & 2.497 (429.5/172) &  \\
  
  \hline
\end{tabular}
\end{table*}
 
  \begin{table*}
  \centering
  Continuation of Tab.~\ref{table4}
  \label{subsec:tables1}
  
\begin{tabular}{llllllllll}
  \hline
  13 & 0158961401 & PL  &  2.562±0.003 & -- & -- & 1.764 (301.6/171) &  &\\
      && \textbf{BPL}  &  2.577±0.004 & 2.366±0.037 & 4.4856±0.295 & 1.418 (239.7/169) &  & \\
      && LP  & --- & 0.039±0.011 & 0.0018±0.0015 & 3.111 (529/170) &  &\\
      
  14 & 0411780101 & PL   & 2.514±0.004 & -- & -- & 1.276 (211.9/166) & &\\
      && \textbf{BPL}  & 2.571±0.012 & 2.498±0.006 & 1.272±0.097 & 1.077 (176.6/164) &  &\\
      && LP & - & 0.052±0.011  & 0.0000±0.0000 & 1.664 (274.6/165) & &\\
      
  15 & 0411780201 & PL  & 2.664±0.002 & -- & -- & 1. 772 (304.8/172) & &\\
      && \textbf{BPL} &  2.666±0.003 & 2.485±0.083 &  6.372±0.698 & 1.732 (294.4/170) & &\\
      && LP & - & 0.053±0.007  & 0.0000±0.00000 & 1.940 (331.9/171) & &\\
      
  16 & 0411780301 & PL  & 2.628±0.002 & -- & -- & 6.838 (1189.9/174) & &\\
      && BPL &  2.628±0.002 & 2.659±-1.00 & 45.694±-1.00 & 6.918 (1189.9/172) & &\\
      && \textbf{LP} & - & 0.172±0.006 & 0.0346±0.0066 & 3.005 (519.9/173) & &\\
      
  17 & 0411780401 & PL  & 2.847±0.003 & -- & -- & 4.012 (682.1/170) & &\\
      && BPL &  2.847±0.003 & -1.755±-1.00 & 1.89382±-1.00 & 4.078 (685.1/168) & &\\
      && \textbf{LP} & - & 0.167±0.008 & 0.0062±0.0018 & 2.004 (338.8/169) & &\\
      
  18 & 0411780501 & PL  & 2.818±0.003 & -- & -- & 2.08 (345.2/166) & &\\
      && \textbf{BPL} &  2.773±0.006 & 2.895±0.011 & 2.126±0.127 & 1.444 (236.8/164) & &\\
     && LP & - & 0.153±0.011 & 0.0045±0.0020 & 1.529 (252.4/165) & &\\
      
  19 & 0411780601 & PL &  2.573±0.002 & -- & -- & 3.434 (590.8/172) & &\\
      && \textbf{BPL} & 2.516±0.004 & 2.681±0.009 & 2.394±0.086 & 1.450 (247/170) & &\\
      && LP & - & 0.144±0.008 & 0.0229±0.0058 & 1.866 (319.2/171) & &\\
      
  20 & 0411780701 & PL   & 2.901±0.007 & -- & -- & 1.025 (153.8/150) & & \\
      && \textbf{BPL}   & 2.836±0.003 & 2.893±0.008 &  0.722±0.106 & 0.975 (149.2/153) & &\\
      && LP & - & -0.189±0.007 & 99.9990±17.7507 & 15.832 (2359.1/149) & &\\
     
 21 & 0411782101 & PL  & 2.799±0.003 & -- & -- & 2.212 (365.1/165) & &\\
      && \textbf{BPL}  &  2.759±0.006 & 2.886±0.013 & 2.288±0.151 & 1.592 (267.5/168) & &\\
      && LP & - & 0.143±0.011 & 0.0034±0.0017 & 1.815 (297.8/164) & &\\
      
  22 & 0727770901 & PL & 2.907±0.004 & -- & -- & 2.246 (370.6/165) & &\\
      && \textbf{BPL}  &  2.860±0.006 & 3.022±0.015 & 2.305±0.126 & 1.431 (229/160) & &\\
      && LP & - & -0.267±0.010 & 99.9560±15.6286 & 9.621 (1549.1/161) & &\\

  \hline
\end{tabular}
\end{table*}


\section{Discussion}
\label{sec:5}
Blazars exhibit multi-timescale and multi-wavelength variability, which are among their most remarkable features, captivating both theorists and observers alike. This section provides a comprehensive summary of our analyses, covering various aspects of blazar behavior, including properties of light curves across different energy bands, hardness ratio, flux histograms, PSD analysis, and source spectra fitted with standard XSPEC models. Together, these analyses offer valuable insights into the complex nature of blazar emission and variability.
\subsection{Timing Analysis}
Variability:  The source displays substantial variability both on intra-day and long-timescale. As indicated in Table \ref{table2}, on intra-day timescales, it exhibits an average fractional variability of $\sim 6\%$, with one of the observations showing variability as high as 20\%. During the entire observational period, the fractional variability reaches as high as $\sim 47\%$. In other words, the variability exhibits significant modulation with maximum flux levels approximately eight times higher than the minimum observed values.

The X-ray emissions of TeV blazars exhibit significant variability at intra-day timescales. This could possibly be attributed to the fact that the observed X-ray emission in HBLs is associated with synchrotron emission by a population of electrons located at the high-energy tail within the synchrotron component of their SEDs.
However, it is also possible that LSPs and FSRQs also show intense X-ray variability. However, the number of these sources detected at TeV energies is much smaller, leading to a sample of TeV blazars that is heavily biased toward HSP blazars. This suggests that strong X-ray variability may be a general characteristic of the broader blazar population.

The IDV can be explained in the scenario of the turbulent jet of the blazar. In this framework, the observed variability can be interpreted as a superposition of fluctuations arising from numerous small-scale inhomogeneous components with varying Doppler boosting. Furthermore, as shocks propagate through the turbulent jet, energy dissipates from the largest to the smallest cells. Emission from some of these dominant smaller cells can contribute to short timescale variability.
\citep{2015JApA...36..255C,2014ApJ...780...87M,2013A&A...558A..92B}. Using the variability timescale and the causality argument, the spatial scale of the inhomogeneity in the jet can be related to the size of the turbulent cells. An upper limit estimate for size of the compact spherical cells at redshift ($z$), corresponding to the minimum variability timescale ($\tau_{\rm var}$) and Doppler factor ($\delta$), can be computed using the following relation:
\begin{equation}
    R\leq\frac{\delta}{(1+z)}\cdot{c\tau_{\rm var}},
\end{equation}
where $\delta$ and $z$ represent bulk Doppler factor and redshift parameter, respectively.
 The lower limit of the Doppler factor associated with the source is of the order $\delta \sim$10 \citep{2014A&ARv..22...73F, 2020ApJS..247...27K}, which corresponds to a size of the emitting central region $R \sim 10^{14}$ cm using an average of minimum variability timescale over all the observations 2.0 ks. Similar results using Suzaku and NuSTAR observations are discussed in \citep{2018A&A...619A..93B, 2021ApJ...909..103Z}.
 Furthermore, the variability timescales can be linked with the characteristics time scales such as cooling timescales of the source. In this particular case, the synchrotron cooling timescale in the observer's frame is
\begin{equation}
  t_{\rm cool}(\gamma) \simeq 7.74 \times 10^{8}\frac{(1+z)}{\delta B^2 \gamma} \ \  {\rm s},
\end{equation}
where $\gamma$ and $B$ represent bulk  electron Lorentz factor and magnetic field in Gauss, respectively. An average cooling timescale of 2 ks, in combination with a magnetic field strength of B = 1 Gauss, typical for the jet's magnetic field, results in an average electron energy of 4.32 × 10$^4$ Lorentz factors, enabling X-ray emission in the {\it XMM-Newton} band.

A linear RMS-flux relation is widely observed in AGNs and microquasars. This correlation is often linked to the multiplicative, non-linear stochastic processes driving the observed variability, which result in a flux distribution skewed toward higher flux levels, such as a lognormal PDF \citep{2005MNRAS.359..345U}. Conventionally, the linear RMS-flux relation, accompanied by a lognormal flux distribution, is attributed to variations in the viscosity parameter, $\alpha$, of the accretion disk. These fluctuations, driven by changes in viscosity at various radii, propagate outward and modulate mass accretion rates on larger scales \citep{1997MNRAS.292..679L}.
In the jet scenario, simulations of multi-frequency light curves demonstrate that this relationship can emerge when internal shocks of varying amplitudes occur. The shock strength is determined by the relative velocities of colliding plasma blobs within the jet \citep{2022MNRAS.510.3688K}.
However, as demonstrated in Figure \ref{rmsflux}, an analysis of the long-term observations reveals neither a linear RMS-flux relation nor any other definitive trend in variability behavior. This departure from a clear trend may be attributed to the complex interplay of multiple processes, including stochastic particle acceleration \citep[e.g.,][]{2010MNRAS.402.1649G,2014ApJ...780...87M}, radiation propagation through curved jet geometries \citep[e.g.,][]{2017Natur.552..374R,2018Galax...6..136B,2010Natur.463..886Y}, and the intricate coupling between jet dynamics and the central engine \citep[e.g.,][]{2012MNRAS.423.3083M,2015MNRAS.447..327T}.

The intra-day flux distributions of the blazar were analyzed using countrate distributions. Of the 22 observations, seven exhibited bimodal Gaussian behavior, while the remaining 15 showed unimodal Gaussian characteristics. The overall X-ray flux histogram revealed a complex structure with multiple peaks, suggesting that the distribution deviates from a standard PDF. In the context of long-term MWL variability studies, these findings can be compared with previous studies where optical and gamma-ray light curves demonstrated a strong correlation, and their respective flux distributions were best characterized by log-normal PDFs \citep{2021ApJ...923....7B,2020ApJ...891..120B}. In the AGN literature, observations following log-normal PDFs are often interpreted as evidence that the observed variability arises from non-linear and multiplicative processes within the source.

On the other hand, the strong correlation between optical and gamma-ray emissions in blazars may be attributed to the fact that optical and Fermi/LAT gamma-ray observations are located near the synchrotron and high-energy peaks of the SED, respectively. However, X-ray emissions lie closer to the tail of the synchrotron component, probing the highest-energy electron population. The absence of a simple PDF in the X-ray regime suggests that the emission likely arises from multiple components and involves various emission processes within the source.

The PSDs of astrophysical time series data are often characterized by various types of noise, which typically follows a PL relationship of the form $S(f) \propto 1/f^\beta$, where f is the frequency and $\beta$ is the PL index.
\citep[see e. g.,][]{2003MNRAS.345.1271V,2004MNRAS.348..783M,2016ApJ...831...92B}. We examined the distribution of the strength of source variability across different timescales, helping to identify the dominant timescales present in a light curve. By employing the PSD method on the long-term optical and gamma-ray light curves a power spectral break at the characteristic timescale of $\sim$ 3 years \citep{2011A&A...531A.123K} and optical and gamma-ray QPO at the timescale of $\sim$ 600 days \citep{2021ApJ...923....7B} have been observed. The PSD analysis performed on the 22 X-ray observation of the source reveals that the intra-day observations exhibit varying PSD slopes across different observation IDs, indicating presence of different noises processes such as flicker-noise, red-noise processes etc. Similar variable slope index were observed during the intra-day observations of various sources including PKS 2155-304 \citep[][]{2023ApJ...955..121D,2022MNRAS.510.5280M,2021ApJ...909..103Z,2010ApJ...718..279G}.

Our analysis reveals that the observed PSD slopes ($\beta_P$) for PKS 2155-304 predominantly fall within the red noise regime $(2 \leq \beta_P \leq 4)$, with occasional instances approaching pink noise $(\beta_P \approx 1)$.  
These variable PSD slopes might suggest distinct local manifestations of stochastic processes occurring within the turbulent jet structure. However, it is also possible that the long-term light curve is actually characterized by a single flatter PL PSD. In this case, the intra-day observations represent only a small segment observed across different flux states, resulting in apparent variations in the local PSD slopes. However, we did not estimate the long-term PSD by combining all the observations, as such a PSD would be affected by the large gaps between observations spanning years. This would result in an inaccurate characterization of the long-term X-ray variability in the source.

Furthermore, steep PSD slopes $(\beta_P > 2)$ are associated with strongly persistent processes or those exhibiting long-term memory effects. These processes show even stronger low-frequency dominance than traditional red noise. In some contexts, they may indicate non-stationary behavior present in the system. It is important to note that as $\beta$ increases beyond 2, the integral of the PSD can diverge, potentially leading to infinite variance. This can violate assumptions of stationarity in some statistical analyses.

\subsection{Spectral Analysis} 
In general, AGN spectra may be composed of various emission components, such as soft X-ray excess, neutral iron emission lines around 6.4 keV, and parts of either a synchrotron hump or a Compton hump. However, blazar spectral emission is primarily jet-dominated due to beamed electromagnetic radiation and the non-thermal emission in blazars usually follows a PL spectral profile \citep{2005A&A...442..895A}. Studying the X-ray spectra of blazars helps us understand the particle acceleration mechanisms at play, the properties of the jet environment, and the overall emission processes that power these luminous objects. The excellent sensitivity of {\it XMM-Newton} provides detailed information about the emission mechanisms, which we obtain by fitting the source X-ray spectra.

A notable characteristic of our observations is the contrast between flux and spectral behavior. While pronounced flux variability was detected throughout the observing period, significant spectral variations, as measured by the HR, were observed in only two out of the total observations. The majority of observations exhibited no substantial spectral changes despite clear flux variations.
To investigate this spectral behavior in greater detail, we performed comprehensive spectral analysis using three widely used {\it XSPEC} spectral models: PL, BPL, and LP. These models are chosen to represent non-thermal emission from relativistic electrons in the magnetic field of the blazar jet.
In the synchrotron emission scenario, a simple PL spectrum can be interpreted as coherent synchrotron emission produced by a simple PL distribution of particles \citep{1986rpa..book.....R}. However, the BPL and LP models suggest more complex scenarios. A BPL might indicate that the observed emission is contributed by multiple populations of particles with different distributions. On the other hand, log-parabolic energy spectra for relativistic electrons emitting synchrotron radiation can be generated through a stochastic acceleration mechanism, where the probability of acceleration varies with energy \citep[e. g., see][]{2024PhRvD.109f3006L,2007A&A...466..521T,2006A&A...448..861M}. 

In the spectral analysis, the model that best represents the observations can be inferred from the reduced chi-squared (\(\chi^2_r\)) values from the spectral fitting. It was found that, in general, the X-ray spectra of the source were best represented by the BPL model (14 observations) and the LP model (8 observations), suggesting that the simple PL model is not favored by observations.

We performed a spectral analysis of PKS 2155-304 using two approaches: a model-independent measure of the spectrum represented by the HR ratio and fitting the spectra with three spectral models commonly used in AGN studies. The HR analysis reveals that, although we observed hints of either a `harder-when-brighter' or `softer-when-bright' trend in some cases, most observations do not follow a simple trend in the HR-flux plane. In the context of AGN, the `harder-when-brighter' trend refers to a phenomenon observed in some blazars where the spectrum of emitted radiation shifts towards higher energies (becomes harder) as the brightness of the blazar increases. Conversely, the `softer-when-bright' trend occurs when the spectrum shifts towards lower energies (becomes softer) as the brightness increases. Such trends have been frequently observed across various frequency bands, e.g., optical, X-ray, and gamma-ray, as blazar flux fluctuates over time \citep[see e. g.,][]{2023MNRAS.524.3797D,2022ApJS..262....4N,2018A&A...619A..93B,2018Galax...6....2B}.

The spectral behavior patterns, specifically the harder-when-brighter and softer-when-brighter trends, can be attributed to distinct physical processes in the jet. In the harder-when-brighter scenario, the correlation likely results from the acceleration of particles to higher energies during enhanced emission states, effectively extending the synchrotron spectrum to higher frequencies. This behavior suggests efficient particle acceleration mechanisms, possibly due to increased magnetic reconnection events or stronger shock acceleration.
Conversely, the softer-when-brighter trend may indicate cooling dominance, where increased emission is accompanied by more efficient radiative cooling of the highest-energy electrons. This could occur when the particle injection rate increases while the maximum electron energy remains relatively constant, leading to a steepening of the electron energy distribution \citep[see][for details]{1997A&A...320...19M,1998A&A...333..452K,2007Ap&SS.309...95B}.

Observations of such definitive trends often point to a co-spatial population of particles contributing to the observed emission, which are seen in several blazars across various EM bands,  Conversely, a complex trend in the HR-flux plane often indicates that the observed emission could be contributed by multiple independent emission regions. Alternatively, complex stochastic particle acceleration in turbulent jets, followed by cooling from emission, can also lead to a complex spectral profile \citep[see e. g.,][]{2014ApJ...780...87M,2023ApJ...949...71Z}.

\section{Conclusion}
The analysis of X-ray emission variability in blazars is an important tool for understanding the fundamental physics governing the central engines of AGN. By studying the temporal and spectral characteristics of X-ray variability, we can gain insights into the structure of their emission regions, the underlying particle acceleration mechanisms, and the extreme environments associated with supermassive black holes in blazars. In this work, we conducted an extensive timing and spectral variability analysis of the TeV blazar PKS 2155-304, using 22-epoch observations spanning 15 years. The key findings of this study can be summarized as follows:

   \begin{enumerate}
      \item The source demonstrates substantial flux variability across multiple timescales, exhibiting significant intra-day fluctuations as well as pronounced long-term variations over the two-decade observational period.

    \item The RMS-flux relation is recognized as a significant characteristic of AGN and micro-quasars, potentially linking accretion disk dynamics to jet properties. Our investigation of this relation in the source does not reveal a straightforward correlation. Complex turbulence patterns extending along the curved jet structure may obscure potential signatures of disk-jet coupling.

      \item Analysis of intra-day flux distributions reveals bimodal patterns in some cases, suggesting the presence of two dominant flux states. However, examination of long-term histograms uncovers more complex substructures that cannot be adequately described by simple PDFs. These findings indicate that the underlying flux dynamics may involve multiple states or processes operating on different timescales.
      \item The PSD analysis yielded a broad spectrum of powerlaw indices, spanning approximately from 1 to 4. This extensive range of indices suggests a scenario involving multiple turbulent regions within relativistic jets.
       \item Despite the wide range of flux states observed, there are only two observations where clear cases of HR variability are detected on intra-day timescales.
       \item Spectral analysis of the observational data indicates the presence of curvature in the spectrum. This feature may be attributed to the stochastic processes inherent in particle acceleration mechanisms.
   \end{enumerate}

\section*{Acknowledgments}
We thank the anonymous referee for their careful review and valuable suggestions, which have significantly improved the presentation of this paper. This work was partially supported by a program of the Polish Ministry of Science under the title ‘Regional Excellence Initiative’, project no. RID/SP/0050/2024/1. TPA acknowledges the support of the National Natural Science --Foundation of China ( grant nos. 12222304, 12192220, and 12192221). MM acknowledges the International Macquarie Research Excellence Scholarship (iMQRES) program for the financial support during the research, and the ARC Centre of Excellence for All Sky Astrophysics in 3 Dimensions (ASTRO 3D), through project CE170100013. RP would like to express his acknowledgement for the institutional support of the Research Centre for Theoretical Physics and Institute of Physics, Silesian University in Opava, SGS/30/2023, and for the support from the project of the Czech Science Foundation GAČR 23-07043S.

\bibliography{samplePKS}{}

\begin{thebibliography}{}
\expandafter\ifx\csname natexlab\endcsname\relax\def\natexlab#1{#1}\fi
\providecommand{\url}[1]{\href{#1}{#1}}
\providecommand{\dodoi}[1]{doi:~\href{http://doi.org/#1}{\nolinkurl{#1}}}
\providecommand{\doeprint}[1]{\href{http://ascl.net/#1}{\nolinkurl{http://ascl.net/#1}}}
\providecommand{\doarXiv}[1]{\href{https://arxiv.org/abs/#1}{\nolinkurl{https://arxiv.org/abs/#1}}}

\bibitem[{{Abdalla} {et~al.}(2020){Abdalla}, {Adam}, {Aharonian}, {Ait
  Benkhali}, {Ang{\"u}ner}, {Arakawa}, {Arcaro}, {Armand}, {Ashkar}, {Backes},
  {Barbosa Martins}, {Barnard}, {Becherini}, {Berge}, {Bernl{\"o}hr},
  {Blackwell}, {B{\"o}ttcher}, {Boisson}, {Bolmont}, {Bonnefoy}, {Bregeon},
  {Breuhaus}, {Brun}, {Brun}, {Bryan}, {B{\"u}chele}, {Bulik}, {Bylund},
  {Caroff}, {Carosi}, {Casanova}, {Cerruti}, {Chand}, {Chandra}, {Chen},
  {Colafrancesco}, {Cury{\l}o}, {Davids}, {Deil}, {Devin}, {deWilt}, {Dirson},
  {Djannati-Ata{\"\i}}, {Dmytriiev}, {Donath}, {Doroshenko}, {Dyks}, {Egberts},
  {Emery}, {Ernenwein}, {Eschbach}, {Feijen}, {Fegan}, {Fiasson}, {Fontaine},
  {Funk}, {F{\"u}{\ss}ling}, {Gabici}, {Gallant}, {Gat{\'e}}, {Giavitto},
  {Giunti}, {Glawion}, {Glicenstein}, {Gottschall}, {Grondin}, {Hahn}, {Haupt},
  {Heinzelmann}, {Henri}, {Hermann}, {Hinton}, {Hofmann}, {Hoischen}, {Holch},
  {Holler}, {Horns}, {Huber}, {Iwasaki}, {Jamrozy}, {Jankowsky}, {Jankowsky},
  {Jardin-Blicq}, {Jung-Richardt}, {Kastendieck}, {Katarzy{\'n}ski},
  {Katsuragawa}, {Katz}, {Khangulyan}, {Kh{\'e}lifi}, {King}, {Klepser},
  {Klu{\'z}niak}, {Komin}, {Kosack}, {Kostunin}, {Kreter}, {Lamanna},
  {Lemi{\`e}re}, {Lemoine-Goumard}, {Lenain}, {Leser}, {Levy}, {Lohse},
  {Lypova}, {Mackey}, {Majumdar}, {Malyshev}, {Marandon}, {Marcowith}, {Mares},
  {Mariaud}, {Mart{\'\i}-Devesa}, {Marx}, {Maurin}, {Meintjes}, {Mitchell},
  {Moderski}, {Mohamed}, {Mohrmann}, {Moore}, {Moulin}, {Muller}, {Murach},
  {Nakashima}, {de Naurois}, {Ndiyavala}, {Niederwanger}, {Niemiec}, {Oakes},
  {O'Brien}, {Odaka}, {Ohm}, {de Ona Wilhelmi}, {Ostrowski}, {Oya}, {Panter},
  {Parsons}, {Perennes}, {Petrucci}, {Peyaud}, {Piel}, {Pita}, {Poireau},
  {Priyana Noel}, {Prokhorov}, {Prokoph}, {P{\"u}hlhofer}, {Punch},
  {Quirrenbach}, {Raab}, {Rauth}, {Reimer}, {Reimer}, {Remy}, {Renaud},
  {Rieger}, {Rinchiuso}, {Romoli}, {Rowell}, {Rudak}, {Ruiz-Velasco},
  {Sahakian}, {Sailer}, {Saito}, {Sanchez}, {Santangelo}, {Sasaki},
  {Schlickeiser}, {Sch{\"u}ssler}, {Schulz}, {Schutte}, {Schwanke},
  {Schwemmer}, {Seglar-Arroyo}, {Senniappan}, {Seyffert}, {Shafi},
  {Shiningayamwe}, {Simoni}, {Sinha}, {Sol}, {Specovius}, {Spir-Jacob},
  {Stawarz}, {Steenkamp}, {Stegmann}, {Steppa}, {Takahashi}, {Tavernier},
  {Taylor}, {Terrier}, {Tiziani}, {Tluczykont}, {Trichard}, {Tsirou}, {Tsuji},
  {Tuffs}, {Uchiyama}, {van der Walt}, {van Eldik}, {van Rensburg}, {van
  Soelen}, {Vasileiadis}, {Veh}, {Venter}, {Vincent}, {Vink}, {V{\"o}lk},
  {Vuillaume}, {Wadiasingh}, {Wagner}, {White}, {Wierzcholska}, {Yang},
  {Yoneda}, {Zacharias}, {Zanin}, {Zdziarski}, {Zech}, {Zorn}, {{\.Z}ywucka},
  {Madejski}, {Nalewajko}, {Madsen}, {Chiang}, {Balokovi{\'c}}, {Paneque},
  {Furniss}, {Hayashida}, {Urry}, {Ajello}, {Harrison}, {Giebels}, {Stern},
  {Forster}, {Giommi}, {Perri}, {Puccetti}, {Zoglauer}, \&
  {Tagliaferri}}]{2020A&A...639A..42A}
{Abdalla}, H., {Adam}, R., {Aharonian}, F., {et~al.} 2020, \aap, 639, A42,
  \dodoi{10.1051/0004-6361/201936900}

\bibitem[{{Abdo} {et~al.}(2010){Abdo}, {Ackermann}, {Agudo}, {Ajello}, {Aller},
  {Aller}, {Angelakis}, {Arkharov}, {Axelsson}, {Bach}, {Baldini}, {Ballet},
  {Barbiellini}, {Bastieri}, {Baughman}, {Bechtol}, {Bellazzini}, {Benitez},
  {Berdyugin}, {Berenji}, {Blandford}, {Bloom}, {Boettcher}, {Bonamente},
  {Borgland}, {Bregeon}, {Brez}, {Brigida}, {Bruel}, {Burnett}, {Burrows},
  {Buson}, {Caliandro}, {Calzoletti}, {Cameron}, {Capalbi}, {Caraveo},
  {Carosati}, {Casandjian}, {Cavazzuti}, {Cecchi}, {{\c{C}}elik}, {Charles},
  {Chaty}, {Chekhtman}, {Chen}, {Chiang}, {Chincarini}, {Ciprini}, {Claus},
  {Cohen-Tanugi}, {Colafrancesco}, {Cominsky}, {Conrad}, {Costamante},
  {Cutini}, {D'ammando}, {Deitrick}, {D'Elia}, {Dermer}, {de Angelis}, {de
  Palma}, {Digel}, {Donnarumma}, {Silva}, {Drell}, {Dubois}, {Dultzin},
  {Dumora}, {Falcone}, {Farnier}, {Favuzzi}, {Fegan}, {Focke}, {Forn{\'e}},
  {Fortin}, {Frailis}, {Fuhrmann}, {Fukazawa}, {Funk}, {Fusco}, {G{\'o}mez},
  {Gargano}, {Gasparrini}, {Gehrels}, {Germani}, {Giebels}, {Giglietto},
  {Giommi}, {Giordano}, {Giuliani}, {Glanzman}, {Godfrey}, {Grenier},
  {Gronwall}, {Grove}, {Guillemot}, {Guiriec}, {Gurwell}, {Hadasch},
  {Hanabata}, {Harding}, {Hayashida}, {Hays}, {Healey}, {Heidt}, {Hiriart},
  {Horan}, {Hoversten}, {Hughes}, {Itoh}, {Jackson}, {J{\'o}hannesson},
  {Johnson}, {Johnson}, {Jorstad}, {Kadler}, {Kamae}, {Katagiri}, {Kataoka},
  {Kawai}, {Kennea}, {Kerr}, {Kimeridze}, {Kn{\"o}dlseder}, {Kocian},
  {Kopatskaya}, {Koptelova}, {Konstantinova}, {Kovalev}, {Kovalev},
  {Kurtanidze}, {Kuss}, {Lande}, {Larionov}, {Latronico}, {Leto}, {Lindfors},
  {Longo}, {Loparco}, {Lott}, {Lovellette}, {Lubrano}, {Madejski}, {Makeev},
  {Marchegiani}, {Marscher}, {Marshall}, {Max-Moerbeck}, {Mazziotta},
  {McConville}, {McEnery}, {Meurer}, {Michelson}, {Mitthumsiri}, {Mizuno},
  {Moiseev}, {Monte}, {Monzani}, {Morselli}, {Moskalenko}, {Murgia},
  {Nestoras}, {Nilsson}, {Nizhelsky}, {Nolan}, {Norris}, {Nuss}, {Ohsugi},
  {Ojha}, {Omodei}, {Orlando}, {Ormes}, {Osborne}, {Ozaki}, {Pacciani},
  {Padovani}, {Pagani}, {Page}, {Paneque}, {Panetta}, {Parent}, {Pasanen},
  {Pavlidou}, {Pelassa}, {Pepe}, {Perri}, {Pesce-Rollins}, {Piranomonte},
  {Piron}, {Pittori}, {Porter}, {Puccetti}, {Rahoui}, {Rain{\`o}}, {Raiteri},
  {Rando}, {Razzano}, {Reimer}, {Reimer}, {Reposeur}, {Richards}, {Ritz},
  {Rochester}, {Rodriguez}, {Romani}, {Ros}, {Roth}, {Roustazadeh}, {Ryde},
  {Sadrozinski}, {Sadun}, {Sanchez}, {Sander}, {Saz Parkinson}, {Scargle},
  {Sellerholm}, {Sgr{\`o}}, {Shaw}, {Sigua}, {Siskind}, {Smith}, {Smith},
  {Spandre}, {Spinelli}, {Starck}, {Stevenson}, {Stratta}, {Strickman},
  {Suson}, {Tajima}, {Takahashi}, {Takahashi}, {Takalo}, {Tanaka}, {Thayer},
  {Thayer}, {Thompson}, {Tibaldo}, {Torres}, {Tosti}, {Tramacere}, {Uchiyama},
  {Usher}, {Vasileiou}, {Verrecchia}, {Vilchez}, {Villata}, {Vitale}, {Waite},
  {Wang}, {Winer}, {Wood}, {Ylinen}, {Zensus}, {Zhekanis}, \&
  {Ziegler}}]{2010ApJ...716...30A}
{Abdo}, A.~A., {Ackermann}, M., {Agudo}, I., {et~al.} 2010, \apj, 716, 30,
  \dodoi{10.1088/0004-637X/716/1/30}

\bibitem[{{Aggrawal} {et~al.}(2018){Aggrawal}, {Pandey}, {Gupta}, {Zhang},
  {Wiita}, {Yadav}, \& {Tiwari}}]{2018MNRAS.480.4873A}
{Aggrawal}, V., {Pandey}, A., {Gupta}, A.~C., {et~al.} 2018, \mnras, 480, 4873,
  \dodoi{10.1093/mnras/sty2173}

\bibitem[{{Aharonian} {et~al.}(2005{\natexlab{a}}){Aharonian}, {Akhperjanian},
  {Aye}, {Bazer-Bachi}, {Beilicke}, {Benbow}, {Berge}, {Berghaus},
  {Bernl{\"o}hr}, {Bolz}, {Boisson}, {Borgmeier}, {Breitling}, {Brown},
  {Bussons Gordo}, {Chadwick}, {Chitnis}, {Chounet}, {Cornils}, {Costamante},
  {Degrange}, {Djannati-Ata{\"\i}}, {Drury}, {Ergin}, {Espigat}, {Feinstein},
  {Fleury}, {Fontaine}, {Funk}, {Gallant}, {Giebels}, {Gillessen}, {Goret},
  {Guy}, {Hadjichristidis}, {Hauser}, {Heinzelmann}, {Henri}, {Hermann},
  {Hinton}, {Hofmann}, {Holleran}, {Horns}, {de Jager}, {Jung I.},
  {Kh{\'e}lifi}, {Komin}, {Konopelko}, {Latham}, {Le Gallou}, {Lemoine},
  {Lemi{\`e}re}, {Leroy}, {Lohse}, {Marcowith}, {Masterson}, {McComb}, {de
  Naurois}, {Nolan}, {Noutsos}, {Orford}, {Osborne}, {Ouchrif}, {Panter},
  {Pelletier}, {Pita}, {Pohl}, {P{\"u}hlhofer}, {Punch}, {Raubenheimer},
  {Raue}, {Raux}, {Rayner}, {Redondo}, {Reimer}, {Reimer}, {Ripken}, {Rivoal},
  {Rob}, {Rolland}, {Rowell}, {Sahakian}, {Saug{\'e}}, {Schlenker},
  {Schlickeiser}, {Schuster}, {Schwanke}, {Siewert}, {Sol}, {Steenkamp},
  {Stegmann}, {Tavernet}, {Th{\'e}oret}, {Tluczykont}, {van der Walt},
  {Vasileiadis}, {Vincent}, {Visser}, {V{\"o}lk}, \&
  {Wagner}}]{2005A&A...430..865A}
{Aharonian}, F., {Akhperjanian}, A.~G., {Aye}, K.~M., {et~al.}
  2005{\natexlab{a}}, \aap, 430, 865, \dodoi{10.1051/0004-6361:20041853}

\bibitem[{{Aharonian} {et~al.}(2005{\natexlab{b}}){Aharonian}, {Akhperjanian},
  {Bazer-Bachi}, {Beilicke}, {Benbow}, {Berge}, {Bernl{\"o}hr}, {Boisson},
  {Bolz}, {Borrel}, {Braun}, {Breitling}, {Brown}, {Chadwick}, {Chounet},
  {Cornils}, {Costamante}, {Degrange}, {Dickinson}, {Djannati-Ata{\"\i}}, {O'C.
  Drury}, {Dubus}, {Emmanoulopoulos}, {Espigat}, {Feinstein}, {Fontaine},
  {Fuchs}, {Funk}, {Gallant}, {Giebels}, {Gillessen}, {Glicenstein}, {Goret},
  {Hadjichristidis}, {Hauser}, {Heinzelmann}, {Henri}, {Hermann}, {Hinton},
  {Hofmann}, {Holleran}, {Horns}, {Jacholkowska}, {de Jager}, {Kh{\'e}lifi},
  {Komin}, {Konopelko}, {Latham}, {Le Gallou}, {Lemi{\`e}re},
  {Lemoine-Goumard}, {Leroy}, {Lohse}, {Martin}, {Martineau-Huynh},
  {Marcowith}, {Masterson}, {McComb}, {de Naurois}, {Nolan}, {Noutsos},
  {Orford}, {Osborne}, {Ouchrif}, {Panter}, {Pelletier}, {Pita},
  {P{\"u}hlhofer}, {Punch}, {Raubenheimer}, {Raue}, {Raux}, {Rayner}, {Reimer},
  {Reimer}, {Ripken}, {Rob}, {Rolland}, {Rowell}, {Sahakian}, {Saug{\'e}},
  {Schlenker}, {Schlickeiser}, {Schuster}, {Schwanke}, {Siewert}, {Sol},
  {Spangler}, {Steenkamp}, {Stegmann}, {Tavernet}, {Terrier}, {Th{\'e}oret},
  {Tluczykont}, {Vasileiadis}, {Venter}, {Vincent}, {V{\"o}lk}, \&
  {Wagner}}]{2005A&A...442..895A}
{Aharonian}, F., {Akhperjanian}, A.~G., {Bazer-Bachi}, A.~R., {et~al.}
  2005{\natexlab{b}}, \aap, 442, 895, \dodoi{10.1051/0004-6361:20053353}

\bibitem[{{Aharonian} {et~al.}(2007){Aharonian}, {Akhperjanian}, {Bazer-Bachi},
  {Behera}, {Beilicke}, {Benbow}, {Berge}, {Bernl{\"o}hr}, {Boisson}, {Bolz},
  {Borrel}, {Boutelier}, {Braun}, {Brion}, {Brown}, {B{\"u}hler},
  {B{\"u}sching}, {Bulik}, {Carrigan}, {Chadwick}, {Clapson}, {Chounet},
  {Coignet}, {Cornils}, {Costamante}, {Degrange}, {Dickinson},
  {Djannati-Ata{\"\i}}, {Domainko}, {Drury}, {Dubus}, {Dyks}, {Egberts},
  {Emmanoulopoulos}, {Espigat}, {Farnier}, {Feinstein}, {Fiasson},
  {F{\"o}rster}, {Fontaine}, {Funk}, {Funk}, {F{\"u}{\ss}ling}, {Gallant},
  {Giebels}, {Glicenstein}, {Gl{\"u}ck}, {Goret}, {Hadjichristidis}, {Hauser},
  {Hauser}, {Heinzelmann}, {Henri}, {Hermann}, {Hinton}, {Hoffmann}, {Hofmann},
  {Holleran}, {Hoppe}, {Horns}, {Jacholkowska}, {de Jager}, {Kendziorra},
  {Kerschhaggl}, {Kh{\'e}lifi}, {Komin}, {Kosack}, {Lamanna}, {Latham}, {Le
  Gallou}, {Lemi{\`e}re}, {Lemoine-Goumard}, {Lenain}, {Lohse}, {Martin},
  {Martineau-Huynh}, {Marcowith}, {Masterson}, {Maurin}, {McComb}, {Moderski},
  {Moulin}, {de Naurois}, {Nedbal}, {Nolan}, {Olive}, {Orford}, {Osborne},
  {Ostrowski}, {Panter}, {Pedaletti}, {Pelletier}, {Petrucci}, {Pita},
  {P{\"u}hlhofer}, {Punch}, {Ranchon}, {Raubenheimer}, {Raue}, {Rayner},
  {Renaud}, {Ripken}, {Rob}, {Rolland}, {Rosier-Lees}, {Rowell}, {Rudak},
  {Ruppel}, {Sahakian}, {Santangelo}, {Saug{\'e}}, {Schlenker}, {Schlickeiser},
  {Schr{\"o}der}, {Schwanke}, {Schwarzburg}, {Schwemmer}, {Shalchi}, {Sol},
  {Spangler}, {Stawarz}, {Steenkamp}, {Stegmann}, {Superina}, {Tam},
  {Tavernet}, {Terrier}, {van Eldik}, {Vasileiadis}, {Venter}, {Vialle},
  {Vincent}, {Vivier}, {V{\"o}lk}, {Volpe}, {Wagner}, {Ward}, \&
  {Zdziarski}}]{2007ApJ...664L..71A}
---. 2007, \apjl, 664, L71, \dodoi{10.1086/520635}

\bibitem[{{Aleksi{\'c}} {et~al.}(2015){Aleksi{\'c}}, {Ansoldi}, {Antonelli},
  {Antoranz}, {Babic}, {Bangale}, {Barres de Almeida}, {Barrio}, {Becerra
  Gonz{\'a}lez}, {Bednarek}, {Berger}, {Bernardini}, {Biland}, {Blanch},
  {Bock}, {Bonnefoy}, {Bonnoli}, {Borracci}, {Bretz}, {Carmona}, {Carosi},
  {Carreto Fidalgo}, {Colin}, {Colombo}, {Contreras}, {Cortina}, {Covino}, {Da
  Vela}, {Dazzi}, {De Angelis}, {De Caneva}, {De Lotto}, {Delgado Mendez},
  {Doert}, {Dom{\'\i}nguez}, {Dominis Prester}, {Dorner}, {Doro}, {Einecke},
  {Eisenacher}, {Elsaesser}, {Farina}, {Ferenc}, {Fonseca}, {Font}, {Frantzen},
  {Fruck}, {Garc{\'\i}a L{\'o}pez}, {Garczarczyk}, {Garrido Terrats}, {Gaug},
  {Giavitto}, {Godinovi{\'c}}, {Gonz{\'a}lez Mu{\~n}oz}, {Gozzini}, {Hadamek},
  {Hadasch}, {Herrero}, {Hildebrand}, {Hose}, {Hrupec}, {Idec}, {Kadenius},
  {Kellermann}, {Knoetig}, {Krause}, {Kushida}, {La Barbera}, {Lelas},
  {Lewandowska}, {Lindfors}, {Longo}, {Lombardi}, {L{\'o}pez},
  {L{\'o}pez-Coto}, {L{\'o}pez-Oramas}, {Lorenz}, {Lozano}, {Makariev},
  {Mallot}, {Maneva}, {Mankuzhiyil}, {Mannheim}, {Maraschi}, {Marcote},
  {Mariotti}, {Mart{\'\i}nez}, {Mazin}, {Menzel}, {Meucci}, {Miranda},
  {Mirzoyan}, {Moralejo}, {Munar-Adrover}, {Nakajima}, {Niedzwiecki},
  {Nilsson}, {Nowak}, {Orito}, {Overkemping}, {Paiano}, {Palatiello},
  {Paneque}, {Paoletti}, {Paredes}, {Paredes-Fortuny}, {Partini}, {Persic},
  {Prada}, {Prada Moroni}, {Prandini}, {Preziuso}, {Puljak}, {Reinthal},
  {Rhode}, {Rib{\'o}}, {Rico}, {RodriguezGarcia}, {R{\"u}gamer}, {Saggion},
  {Saito}, {Salvati}, {Satalecka}, {Scalzotto}, {Scapin}, {Schultz},
  {Schweizer}, {Shore}, {Sillanp{\"a}{\"a}}, {Sitarek}, {Snidaric},
  {Sobczynska}, {Spanier}, {Stamatescu}, {Stamerra}, {Steinbring}, {Storz},
  {Sun}, {Suri{\'c}}, {Takalo}, {Tavecchio}, {Temnikov}, {Terzi{\'c}},
  {Tescaro}, {Teshima}, {Thaele}, {Tibolla}, {Torres}, {Toyama}, {Treves},
  {Uellenbeck}, {Vogler}, {Wagner}, {Zandanel}, {Zanin}, {MAGIC Collaboration},
  {Archambault}, {Behera}, {Beilicke}, {Benbow}, {Bird}, {Buckley}, {Bugaev},
  {Cerruti}, {Chen}, {Ciupik}, {Collins-Hughes}, {Cui}, {Dumm}, {Eisch},
  {Falcone}, {Federici}, {Feng}, {Finley}, {Fleischhack}, {Fortin}, {Fortson},
  {Furniss}, {Griffin}, {Griffiths}, {Grube}, {Gyuk}, {Hanna}, {Holder},
  {Hughes}, {Humensky}, {Johnson}, {Kaaret}, {Kertzman}, {Khassen}, {Kieda},
  {Krawczynski}, {Krennrich}, {Kumar}, {Lang}, {Maier}, {McArthur}, {Meagher},
  {Moriarty}, \& {Mukherjee}}]{2015A&A...576A.126A}
{Aleksi{\'c}}, J., {Ansoldi}, S., {Antonelli}, L.~A., {et~al.} 2015, \aap, 576,
  A126, \dodoi{10.1051/0004-6361/201424216}

\bibitem[{{Alston} {et~al.}(2019){Alston}, {Fabian}, {Buisson}, {Kara},
  {Parker}, {Lohfink}, {Uttley}, {Wilkins}, {Pinto}, {De Marco}, {Cackett},
  {Middleton}, {Walton}, {Reynolds}, {Jiang}, {Gallo}, {Zogbhi}, {Miniutti},
  {Dovciak}, \& {Young}}]{2019MNRAS.482.2088A}
{Alston}, W.~N., {Fabian}, A.~C., {Buisson}, D.~J.~K., {et~al.} 2019, \mnras,
  482, 2088, \dodoi{10.1093/mnras/sty2527}

\bibitem[{{Arnaud}(1996)}]{1996ASPC..101...17A}
{Arnaud}, K.~A. 1996, in Astronomical Society of the Pacific Conference Series,
  Vol. 101, Astronomical Data Analysis Software and Systems V, ed. G.~H.
  {Jacoby} \& J.~{Barnes}, 17

\bibitem[{{Bhagwan} {et~al.}(2014){Bhagwan}, {Gupta}, {Papadakis}, \&
  {Wiita}}]{2014MNRAS.444.3647B}
{Bhagwan}, J., {Gupta}, A.~C., {Papadakis}, I.~E., \& {Wiita}, P.~J. 2014,
  \mnras, 444, 3647, \dodoi{10.1093/mnras/stu1703}

\bibitem[{{Bhatta}(2018)}]{2018Galax...6..136B}
{Bhatta}, G. 2018, Galaxies, 6, 136, \dodoi{10.3390/galaxies6040136}

\bibitem[{{Bhatta}(2021)}]{2021ApJ...923....7B}
---. 2021, \apj, 923, 7, \dodoi{10.3847/1538-4357/ac2819}

\bibitem[{{Bhatta} \& {Dhital}(2020)}]{2020ApJ...891..120B}
{Bhatta}, G., \& {Dhital}, N. 2020, \apj, 891, 120,
  \dodoi{10.3847/1538-4357/ab7455}

\bibitem[{{Bhatta} {et~al.}(2024){Bhatta}, {Gharat}, {Borthakur}, \&
  {Kumar}}]{2024MNRAS.528..976B}
{Bhatta}, G., {Gharat}, S., {Borthakur}, A., \& {Kumar}, A. 2024, \mnras, 528,
  976, \dodoi{10.1093/mnras/stae028}

\bibitem[{{Bhatta} {et~al.}(2018){Bhatta}, {Mohorian}, \&
  {Bilinsky}}]{2018A&A...619A..93B}
{Bhatta}, G., {Mohorian}, M., \& {Bilinsky}, I. 2018, \aap, 619, A93,
  \dodoi{10.1051/0004-6361/201833628}

\bibitem[{{Bhatta} \& {Webb}(2018)}]{2018Galax...6....2B}
{Bhatta}, G., \& {Webb}, J. 2018, Galaxies, 6, 2,
  \dodoi{10.3390/galaxies6010002}

\bibitem[{{Bhatta} {et~al.}(2013){Bhatta}, {Webb}, {Hollingsworth}, {Dhalla},
  {Khanuja}, {Bachev}, {Blinov}, {B{\"o}ttcher}, {Bravo Calle}, {Calcidese},
  {Capezzali}, {Carosati}, {Chigladze}, {Collins}, {Coloma}, {Efimov}, {Gupta},
  {Hu}, {Kurtanidze}, {Lamerato}, {Larionov}, {Lee}, {Lindfors}, {Murphy},
  {Nilsson}, {Ohlert}, {Oksanen}, {P{\"a}{\"a}kk{\"o}nen}, {Pollock}, {Rani},
  {Reinthal}, {Rodriguez}, {Ros}, {Roustazadeh}, {Sagar}, {Sanchez}, {Shastri},
  {Sillanp{\"a}{\"a}}, {Strigachev}, {Takalo}, {Vennes}, {Villata},
  {Villforth}, {Wu}, \& {Zhou}}]{2013A&A...558A..92B}
{Bhatta}, G., {Webb}, J.~R., {Hollingsworth}, H., {et~al.} 2013, \aap, 558,
  A92, \dodoi{10.1051/0004-6361/201220236}

\bibitem[{{Bhatta} {et~al.}(2016){Bhatta}, {Stawarz}, {Ostrowski}, {Markowitz},
  {Akitaya}, {Arkharov}, {Bachev}, {Ben{\'\i}tez}, {Borman}, {Carosati},
  {Cason}, {Chanishvili}, {Damljanovic}, {Dhalla}, {Frasca}, {Hiriart}, {Hu},
  {Itoh}, {Jableka}, {Jorstad}, {Jovanovic}, {Kawabata}, {Klimanov},
  {Kurtanidze}, {Larionov}, {Laurence}, {Leto}, {Marscher}, {Moody},
  {Moritani}, {Ohlert}, {Di Paola}, {Raiteri}, {Rizzi}, {Sadun}, {Sasada},
  {Sergeev}, {Strigachev}, {Takaki}, {Troitsky}, {Ui}, {Villata}, {Vince},
  {Webb}, {Yoshida}, \& {Zola}}]{2016ApJ...831...92B}
{Bhatta}, G., {Stawarz}, {\L}., {Ostrowski}, M., {et~al.} 2016, \apj, 831, 92,
  \dodoi{10.3847/0004-637X/831/1/92}

\bibitem[{{Bhatta} {et~al.}(2023){Bhatta}, {Zola}, {Drozdz}, {Reichart},
  {Haislip}, {Kouprianov}, {Matsumoto}, {Sonbas}, {Caton},
  {Pajdosz-{\'S}mierciak}, {Simon}, {Provencal}, {G{\'o}ra}, \&
  {Stachowski}}]{2023MNRAS.520.2633B}
{Bhatta}, G., {Zola}, S., {Drozdz}, M., {et~al.} 2023, \mnras, 520, 2633,
  \dodoi{10.1093/mnras/stad280}

\bibitem[{{Bhattacharyya} {et~al.}(2020){Bhattacharyya}, {Ghosh}, {Chatterjee},
  \& {Das}}]{2020ApJ...897...25B}
{Bhattacharyya}, S., {Ghosh}, R., {Chatterjee}, R., \& {Das}, N. 2020, \apj,
  897, 25, \dodoi{10.3847/1538-4357/ab91a8}

\bibitem[{{B{\l}a{\.z}ejowski} {et~al.}(2000){B{\l}a{\.z}ejowski}, {Sikora},
  {Moderski}, \& {Madejski}}]{2000ApJ...545..107B}
{B{\l}a{\.z}ejowski}, M., {Sikora}, M., {Moderski}, R., \& {Madejski}, G.~M.
  2000, \apj, 545, 107, \dodoi{10.1086/317791}

\bibitem[{{B{\"o}ttcher}(2007)}]{2007Ap&SS.309...95B}
{B{\"o}ttcher}, M. 2007, \apss, 309, 95, \dodoi{10.1007/s10509-007-9404-0}

\bibitem[{{B{\"o}ttcher}(2019)}]{2019Galax...7...20B}
---. 2019, Galaxies, 7, 20, \dodoi{10.3390/galaxies7010020}

\bibitem[{{B{\"o}ttcher} \& {Dermer}(2010)}]{2010ApJ...711..445B}
{B{\"o}ttcher}, M., \& {Dermer}, C.~D. 2010, \apj, 711, 445,
  \dodoi{10.1088/0004-637X/711/1/445}

\bibitem[{{Briel} {et~al.}(2000){Briel}, {Aschenbach}, {Balasini},
  {Braeuninger}, {Burkert}, {Dennerl}, {Ehle}, {Haberl}, {Hartmann}, {Hartner},
  {Holl}, {Massa}, {Meidinger}, {Kemmer}, {Kendziorra}, {Kirsch}, {Krause},
  {Kuster}, {Lumb}, {Pfeffermann}, {Pal}, {Pietsch}, {Popp}, {Read}, {Reppin},
  {Soltau}, {Staubert}, {Strueder}, {Truemper}, {Villa}, {van Zanthier}, \&
  {Zavlin}}]{2000SPIE.4012..154B}
{Briel}, U.~G., {Aschenbach}, B., {Balasini}, M., {et~al.} 2000, in Society of
  Photo-Optical Instrumentation Engineers (SPIE) Conference Series, Vol. 4012,
  X-Ray Optics, Instruments, and Missions III, ed. J.~E. {Truemper} \&
  B.~{Aschenbach}, 154--164, \dodoi{10.1117/12.391613}

\bibitem[{{Burbidge} {et~al.}(1974){Burbidge}, {Jones}, \&
  {O'Dell}}]{1974ApJ...193...43B}
{Burbidge}, G.~R., {Jones}, T.~W., \& {O'Dell}, S.~L. 1974, \apj, 193, 43,
  \dodoi{10.1086/153125}

\bibitem[{{Calafut} \& {Wiita}(2015)}]{2015JApA...36..255C}
{Calafut}, V., \& {Wiita}, P.~J. 2015, Journal of Astrophysics and Astronomy,
  36, 255, \dodoi{10.1007/s12036-015-9324-2}

\bibitem[{{Chang} {et~al.}(2019){Chang}, {Arsioli}, {Giommi}, {Padovani}, \&
  {Brandt}}]{2019A&A...632A..77C}
{Chang}, Y.~L., {Arsioli}, B., {Giommi}, P., {Padovani}, P., \& {Brandt}, C.~H.
  2019, \aap, 632, A77, \dodoi{10.1051/0004-6361/201834526}

\bibitem[{{Chevalier} {et~al.}(2019){Chevalier}, {Sanchez}, {Serpico},
  {Lenain}, \& {Maurin}}]{2019MNRAS.484..749C}
{Chevalier}, J., {Sanchez}, D.~A., {Serpico}, P.~D., {Lenain}, J.~P., \&
  {Maurin}, G. 2019, \mnras, 484, 749, \dodoi{10.1093/mnras/stz027}

\bibitem[{{Christie} {et~al.}(2019){Christie}, {Petropoulou}, {Sironi}, \&
  {Giannios}}]{2019MNRAS.482...65C}
{Christie}, I.~M., {Petropoulou}, M., {Sironi}, L., \& {Giannios}, D. 2019,
  \mnras, 482, 65, \dodoi{10.1093/mnras/sty2636}

\bibitem[{{Das} \& {Chatterjee}(2023)}]{2023MNRAS.524.3797D}
{Das}, S., \& {Chatterjee}, R. 2023, \mnras, 524, 3797,
  \dodoi{10.1093/mnras/stad2131}

\bibitem[{{De Luca} \& {Molendi}(2004)}]{2004A&A...419..837D}
{De Luca}, A., \& {Molendi}, S. 2004, \aap, 419, 837,
  \dodoi{10.1051/0004-6361:20034421}

\bibitem[{{Dermer} \& {Schlickeiser}(1993)}]{1993ApJ...416..458D}
{Dermer}, C.~D., \& {Schlickeiser}, R. 1993, \apj, 416, 458,
  \dodoi{10.1086/173251}

\bibitem[{{Dhiman} {et~al.}(2021){Dhiman}, {Gupta}, {Gaur}, \&
  {Wiita}}]{2021MNRAS.506.1198D}
{Dhiman}, V., {Gupta}, A.~C., {Gaur}, H., \& {Wiita}, P.~J. 2021, \mnras, 506,
  1198, \dodoi{10.1093/mnras/stab1743}

\bibitem[{{Dinesh} {et~al.}(2023){Dinesh}, {Bhatta}, {Adhikari}, {Mohorian},
  {Dhital}, {Chaudhary}, {P{\'a}nis}, \& {G{\'o}ra}}]{2023ApJ...955..121D}
{Dinesh}, A., {Bhatta}, G., {Adhikari}, T.~P., {et~al.} 2023, \apj, 955, 121,
  \dodoi{10.3847/1538-4357/acf316}

\bibitem[{{Edelson} {et~al.}(2013){Edelson}, {Mushotzky}, {Vaughan}, {Scargle},
  {Gandhi}, {Malkan}, \& {Baumgartner}}]{2013ApJ...766...16E}
{Edelson}, R., {Mushotzky}, R., {Vaughan}, S., {et~al.} 2013, \apj, 766, 16,
  \dodoi{10.1088/0004-637X/766/1/16}

\bibitem[{{Edelson} {et~al.}(2002){Edelson}, {Turner}, {Pounds}, {Vaughan},
  {Markowitz}, {Marshall}, {Dobbie}, \& {Warwick}}]{2002ApJ...568..610E}
{Edelson}, R., {Turner}, T.~J., {Pounds}, K., {et~al.} 2002, \apj, 568, 610,
  \dodoi{10.1086/323779}

\bibitem[{{Edelson} {et~al.}(1990){Edelson}, {Krolik}, \&
  {Pike}}]{1990ApJ...359...86E}
{Edelson}, R.~A., {Krolik}, J.~H., \& {Pike}, G.~F. 1990, \apj, 359, 86,
  \dodoi{10.1086/169036}

\bibitem[{{Falomo} {et~al.}(1993){Falomo}, {Pesce}, \&
  {Treves}}]{1993ApJ...411L..63F}
{Falomo}, R., {Pesce}, J.~E., \& {Treves}, A. 1993, \apjl, 411, L63,
  \dodoi{10.1086/186913}

\bibitem[{{Falomo} {et~al.}(2014){Falomo}, {Pian}, \&
  {Treves}}]{2014A&ARv..22...73F}
{Falomo}, R., {Pian}, E., \& {Treves}, A. 2014, \aapr, 22, 73,
  \dodoi{10.1007/s00159-014-0073-z}

\bibitem[{{Foschini} {et~al.}(2006){Foschini}, {Tagliaferri}, {Pian},
  {Ghisellini}, {Treves}, {Maraschi}, {Tavecchio}, {Di Cocco}, \&
  {Rosen}}]{2006A&A...455..871F}
{Foschini}, L., {Tagliaferri}, G., {Pian}, E., {et~al.} 2006, \aap, 455, 871,
  \dodoi{10.1051/0004-6361:20064959}

\bibitem[{{Gaskell}(2004)}]{2004ApJ...612L..21G}
{Gaskell}, C.~M. 2004, \apjl, 612, L21, \dodoi{10.1086/424565}

\bibitem[{{Gaur} {et~al.}(2017){Gaur}, {Chen}, {Misra}, {Sahayanathan}, {Gu},
  {Kushwaha}, \& {Dewangan}}]{2017ApJ...850..209G}
{Gaur}, H., {Chen}, L., {Misra}, R., {et~al.} 2017, \apj, 850, 209,
  \dodoi{10.3847/1538-4357/aa95bc}

\bibitem[{{Gaur} {et~al.}(2010){Gaur}, {Gupta}, {Lachowicz}, \&
  {Wiita}}]{2010ApJ...718..279G}
{Gaur}, H., {Gupta}, A.~C., {Lachowicz}, P., \& {Wiita}, P.~J. 2010, \apj, 718,
  279, \dodoi{10.1088/0004-637X/718/1/279}

\bibitem[{{Ghisellini} {et~al.}(1985){Ghisellini}, {Maraschi}, \&
  {Treves}}]{1985A&A...146..204G}
{Ghisellini}, G., {Maraschi}, L., \& {Treves}, A. 1985, \aap, 146, 204

\bibitem[{{Giannios} {et~al.}(2010){Giannios}, {Uzdensky}, \&
  {Begelman}}]{2010MNRAS.402.1649G}
{Giannios}, D., {Uzdensky}, D.~A., \& {Begelman}, M.~C. 2010, \mnras, 402,
  1649, \dodoi{10.1111/j.1365-2966.2009.16045.x}

\bibitem[{{Gleissner} {et~al.}(2004){Gleissner}, {Wilms}, {Pottschmidt},
  {Uttley}, {Nowak}, \& {Staubert}}]{2004A&A...414.1091G}
{Gleissner}, T., {Wilms}, J., {Pottschmidt}, K., {et~al.} 2004, \aap, 414,
  1091, \dodoi{10.1051/0004-6361:20031684}

\bibitem[{{Gonz{\'a}lez-Mart{\'\i}n} \& {Vaughan}(2012)}]{2012A&A...544A..80G}
{Gonz{\'a}lez-Mart{\'\i}n}, O., \& {Vaughan}, S. 2012, \aap, 544, A80,
  \dodoi{10.1051/0004-6361/201219008}

\bibitem[{{Goswami} {et~al.}(2024){Goswami}, {Zacharias}, {Zech}, {Chandra},
  {Boettcher}, \& {Sushch}}]{2024A&A...682A.134G}
{Goswami}, P., {Zacharias}, M., {Zech}, A., {et~al.} 2024, \aap, 682, A134,
  \dodoi{10.1051/0004-6361/202348121}

\bibitem[{{Griffiths} {et~al.}(1979){Griffiths}, {Tapia}, {Briel}, \&
  {Chaisson}}]{1979ApJ...234..810G}
{Griffiths}, R.~E., {Tapia}, S., {Briel}, U., \& {Chaisson}, L. 1979, \apj,
  234, 810, \dodoi{10.1086/157560}

\bibitem[{{Gupta}(2020)}]{2020Galax...8...64G}
{Gupta}, A.~C. 2020, Galaxies, 8, 64, \dodoi{10.3390/galaxies8030064}

\bibitem[{{H.~E.~S.~S. Collaboration} {et~al.}(2010){H.~E.~S.~S.
  Collaboration}, {Abramowski}, {Acero}, {Aharonian}, {Akhperjanian}, {Anton},
  {Barres de Almeida}, {Bazer-Bachi}, {Becherini}, {Benbow}, {Bernl{\"o}hr},
  {Bochow}, {Boisson}, {Bolmont}, {Borrel}, {Brucker}, {Brun}, {Brun},
  {B{\"u}hler}, {Bulik}, {B{\"u}sching}, {Boutelier}, {Chadwick},
  {Charbonnier}, {Chaves}, {Cheesebrough}, {Chounet}, {Clapson}, {Coignet},
  {Conrad}, {Costamante}, {Dalton}, {Daniel}, {Davids}, {Degrange}, {Deil},
  {Dickinson}, {Djannati-Ata{\"\i}}, {Domainko}, {O'C. Drury}, {Dubois},
  {Dubus}, {Dyks}, {Dyrda}, {Egberts}, {Eger}, {Espigat}, {Fallon}, {Farnier},
  {Fegan}, {Feinstein}, {Fernandes}, {Fiasson}, {F{\"o}rster}, {Fontaine},
  {F{\"u}{\ss}ling}, {Gabici}, {Gallant}, {G{\'e}rard}, {Gerbig}, {Giebels},
  {Glicenstein}, {Gl{\"u}ck}, {Goret}, {G{\"o}ring}, {Hampf}, {Hauser},
  {Heinz}, {Heinzelmann}, {Henri}, {Hermann}, {Hinton}, {Hoffmann}, {Hofmann},
  {Hofverberg}, {Holleran}, {Hoppe}, {Horns}, {Jacholkowska}, {de Jager},
  {Jahn}, {Jung}, {Katarzy{\'n}ski}, {Katz}, {Kaufmann}, {Kerschhaggl},
  {Khangulyan}, {Kh{\'e}lifi}, {Keogh}, {Klochkov}, {Klu{\'z}niak}, {Kneiske},
  {Komin}, {Kosack}, {Kossakowski}, {Lamanna}, {Lenain}, {Lohse}, {Lu},
  {Marandon}, {Marcowith}, {Masbou}, {Maurin}, {McComb}, {Medina},
  {M{\'e}hault}, {Moderski}, {Moulin}, {Naumann-Godo}, {de Naurois}, {Nedbal},
  {Nekrassov}, {Nguyen}, {Nicholas}, {Niemiec}, {Nolan}, {Ohm}, {Olive}, {de
  O{\~n}a Wilhelmi}, {Opitz}, {Orford}, {Ostrowski}, {Panter}, {Paz Arribas},
  {Pedaletti}, {Pelletier}, {Petrucci}, {Pita}, {P{\"u}hlhofer}, {Punch},
  {Quirrenbach}, {Raubenheimer}, {Raue}, {Rayner}, {Reimer}, {Renaud}, {de los
  Reyes}, {Rieger}, {Ripken}, {Rob}, {Rosier-Lees}, {Rowell}, {Rudak},
  {Rulten}, {Ruppel}, {Ryde}, {Sahakian}, {Santangelo}, {Schlickeiser},
  {Sch{\"o}ck}, {Sch{\"o}nwald}, {Schwanke}, {Schwarzburg}, {Schwemmer},
  {Shalchi}, {Sushch}, {Sikora}, {Skilton}, {Sol}, {Stawarz}, {Steenkamp},
  {Stegmann}, {Stinzing}, {Superina}, {Szostek}, {Tam}, {Tavernet}, {Terrier},
  {Tibolla}, {Tluczykont}, {Valerius}, {van Eldik}, {Vasileiadis}, {Venter},
  {Venter}, {Vialle}, {Viana}, {Vincent}, {Vivier}, {V{\"o}lk}, {Volpe},
  {Vorobiov}, {Wagner}, {Ward}, {Zdziarski}, {Zech}, \&
  {Zechlin}}]{2010A&A...520A..83H}
{H.~E.~S.~S. Collaboration}, {Abramowski}, A., {Acero}, F., {et~al.} 2010,
  \aap, 520, A83, \dodoi{10.1051/0004-6361/201014484}

\bibitem[{{Heil} {et~al.}(2012){Heil}, {Vaughan}, \&
  {Uttley}}]{2012MNRAS.422.2620H}
{Heil}, L.~M., {Vaughan}, S., \& {Uttley}, P. 2012, \mnras, 422, 2620,
  \dodoi{10.1111/j.1365-2966.2012.20824.x}

\bibitem[{{Jansen} {et~al.}(2001){Jansen}, {Lumb}, {Altieri}, {Clavel}, {Ehle},
  {Erd}, {Gabriel}, {Guainazzi}, {Gondoin}, {Much}, {Munoz}, {Santos},
  {Schartel}, {Texier}, \& {Vacanti}}]{2001A&A...365L...1J}
{Jansen}, F., {Lumb}, D., {Altieri}, B., {et~al.} 2001, \aap, 365, L1,
  \dodoi{10.1051/0004-6361:20000036}

\bibitem[{{Jolley} {et~al.}(2009){Jolley}, {Kuncic}, {Bicknell}, \&
  {Wagner}}]{2009MNRAS.400.1521J}
{Jolley}, E.~J.~D., {Kuncic}, Z., {Bicknell}, G.~V., \& {Wagner}, S. 2009,
  \mnras, 400, 1521, \dodoi{10.1111/j.1365-2966.2009.15554.x}

\bibitem[{{Kapanadze} {et~al.}(2014){Kapanadze}, {Romano}, {Vercellone}, \&
  {Kapanadze}}]{2014MNRAS.444.1077K}
{Kapanadze}, B., {Romano}, P., {Vercellone}, S., \& {Kapanadze}, S. 2014,
  \mnras, 444, 1077, \dodoi{10.1093/mnras/stu1504}

\bibitem[{{Kapanadze} {et~al.}(2020){Kapanadze}, {Gurchumelia}, {Dorner},
  {Vercellone}, {Romano}, {Hughes}, {Aller}, {Aller}, \&
  {Kharshiladze}}]{2020ApJS..247...27K}
{Kapanadze}, B., {Gurchumelia}, A., {Dorner}, D., {et~al.} 2020, \apjs, 247,
  27, \dodoi{10.3847/1538-4365/ab6322}

\bibitem[{{Kastendieck} {et~al.}(2011){Kastendieck}, {Ashley}, \&
  {Horns}}]{2011A&A...531A.123K}
{Kastendieck}, M.~A., {Ashley}, M.~C.~B., \& {Horns}, D. 2011, \aap, 531, A123,
  \dodoi{10.1051/0004-6361/201015918}

\bibitem[{{Kirk} {et~al.}(1998){Kirk}, {Rieger}, \&
  {Mastichiadis}}]{1998A&A...333..452K}
{Kirk}, J.~G., {Rieger}, F.~M., \& {Mastichiadis}, A. 1998, \aap, 333, 452,
  \dodoi{10.48550/arXiv.astro-ph/9801265}

\bibitem[{{Kundu} {et~al.}(2022){Kundu}, {Chatterjee}, {Mitra}, \&
  {Mondal}}]{2022MNRAS.510.3688K}
{Kundu}, A., {Chatterjee}, R., {Mitra}, K., \& {Mondal}, S. 2022, \mnras, 510,
  3688, \dodoi{10.1093/mnras/stab3750}

\bibitem[{{Kushwaha} {et~al.}(2017){Kushwaha}, {Sinha}, {Misra}, {Singh}, \&
  {de Gouveia Dal Pino}}]{2017ApJ...849..138K}
{Kushwaha}, P., {Sinha}, A., {Misra}, R., {Singh}, K.~P., \& {de Gouveia Dal
  Pino}, E.~M. 2017, \apj, 849, 138, \dodoi{10.3847/1538-4357/aa8ef5}

\bibitem[{{Lemoine} {et~al.}(2024){Lemoine}, {Murase}, \&
  {Rieger}}]{2024PhRvD.109f3006L}
{Lemoine}, M., {Murase}, K., \& {Rieger}, F. 2024, \prd, 109, 063006,
  \dodoi{10.1103/PhysRevD.109.063006}

\bibitem[{{Liuzzo} {et~al.}(2013){Liuzzo}, {Falomo}, {Treves}, {Arcidiacono},
  {Torresi}, {Uslenghi}, {Farinato}, {Moretti}, {Ragazzoni}, {Diolaiti},
  {Lombini}, {Brast}, {Donaldson}, {Kolb}, {Marchetti}, \&
  {Tordo}}]{2013AJ....145...73L}
{Liuzzo}, E., {Falomo}, R., {Treves}, A., {et~al.} 2013, \aj, 145, 73,
  \dodoi{10.1088/0004-6256/145/3/73}

\bibitem[{{Lyubarskii}(1997)}]{1997MNRAS.292..679L}
{Lyubarskii}, Y.~E. 1997, \mnras, 292, 679, \dodoi{10.1093/mnras/292.3.679}

\bibitem[{{Mannheim}(1993)}]{1993A&A...269...67M}
{Mannheim}, K. 1993, \aap, 269, 67, \dodoi{10.48550/arXiv.astro-ph/9302006}

\bibitem[{{Maraschi} {et~al.}(1992){Maraschi}, {Ghisellini}, \&
  {Celotti}}]{1992ApJ...397L...5M}
{Maraschi}, L., {Ghisellini}, G., \& {Celotti}, A. 1992, \apjl, 397, L5,
  \dodoi{10.1086/186531}

\bibitem[{{Markowitz} {et~al.}(2022){Markowitz}, {Nalewajko}, {Bhatta},
  {Dewangan}, {Chandra}, {Dorner}, {Schleicher}, {Pajdosz-{\'S}mierciak},
  {Stawarz}, {Zola}, {Ostrowski}, {Carosati}, {Krishnan}, {Bachev},
  {Ben{\'\i}tez}, {Gazeas}, {Hiriart}, {Hu}, {Larionov}, {Marchini},
  {Matsumoto}, {Nikiforova}, {Pursimo}, {Raiteri}, {Reichart}, {Rodriguez},
  {Semkov}, {Strigachev}, {Sugiura}, {Villata}, {Webb}, {Arbet-Engels},
  {Baack}, {Balbo}, {Biland}, {Bretz}, {Buss}, {Eisenberger}, {Elsaesser},
  {Hildebrand}, {Iotov}, {Kalenski}, {Mannheim}, {Mitchell}, {Neise}, {Noethe},
  {Paravac}, {Rhode}, {Sliusar}, \& {Walter}}]{2022MNRAS.513.1662M}
{Markowitz}, A.~G., {Nalewajko}, K., {Bhatta}, G., {et~al.} 2022, \mnras, 513,
  1662, \dodoi{10.1093/mnras/stac917}

\bibitem[{{Marscher}(2014)}]{2014ApJ...780...87M}
{Marscher}, A.~P. 2014, \apj, 780, 87, \dodoi{10.1088/0004-637X/780/1/87}

\bibitem[{{Marscher} \& {Jorstad}(2021)}]{2021Galax...9...27M}
{Marscher}, A.~P., \& {Jorstad}, S.~G. 2021, Galaxies, 9, 27,
  \dodoi{10.3390/galaxies9020027}

\bibitem[{{Massaro} {et~al.}(2004){Massaro}, {Perri}, {Giommi}, \&
  {Nesci}}]{2004A&A...413..489M}
{Massaro}, E., {Perri}, M., {Giommi}, P., \& {Nesci}, R. 2004, \aap, 413, 489,
  \dodoi{10.1051/0004-6361:20031558}

\bibitem[{{Massaro} {et~al.}(2006){Massaro}, {Tramacere}, {Perri}, {Giommi}, \&
  {Tosti}}]{2006A&A...448..861M}
{Massaro}, E., {Tramacere}, A., {Perri}, M., {Giommi}, P., \& {Tosti}, G. 2006,
  \aap, 448, 861, \dodoi{10.1051/0004-6361:20053644}

\bibitem[{{Mastichiadis} \& {Kirk}(1997)}]{1997A&A...320...19M}
{Mastichiadis}, A., \& {Kirk}, J.~G. 1997, \aap, 320, 19,
  \dodoi{10.48550/arXiv.astro-ph/9610058}

\bibitem[{{McHardy} {et~al.}(2004){McHardy}, {Papadakis}, {Uttley}, {Page}, \&
  {Mason}}]{2004MNRAS.348..783M}
{McHardy}, I.~M., {Papadakis}, I.~E., {Uttley}, P., {Page}, M.~J., \& {Mason},
  K.~O. 2004, \mnras, 348, 783, \dodoi{10.1111/j.1365-2966.2004.07376.x}

\bibitem[{{McKinney} {et~al.}(2012){McKinney}, {Tchekhovskoy}, \&
  {Blandford}}]{2012MNRAS.423.3083M}
{McKinney}, J.~C., {Tchekhovskoy}, A., \& {Blandford}, R.~D. 2012, \mnras, 423,
  3083, \dodoi{10.1111/j.1365-2966.2012.21074.x}

\bibitem[{{Mohorian} {et~al.}(2022){Mohorian}, {Bhatta}, {Adhikari}, {Dhital},
  {P{\'a}nis}, {Dinesh}, {Chaudhary}, {Bachchan}, \&
  {Stuchl{\'\i}k}}]{2022MNRAS.510.5280M}
{Mohorian}, M., {Bhatta}, G., {Adhikari}, T.~P., {et~al.} 2022, \mnras, 510,
  5280, \dodoi{10.1093/mnras/stab3738}

\bibitem[{{M{\"u}cke} {et~al.}(2003){M{\"u}cke}, {Protheroe}, {Engel},
  {Rachen}, \& {Stanev}}]{2003APh....18..593M}
{M{\"u}cke}, A., {Protheroe}, R.~J., {Engel}, R., {Rachen}, J.~P., \& {Stanev},
  T. 2003, Astroparticle Physics, 18, 593,
  \dodoi{10.1016/S0927-6505(02)00185-8}

\bibitem[{{Nandra} {et~al.}(1997){Nandra}, {George}, {Mushotzky}, {Turner}, \&
  {Yaqoob}}]{1997ApJ...476...70N}
{Nandra}, K., {George}, I.~M., {Mushotzky}, R.~F., {Turner}, T.~J., \&
  {Yaqoob}, T. 1997, \apj, 476, 70, \dodoi{10.1086/303600}

\bibitem[{{Nilsson} {et~al.}(2018){Nilsson}, {Lindfors}, {Takalo}, {Reinthal},
  {Berdyugin}, {Sillanp{\"a}{\"a}}, {Ciprini}, {Halkola}, {Hein{\"a}m{\"a}ki},
  {Hovatta}, {Kadenius}, {Nurmi}, {Ostorero}, {Pasanen}, {Rekola}, {Saarinen},
  {Sainio}, {Tuominen}, {Villforth}, {Vornanen}, \&
  {Zaprudin}}]{2018A&A...620A.185N}
{Nilsson}, K., {Lindfors}, E., {Takalo}, L.~O., {et~al.} 2018, \aap, 620, A185,
  \dodoi{10.1051/0004-6361/201833621}

\bibitem[{{Noel} {et~al.}(2022){Noel}, {Gaur}, {Gupta}, {Wierzcholska},
  {Ostrowski}, {Dhiman}, \& {Bhatta}}]{2022ApJS..262....4N}
{Noel}, A.~P., {Gaur}, H., {Gupta}, A.~C., {et~al.} 2022, \apjs, 262, 4,
  \dodoi{10.3847/1538-4365/ac7799}

\bibitem[{{Otero-Santos} {et~al.}(2024){Otero-Santos}, {Raiteri},
  {Acosta-Pulido}, {Carnerero}, {Villata}, {Savchenko}, {Carosati}, {Chen},
  {Kurtanidze}, {Joner}, {Semkov}, {Pursimo}, {Ben{\'\i}tez}, {Damljanovic},
  {Apolonio}, {Borman}, {Bozhilov}, {Galindo-Guil}, {Grishina}, {Hagen-Thorn},
  {Hiriart}, {Hsiao}, {Ibryamov}, {Ivanidze}, {Kimeridze}, {Kopatskaya},
  {Kurtanidze}, {Larionov}, {Larionova}, {Larionova}, {Minev}, {Morozova},
  {Nikolashvili}, {Ovcharov}, {Sigua}, {Stojanovic}, {Troitskiy}, {Troitskaya},
  {Tsai}, {Valcheva}, {Vasilyev}, {Vince}, {Zaharieva}, \&
  {Zhovtan}}]{2024A&A...686A.228O}
{Otero-Santos}, J., {Raiteri}, C.~M., {Acosta-Pulido}, J.~A., {et~al.} 2024,
  \aap, 686, A228, \dodoi{10.1051/0004-6361/202449647}

\bibitem[{{Paltani} {et~al.}(1998){Paltani}, {Courvoisier}, \&
  {Walter}}]{1998A&A...340...47P}
{Paltani}, S., {Courvoisier}, T.~J.~L., \& {Walter}, R. 1998, \aap, 340, 47,
  \dodoi{10.48550/arXiv.astro-ph/9809113}

\bibitem[{{Pe{\~n}il} {et~al.}(2020){Pe{\~n}il}, {Dom{\'\i}nguez}, {Buson},
  {Ajello}, {Otero-Santos}, {Barrio}, {Nemmen}, {Cutini}, {Rani},
  {Franckowiak}, \& {Cavazzuti}}]{2020ApJ...896..134P}
{Pe{\~n}il}, P., {Dom{\'\i}nguez}, A., {Buson}, S., {et~al.} 2020, \apj, 896,
  134, \dodoi{10.3847/1538-4357/ab910d}

\bibitem[{{Piner} {et~al.}(2010){Piner}, {Pant}, \&
  {Edwards}}]{2010ApJ...723.1150P}
{Piner}, B.~G., {Pant}, N., \& {Edwards}, P.~G. 2010, \apj, 723, 1150,
  \dodoi{10.1088/0004-637X/723/2/1150}

\bibitem[{{Pininti} {et~al.}(2023){Pininti}, {Bhatta}, {Paul}, {Kumar},
  {Rajgor}, {Barnwal}, \& {Gharat}}]{2023MNRAS.518.1459P}
{Pininti}, V.~R., {Bhatta}, G., {Paul}, S., {et~al.} 2023, \mnras, 518, 1459,
  \dodoi{10.1093/mnras/stac3125}

\bibitem[{{Press}(1978)}]{1978ComAp...7..103P}
{Press}, W.~H. 1978, Comments on Astrophysics, 7, 103

\bibitem[{{Raiteri} {et~al.}(2017){Raiteri}, {Villata}, {Acosta-Pulido},
  {Agudo}, {Arkharov}, {Bachev}, {Baida}, {Ben{\'\i}tez}, {Borman}, {Boschin},
  {Bozhilov}, {Butuzova}, {Calcidese}, {Carnerero}, {Carosati}, {Casadio},
  {Castro-Segura}, {Chen}, {Damljanovic}, {D'Ammando}, {di Paola},
  {Echevarr{\'\i}a}, {Efimova}, {Ehgamberdiev}, {Espinosa}, {Fuentes},
  {Giunta}, {G{\'o}mez}, {Grishina}, {Gurwell}, {Hiriart}, {Jermak}, {Jordan},
  {Jorstad}, {Joshi}, {Kopatskaya}, {Kuratov}, {Kurtanidze}, {Kurtanidze},
  {L{\"a}hteenm{\"a}ki}, {Larionov}, {Larionova}, {Larionova}, {L{\'a}zaro},
  {Lin}, {Malmrose}, {Marscher}, {Matsumoto}, {McBreen}, {Michel}, {Mihov},
  {Minev}, {Mirzaqulov}, {Mokrushina}, {Molina}, {Moody}, {Morozova},
  {Nazarov}, {Nikolashvili}, {Ohlert}, {Okhmat}, {Ovcharov}, {Pinna},
  {Polakis}, {Protasio}, {Pursimo}, {Redondo-Lorenzo}, {Rizzi},
  {Rodriguez-Coira}, {Sadakane}, {Sadun}, {Samal}, {Savchenko}, {Semkov},
  {Skiff}, {Slavcheva-Mihova}, {Smith}, {Steele}, {Strigachev}, {Tammi},
  {Thum}, {Tornikoski}, {Troitskaya}, {Troitsky}, {Vasilyev}, \&
  {Vince}}]{2017Natur.552..374R}
{Raiteri}, C.~M., {Villata}, M., {Acosta-Pulido}, J.~A., {et~al.} 2017, \nat,
  552, 374, \dodoi{10.1038/nature24623}

\bibitem[{{Raiteri} {et~al.}(2023){Raiteri}, {Villata}, {Jorstad}, {Marscher},
  {Acosta Pulido}, {Carosati}, {Chen}, {Joner}, {Kurtanidze}, {Lorey},
  {Marchini}, {Matsumoto}, {Mirzaqulov}, {Savchenko}, {Strigachev}, {Vince},
  {Aceti}, {Apolonio}, {Arena}, {Arkharov}, {Bachev}, {Bader}, {Banfi},
  {Bonnoli}, {Borman}, {Bozhilov}, {Brown}, {Carbonell}, {Carnerero},
  {Damljanovic}, {Dhiman}, {Ehgamberdiev}, {Elsaesser}, {Feige}, {Gabellini},
  {Gal{\'a}n}, {Galli}, {Gaur}, {Gazeas}, {Grishina}, {Gupta}, {Hagen-Thorn},
  {Hallum}, {Hart}, {Hasuda}, {Heidemann}, {Horst}, {Hou}, {Ibryamov},
  {Ivanidze}, {Jovanovic}, {Kimeridze}, {Kishore}, {Klimanov}, {Kopatskaya},
  {Kurtanidze}, {Kushwaha}, {Lane}, {Larionova}, {Leonini}, {Lin}, {Mannheim},
  {Marino}, {Minev}, {Modaressi}, {Morozova}, {Mortari}, {Nazarov},
  {Nikolashvili}, {Otero Santos}, {Ovcharov}, {Papini}, {Pinter}, {Privitera},
  {Pursimo}, {Reinhart}, {Roberts}, {Romanov}, {Rosenlehner}, {Sakamoto},
  {Salvaggio}, {Schoch}, {Semkov}, {Seufert}, {Shakhovskoy}, {Sigua}, {Singh},
  {Steineke}, {Stojanovic}, {Tripathi}, {Troitskaya}, {Troitskiy}, {Tsai},
  {Valcheva}, {Vasilyev}, {Vrontaki}, {Weaver}, {Wooley}, {Zaharieva}, \&
  {Zhovtan}}]{2023MNRAS.522..102R}
{Raiteri}, C.~M., {Villata}, M., {Jorstad}, S.~G., {et~al.} 2023, \mnras, 522,
  102, \dodoi{10.1093/mnras/stad942}

\bibitem[{{Rajput} {et~al.}(2021){Rajput}, {Shah}, {Stalin}, {Sahayanathan}, \&
  {Rakshit}}]{2021MNRAS.504.1772R}
{Rajput}, B., {Shah}, Z., {Stalin}, C.~S., {Sahayanathan}, S., \& {Rakshit}, S.
  2021, \mnras, 504, 1772, \dodoi{10.1093/mnras/stab970}

\bibitem[{{Rajput} {et~al.}(2020){Rajput}, {Stalin}, \&
  {Rakshit}}]{2020A&A...634A..80R}
{Rajput}, B., {Stalin}, C.~S., \& {Rakshit}, S. 2020, \aap, 634, A80,
  \dodoi{10.1051/0004-6361/201936769}

\bibitem[{{Rybicki} \& {Lightman}(1986)}]{1986rpa..book.....R}
{Rybicki}, G.~B., \& {Lightman}, A.~P. 1986, {Radiative Processes in
  Astrophysics}

\bibitem[{{Sandrinelli} {et~al.}(2014){Sandrinelli}, {Covino}, \&
  {Treves}}]{2014ApJ...793L...1S}
{Sandrinelli}, A., {Covino}, S., \& {Treves}, A. 2014, \apjl, 793, L1,
  \dodoi{10.1088/2041-8205/793/1/L1}

\bibitem[{{Schwartz} {et~al.}(1979){Schwartz}, {Doxsey}, {Griffiths},
  {Johnston}, \& {Schwarz}}]{1979ApJ...229L..53S}
{Schwartz}, D.~A., {Doxsey}, R.~E., {Griffiths}, R.~E., {Johnston}, M.~D., \&
  {Schwarz}, J. 1979, \apjl, 229, L53, \dodoi{10.1086/182929}

\bibitem[{{Shah} {et~al.}(2018){Shah}, {Mankuzhiyil}, {Sinha}, {Misra},
  {Sahayanathan}, \& {Iqbal}}]{2018RAA....18..141S}
{Shah}, Z., {Mankuzhiyil}, N., {Sinha}, A., {et~al.} 2018, Research in
  Astronomy and Astrophysics, 18, 141, \dodoi{10.1088/1674-4527/18/11/141}

\bibitem[{{Sikora} {et~al.}(1994){Sikora}, {Begelman}, \&
  {Rees}}]{1994ApJ...421..153S}
{Sikora}, M., {Begelman}, M.~C., \& {Rees}, M.~J. 1994, \apj, 421, 153,
  \dodoi{10.1086/173633}

\bibitem[{{Stickel} {et~al.}(1991){Stickel}, {Padovani}, {Urry}, {Fried}, \&
  {Kuehr}}]{1991ApJ...374..431S}
{Stickel}, M., {Padovani}, P., {Urry}, C.~M., {Fried}, J.~W., \& {Kuehr}, H.
  1991, \apj, 374, 431, \dodoi{10.1086/170133}

\bibitem[{{Stocke} {et~al.}(1991){Stocke}, {Morris}, {Gioia}, {Maccacaro},
  {Schild}, {Wolter}, {Fleming}, \& {Henry}}]{1991ApJS...76..813S}
{Stocke}, J.~T., {Morris}, S.~L., {Gioia}, I.~M., {et~al.} 1991, \apjs, 76,
  813, \dodoi{10.1086/191582}

\bibitem[{{Str{\"u}der} {et~al.}(2001){Str{\"u}der}, {Briel}, {Dennerl},
  {Hartmann}, {Kendziorra}, {Meidinger}, {Pfeffermann}, {Reppin}, {Aschenbach},
  {Bornemann}, {Br{\"a}uninger}, {Burkert}, {Elender}, {Freyberg}, {Haberl},
  {Hartner}, {Heuschmann}, {Hippmann}, {Kastelic}, {Kemmer}, {Kettenring},
  {Kink}, {Krause}, {M{\"u}ller}, {Oppitz}, {Pietsch}, {Popp}, {Predehl},
  {Read}, {Stephan}, {St{\"o}tter}, {Tr{\"u}mper}, {Holl}, {Kemmer}, {Soltau},
  {St{\"o}tter}, {Weber}, {Weichert}, {von Zanthier}, {Carathanassis}, {Lutz},
  {Richter}, {Solc}, {B{\"o}ttcher}, {Kuster}, {Staubert}, {Abbey}, {Holland},
  {Turner}, {Balasini}, {Bignami}, {La Palombara}, {Villa}, {Buttler},
  {Gianini}, {Lain{\'e}}, {Lumb}, \& {Dhez}}]{2001A&A...365L..18S}
{Str{\"u}der}, L., {Briel}, U., {Dennerl}, K., {et~al.} 2001, \aap, 365, L18,
  \dodoi{10.1051/0004-6361:20000066}

\bibitem[{{Tchekhovskoy} \& {Giannios}(2015)}]{2015MNRAS.447..327T}
{Tchekhovskoy}, A., \& {Giannios}, D. 2015, \mnras, 447, 327,
  \dodoi{10.1093/mnras/stu2229}

\bibitem[{{Tramacere} {et~al.}(2007){Tramacere}, {Massaro}, \&
  {Cavaliere}}]{2007A&A...466..521T}
{Tramacere}, A., {Massaro}, F., \& {Cavaliere}, A. 2007, \aap, 466, 521,
  \dodoi{10.1051/0004-6361:20066723}

\bibitem[{{Urry} \& {Padovani}(1995)}]{urry1995}
{Urry}, C.~M., \& {Padovani}, P. 1995, \pasp, 107, 803, \dodoi{10.1086/133630}

\bibitem[{{Uttley} \& {McHardy}(2001)}]{2001MNRAS.323L..26U}
{Uttley}, P., \& {McHardy}, I.~M. 2001, \mnras, 323, L26,
  \dodoi{10.1046/j.1365-8711.2001.04496.x}

\bibitem[{{Uttley} {et~al.}(2002){Uttley}, {McHardy}, \&
  {Papadakis}}]{2002MNRAS.332..231U}
{Uttley}, P., {McHardy}, I.~M., \& {Papadakis}, I.~E. 2002, \mnras, 332, 231,
  \dodoi{10.1046/j.1365-8711.2002.05298.x}

\bibitem[{{Uttley} {et~al.}(2005){Uttley}, {McHardy}, \&
  {Vaughan}}]{2005MNRAS.359..345U}
{Uttley}, P., {McHardy}, I.~M., \& {Vaughan}, S. 2005, \mnras, 359, 345,
  \dodoi{10.1111/j.1365-2966.2005.08886.x}

\bibitem[{{Vaughan} {et~al.}(2003){Vaughan}, {Edelson}, {Warwick}, \&
  {Uttley}}]{2003MNRAS.345.1271V}
{Vaughan}, S., {Edelson}, R., {Warwick}, R.~S., \& {Uttley}, P. 2003, \mnras,
  345, 1271, \dodoi{10.1046/j.1365-2966.2003.07042.x}

\bibitem[{Wang {et~al.}(2023)Wang, Yi, Wang, Mao, Pu, Ning, Huang, Lu, Zhang,
  Chen, \& Dong}]{wang2023comprehensive}
Wang, N., Yi, T.-F., Wang, L., {et~al.} 2023, Comprehensive study of the
  blazars from Fermi-LAT LCR: The log-normal flux distribution and linear
  RMS-Flux relation.
\newblock \doarXiv{2307.10547}

\bibitem[{{Wang} {et~al.}(2018){Wang}, {Xue}, {Zhu}, \&
  {Fan}}]{2018ApJ...867...68W}
{Wang}, Y., {Xue}, Y., {Zhu}, S., \& {Fan}, J. 2018, \apj, 867, 68,
  \dodoi{10.3847/1538-4357/aae307}

\bibitem[{{Webb} {et~al.}(2021){Webb}, {Arroyave}, {Laurence}, {Revesz},
  {Bhatta}, {Hollingsworth}, {Dhalla}, {Howard}, \&
  {Cioffi}}]{2021Galax...9..114W}
{Webb}, J.~R., {Arroyave}, V., {Laurence}, D., {et~al.} 2021, Galaxies, 9, 114,
  \dodoi{10.3390/galaxies9040114}

\bibitem[{{Young}(2010)}]{2010Natur.463..886Y}
{Young}, A. 2010, \nat, 463, 886, \dodoi{10.1038/463886a}

\bibitem[{{Zhang} {et~al.}(2023){Zhang}, {Marscher}, {Guo}, {Giannios}, {Li},
  \& {Negro}}]{2023ApJ...949...71Z}
{Zhang}, H., {Marscher}, A.~P., {Guo}, F., {et~al.} 2023, \apj, 949, 71,
  \dodoi{10.3847/1538-4357/acc657}

\bibitem[{{Zhang} {et~al.}(2021){Zhang}, {Gupta}, {Gaur}, {Wiita}, {An}, {Lu},
  {Fan}, \& {Xu}}]{2021ApJ...909..103Z}
{Zhang}, Z., {Gupta}, A.~C., {Gaur}, H., {et~al.} 2021, \apj, 909, 103,
  \dodoi{10.3847/1538-4357/abdd38}

\end{thebibliography}
\bibliographystyle{aasjournal}

\clearpage
\appendix
\section{Light curves, hardness ratio, X-ray spectra and PSD plots and flux histograms for the rest observations of  {\it XMM-Newton} observations of PKS 2155-304.}

\begin{figure*}
    \centering{
    \includegraphics[width=5truecm]{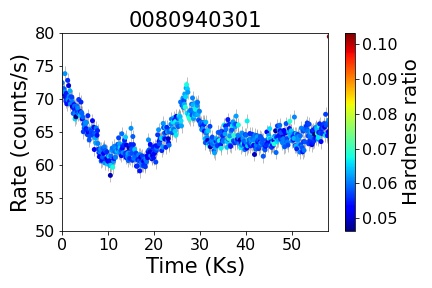}
    \includegraphics[width=5truecm]{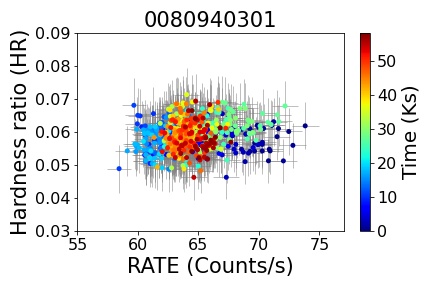}
    \includegraphics[width=5truecm]{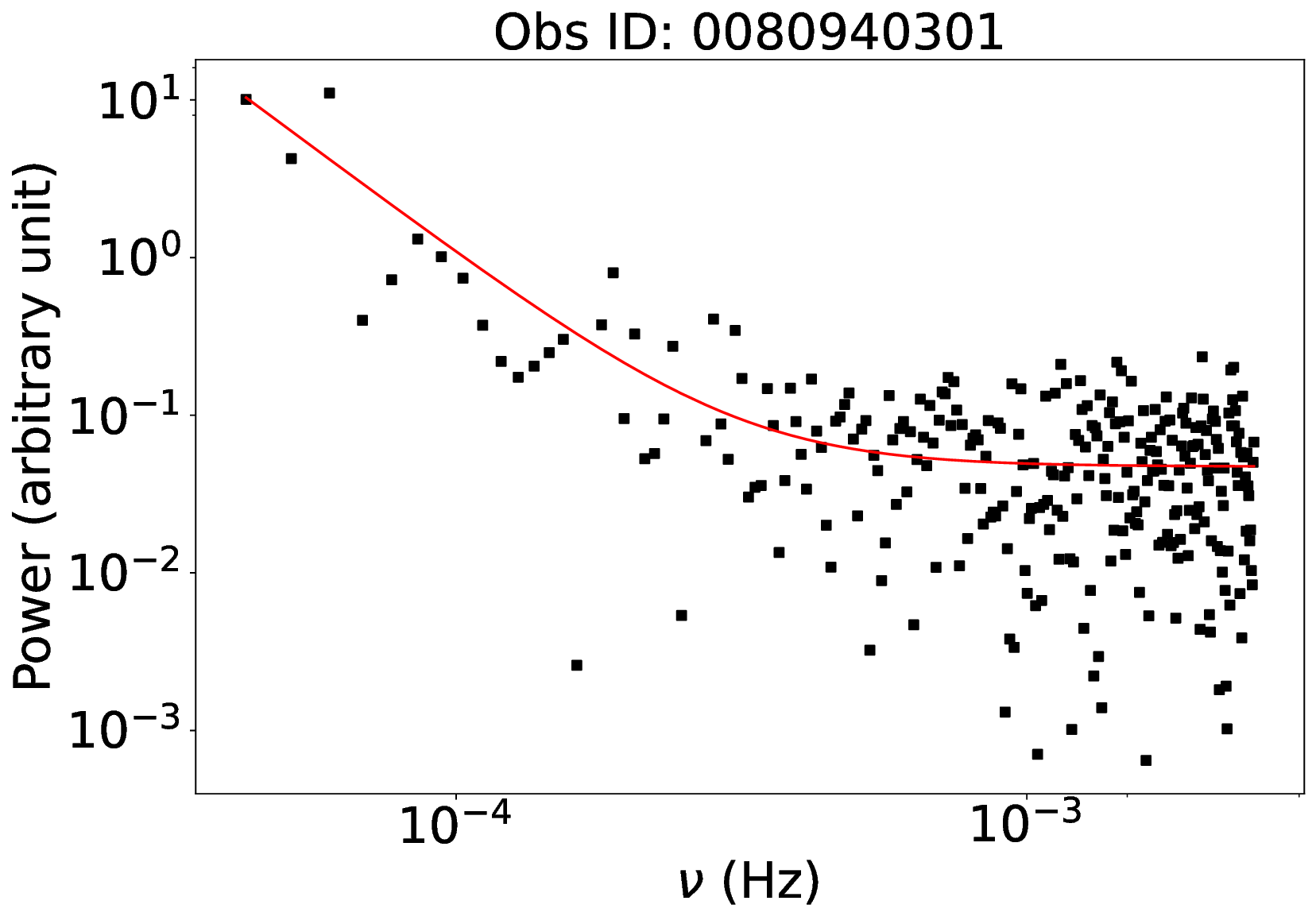}

    \includegraphics[width=5truecm]{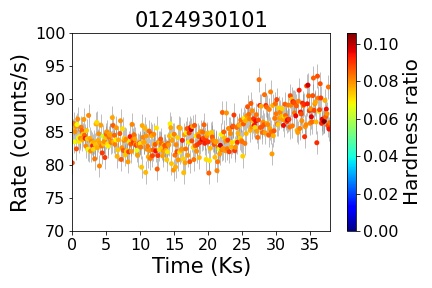}
    \includegraphics[width=5truecm]{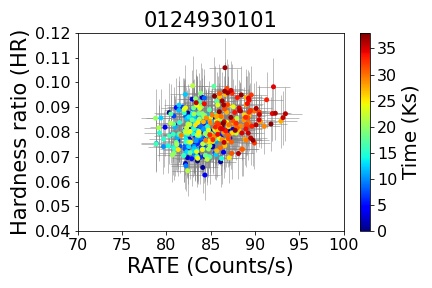}
    \includegraphics[width=5truecm]{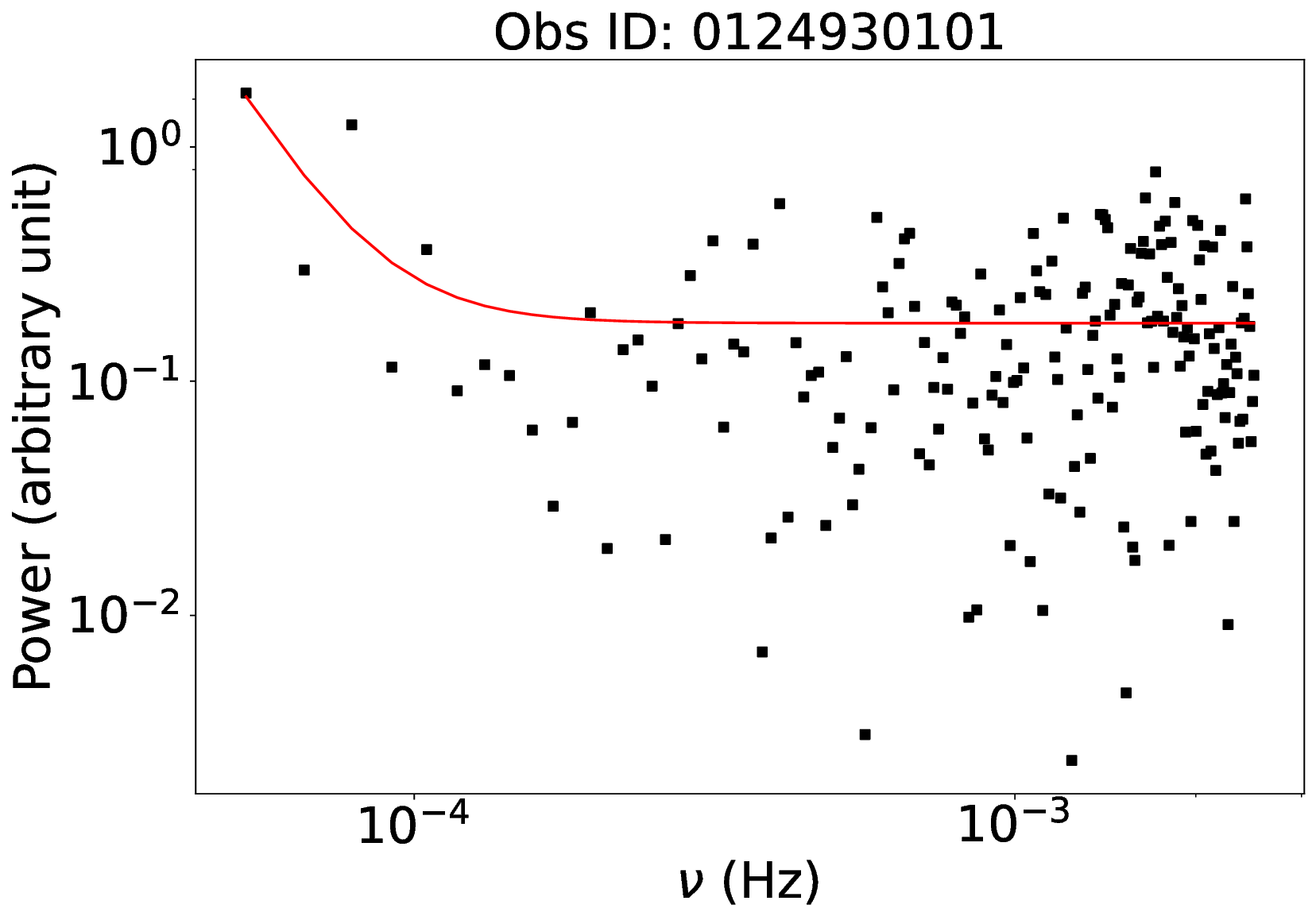}
    
    \includegraphics[width=5truecm]{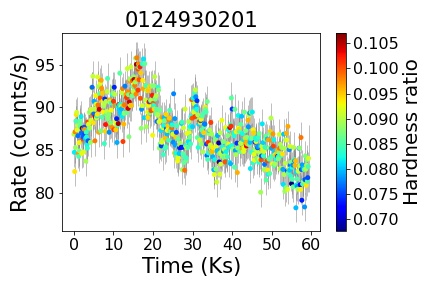}
    \includegraphics[width=5truecm]{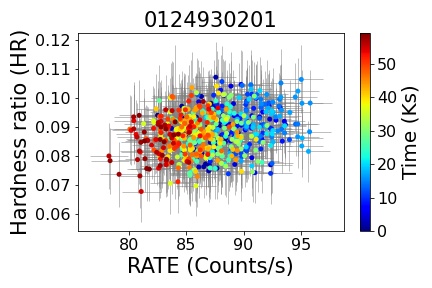}
    \includegraphics[width=5truecm]{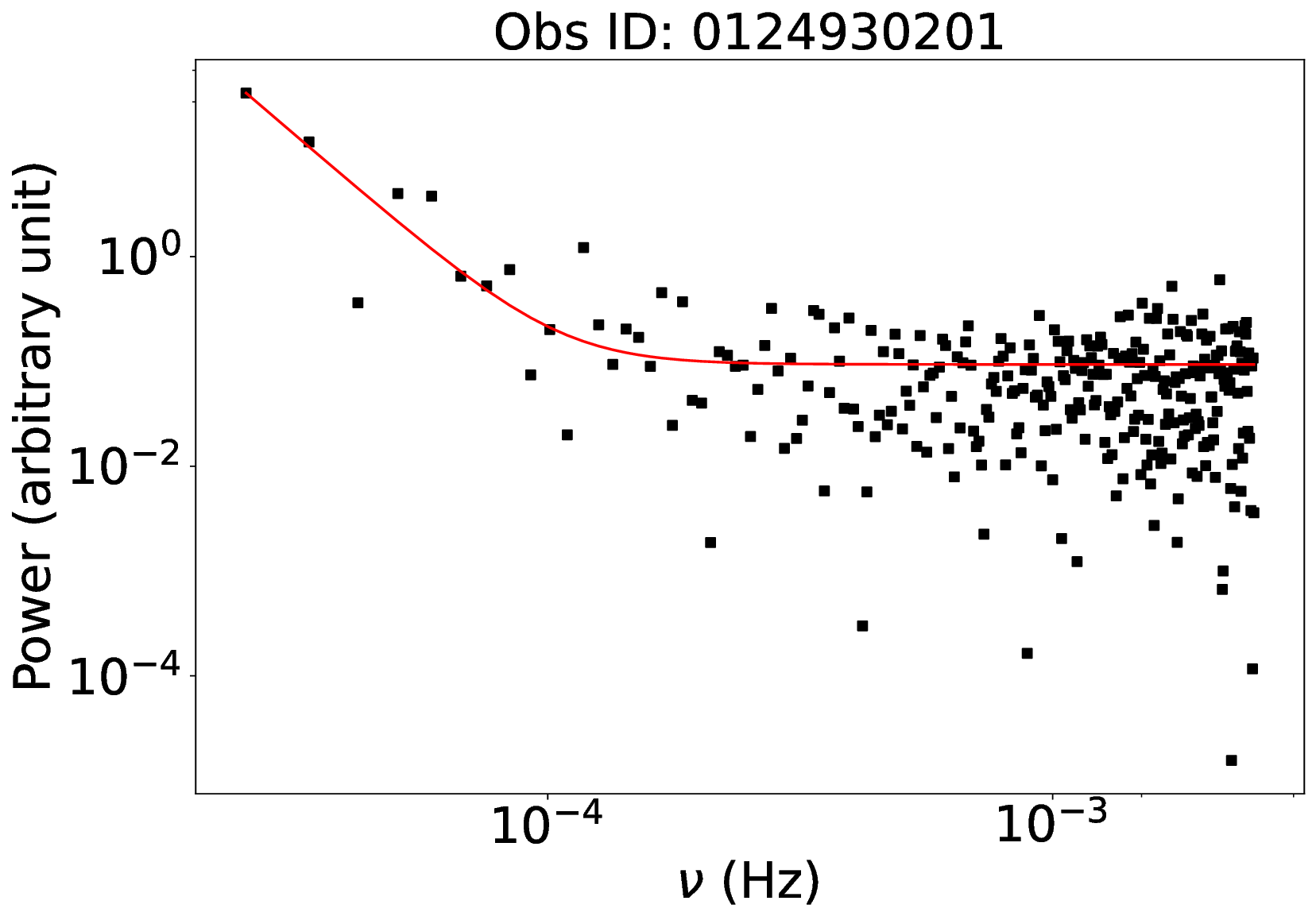}
    
    \includegraphics[width=5truecm]{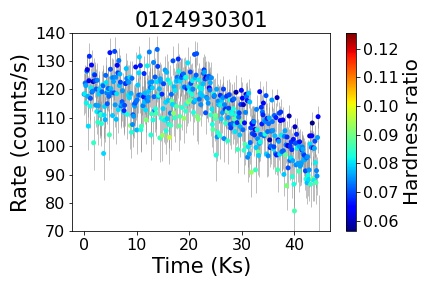}
    \includegraphics[width=5truecm]{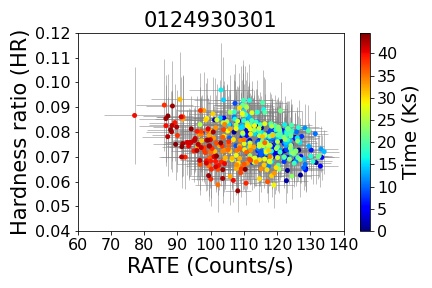}
    \includegraphics[width=5truecm]{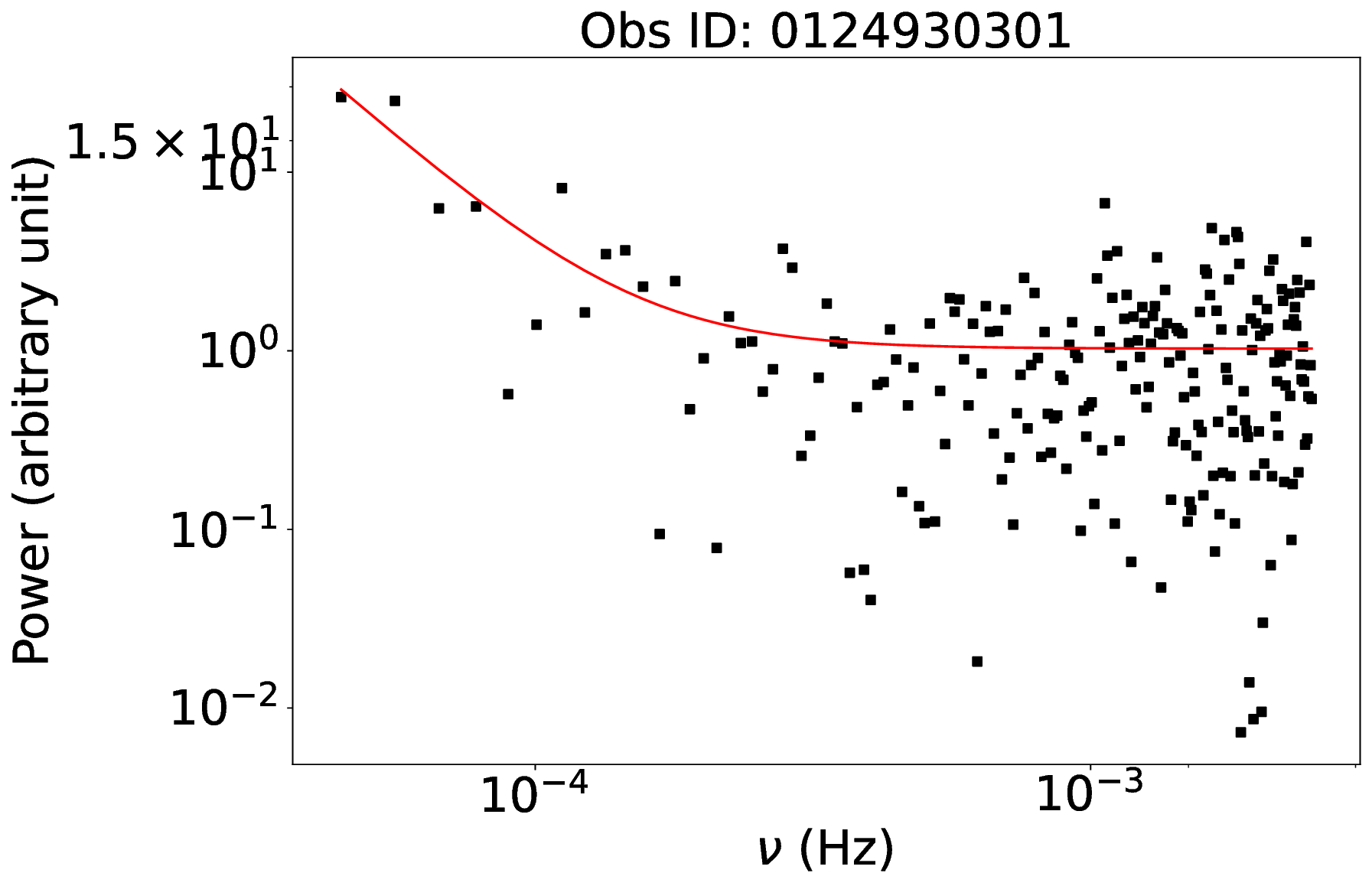}
    
    \includegraphics[width=5truecm]{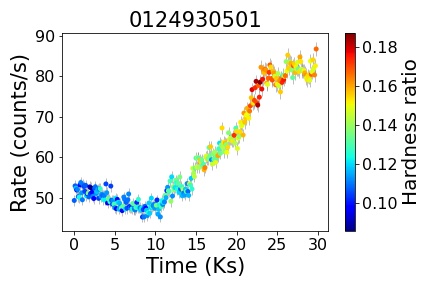}
    \includegraphics[width=5truecm]{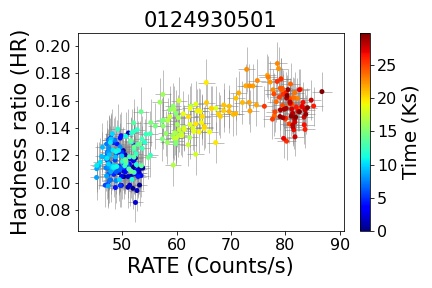}
    \includegraphics[width=5truecm]{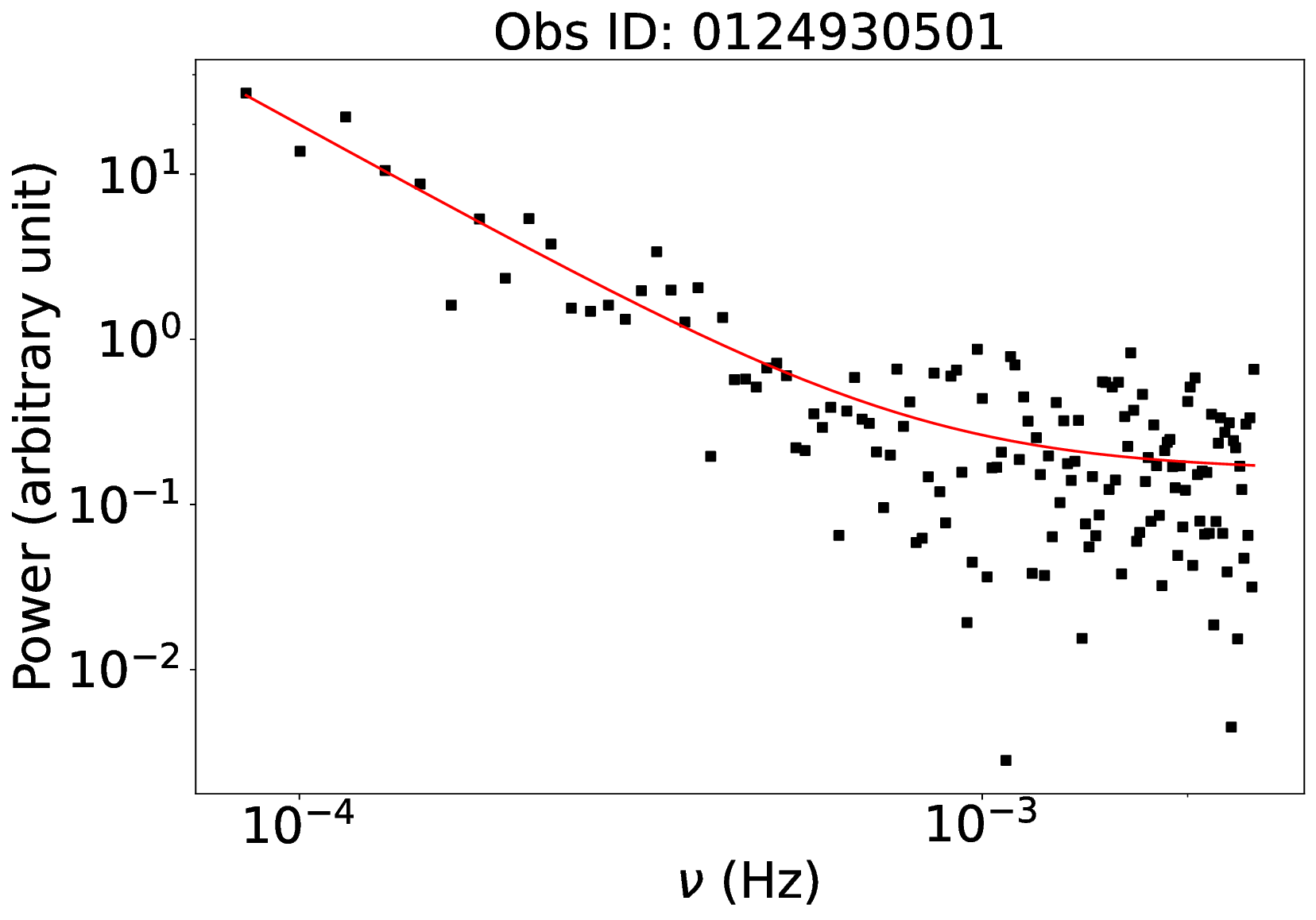}
    \includegraphics[width=5truecm]{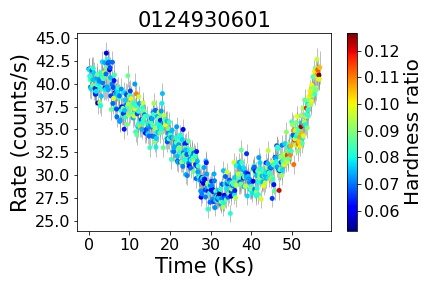}
    \includegraphics[width=5truecm]{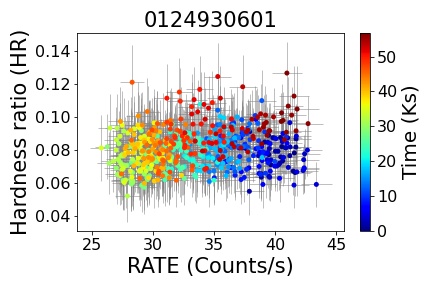}
    \includegraphics[width=5truecm]{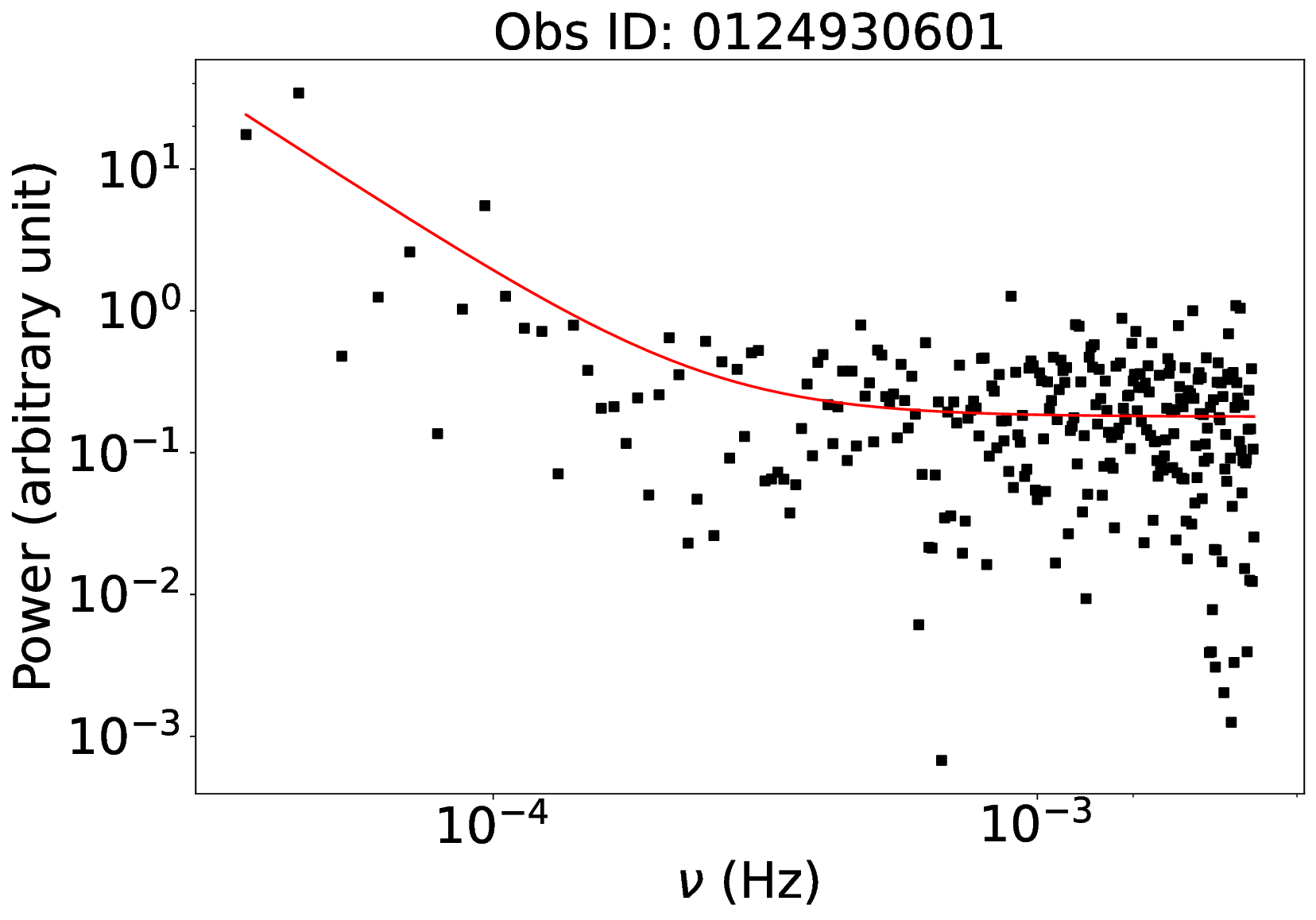}}
     \caption{LCs (Left), Hardness ratio (Middle) and PSD (Right) of Blazar PKS2155-304 from 2000 to 2014}
    \label{fig:0test}
\end{figure*}

\begin{figure*} 
    \centering{
    
    \includegraphics[width=5truecm]{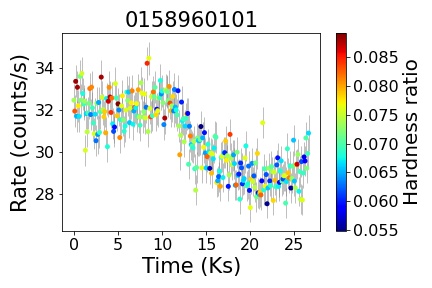}
    \includegraphics[width=5truecm]{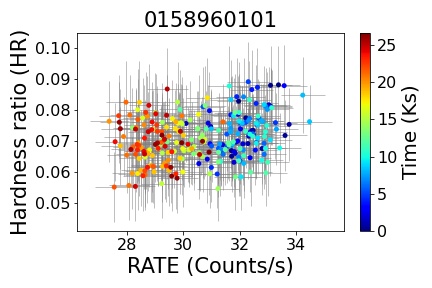}
    \includegraphics[width=5truecm]{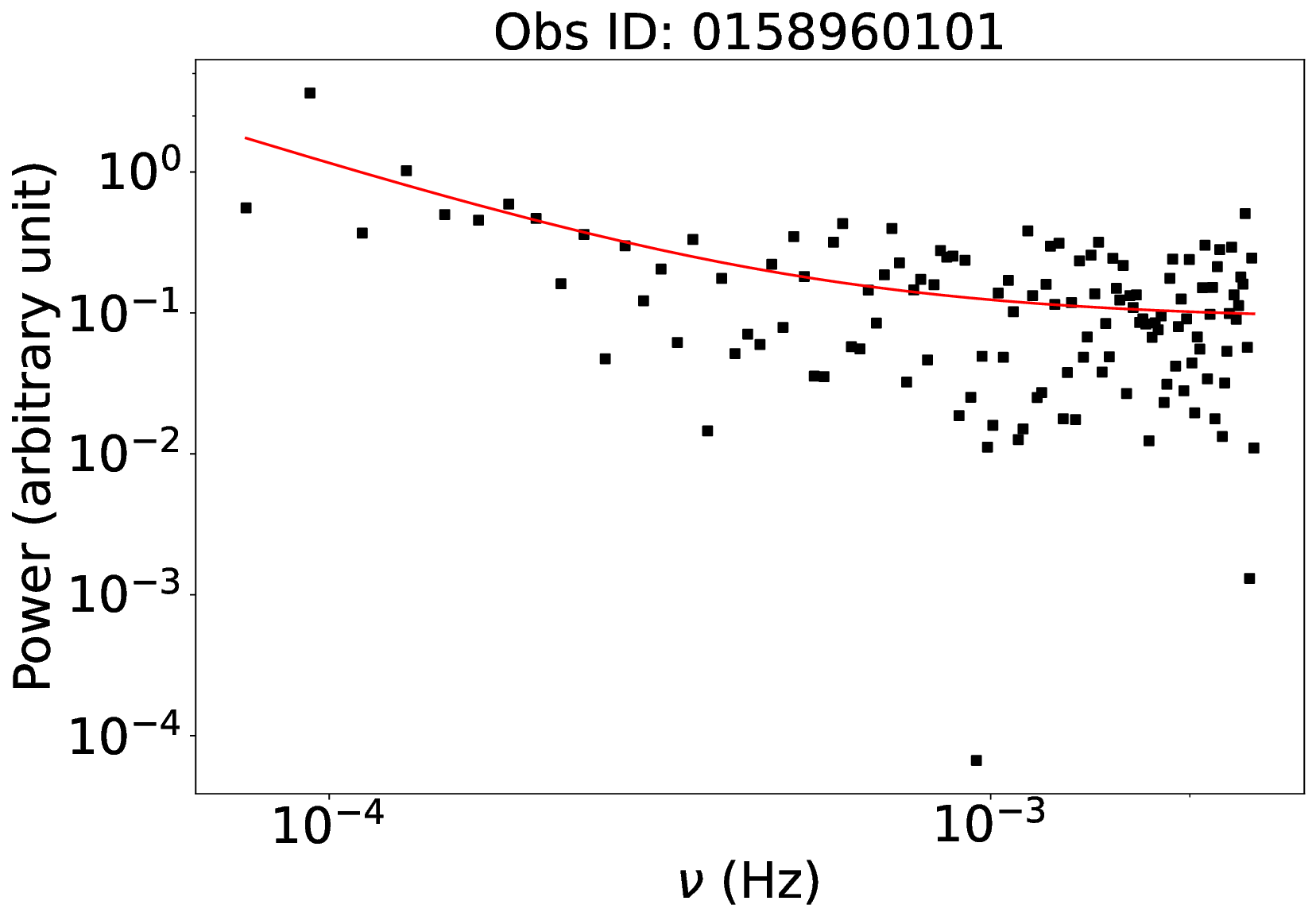}
    \includegraphics[width=5truecm]{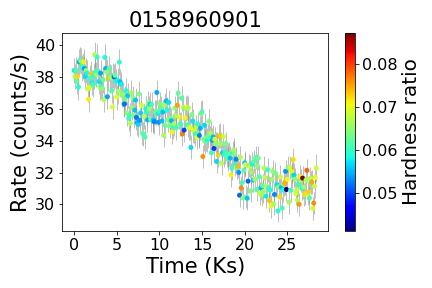}
    \includegraphics[width=5truecm]{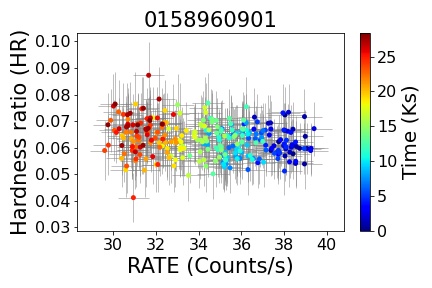}
    \includegraphics[width=5truecm]{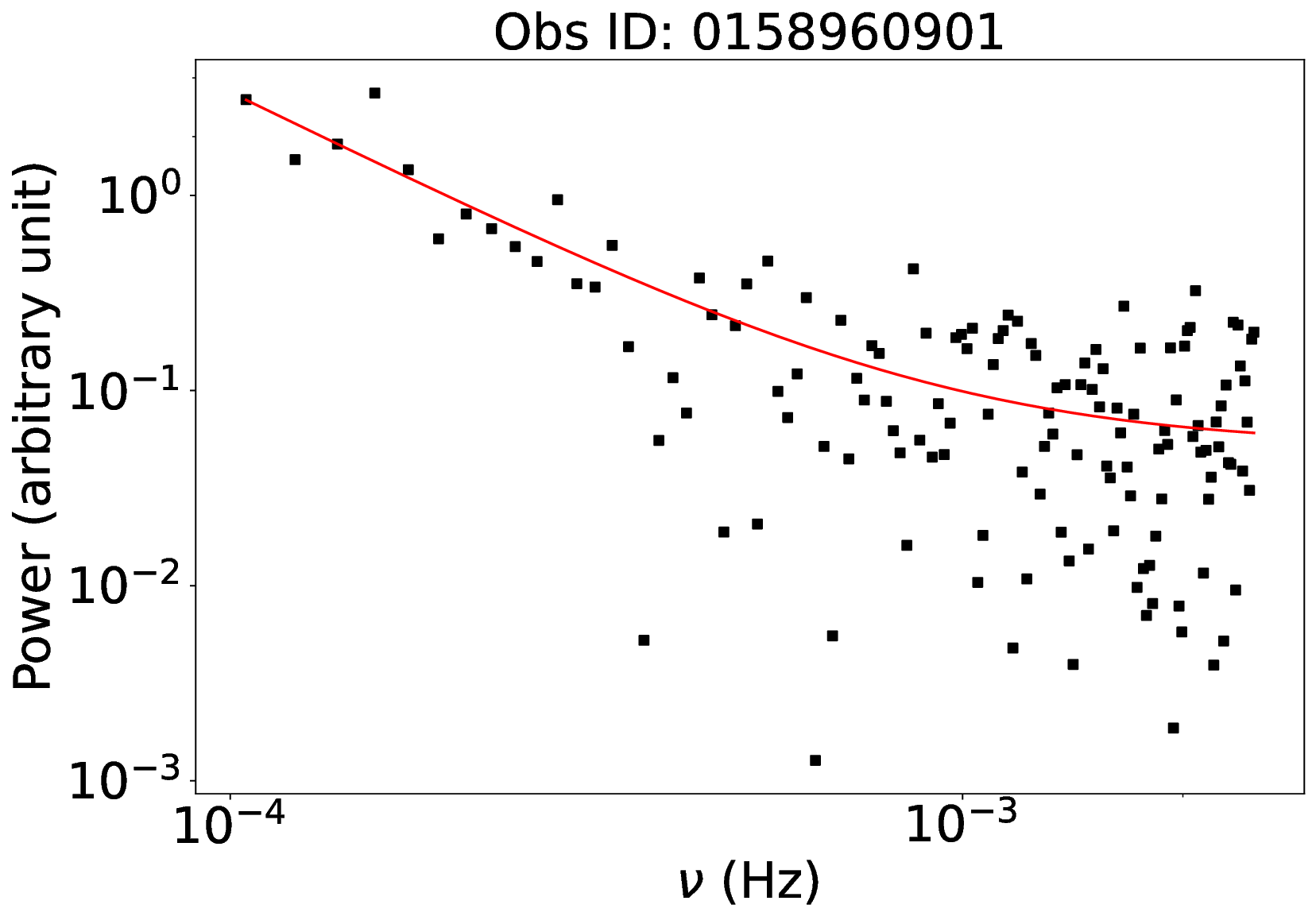}
    \includegraphics[width=5truecm]{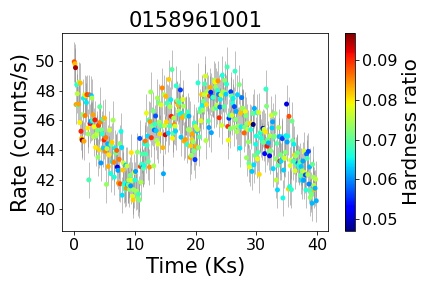}
    \includegraphics[width=5truecm]{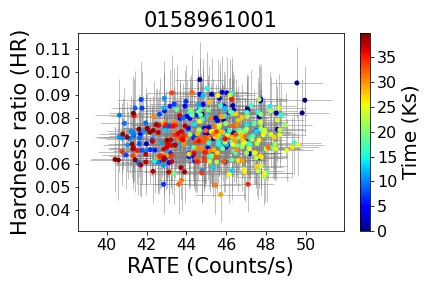}
    \includegraphics[width=5truecm]{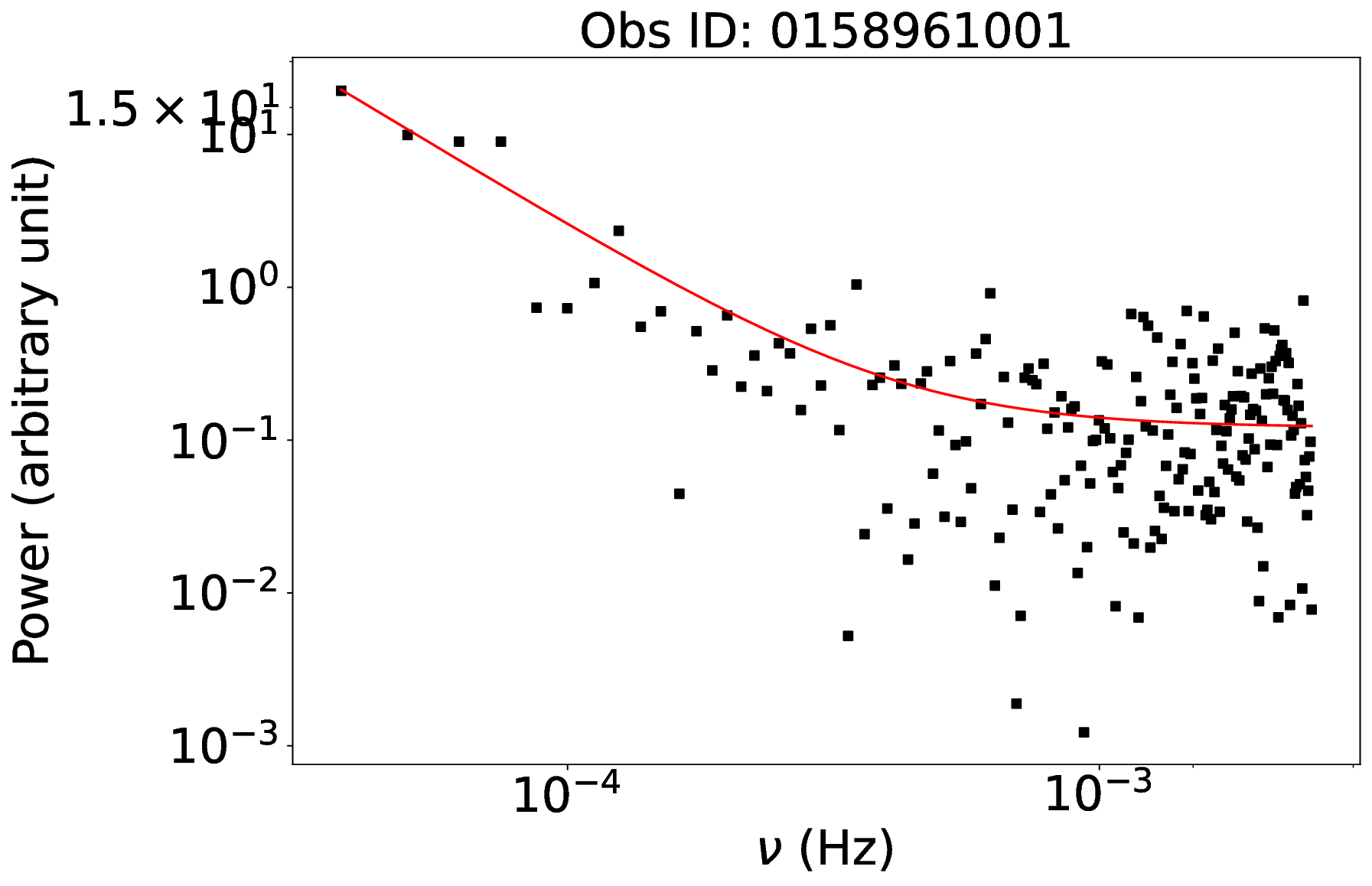}
    \includegraphics[width=5truecm]{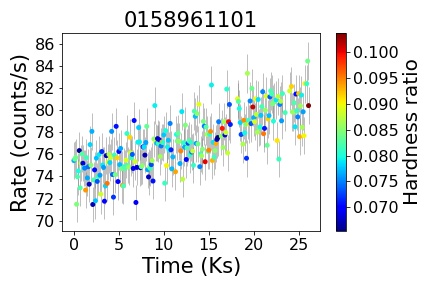}
    \includegraphics[width=5truecm]{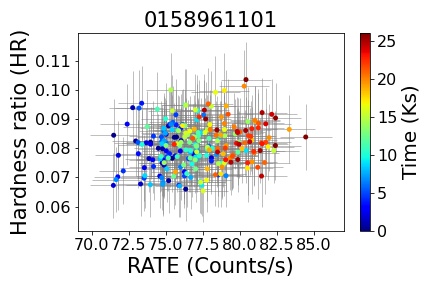}
    \includegraphics[width=5truecm]{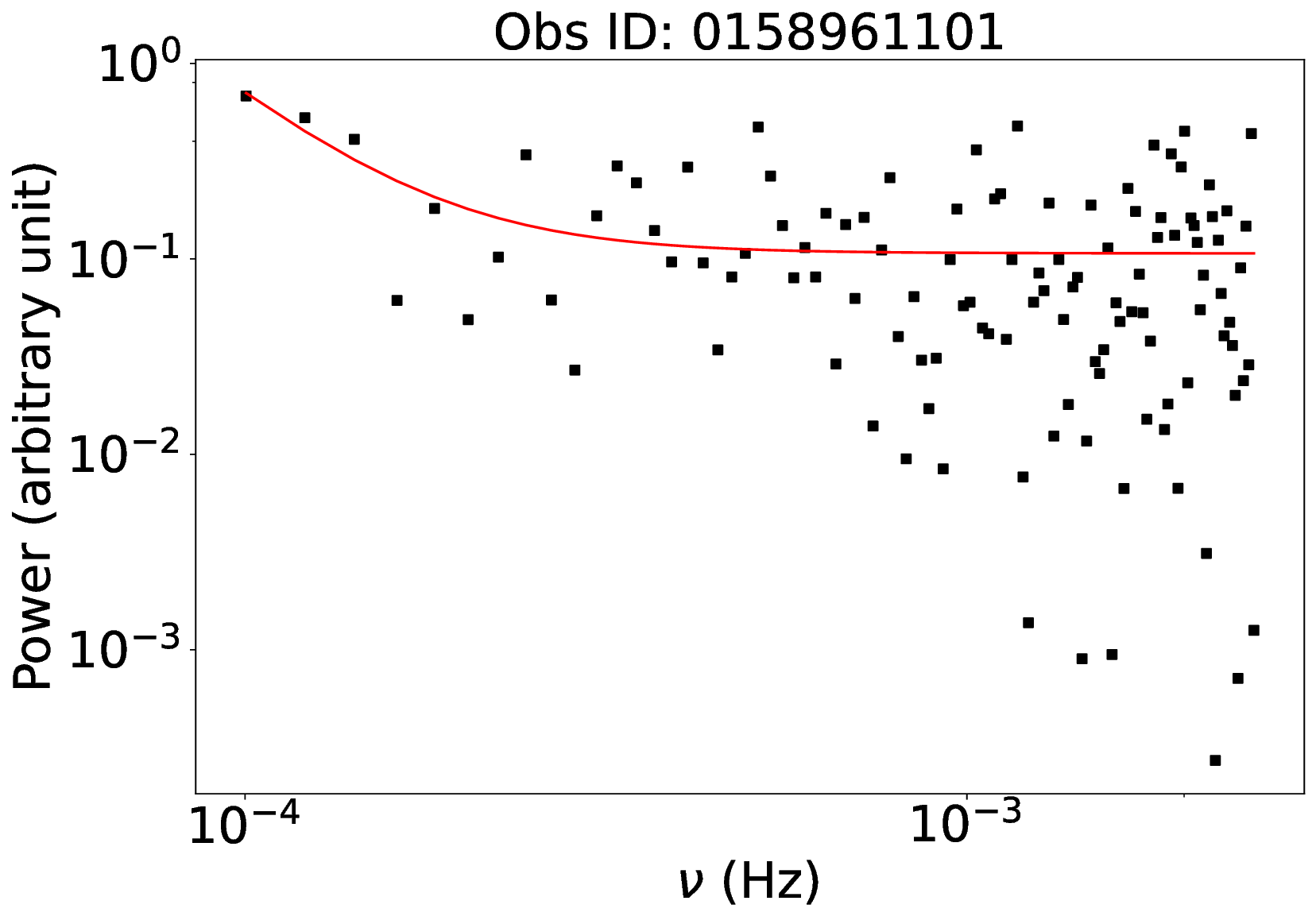}
    \includegraphics[width=5truecm]{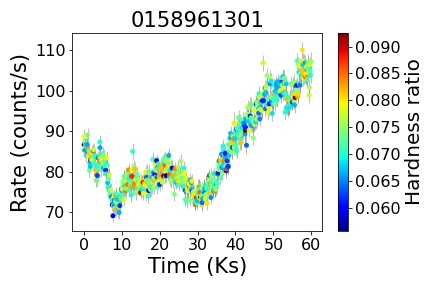}
    \includegraphics[width=5truecm]{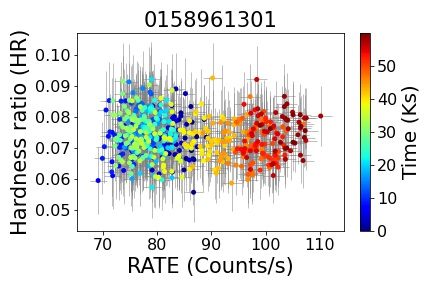}
    \includegraphics[width=5truecm]{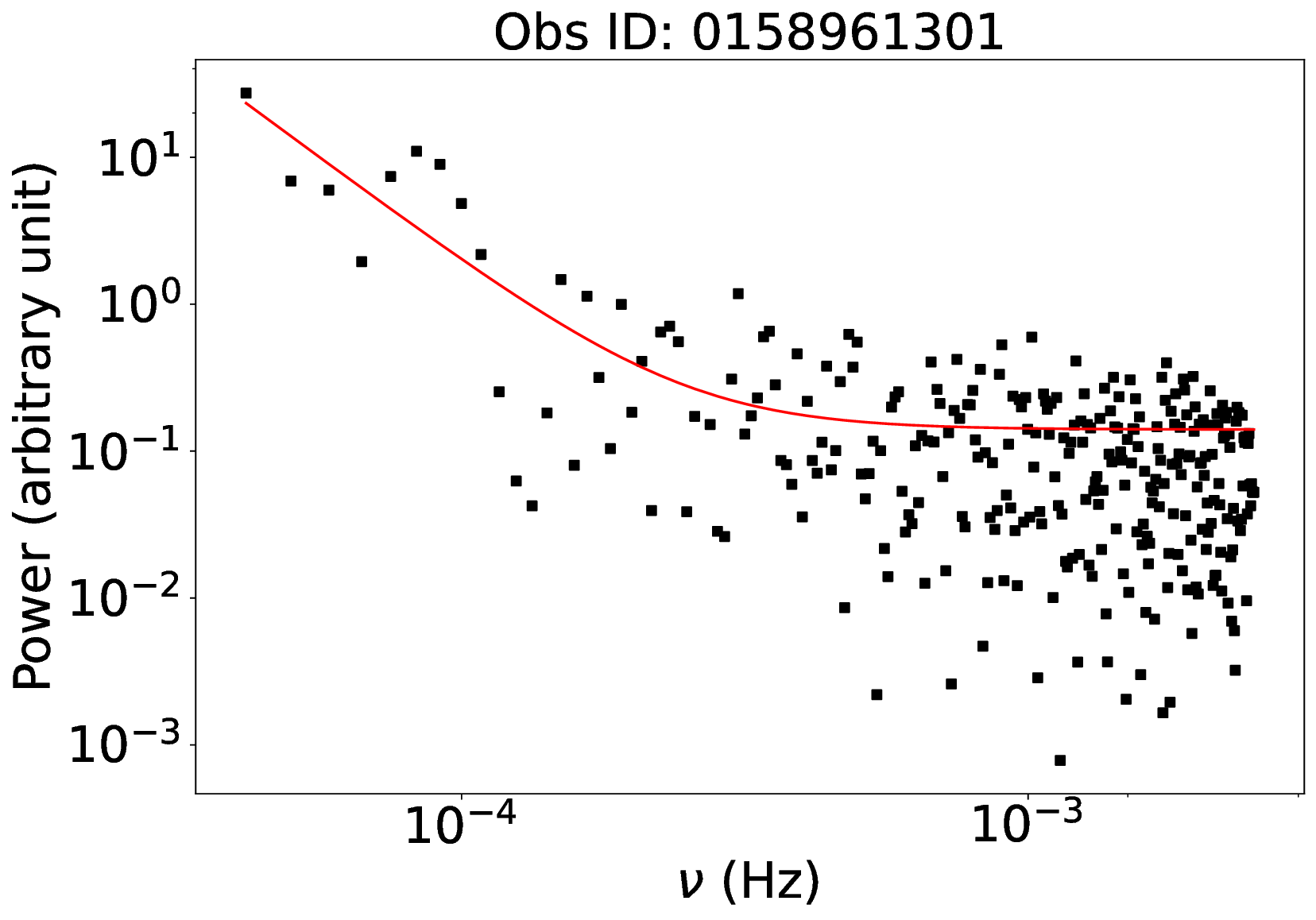}}
     \caption{LCs (Left), Hardness ratio (Middle) and PSD (Right) of Blazar PKS2155-304 from 2000 to 2014}
    \label{fig:1test}
\end{figure*}

\begin{figure*} 
    \centering{
    
    \includegraphics[width=5.8truecm]{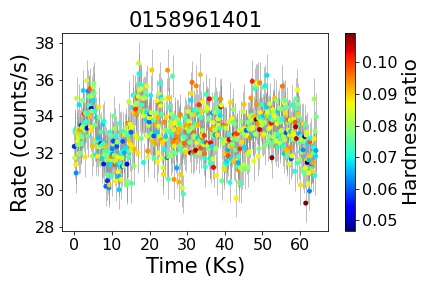}
    \includegraphics[width=5.8truecm]{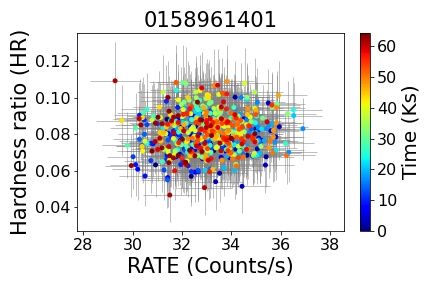}
    \includegraphics[width=5.8truecm]{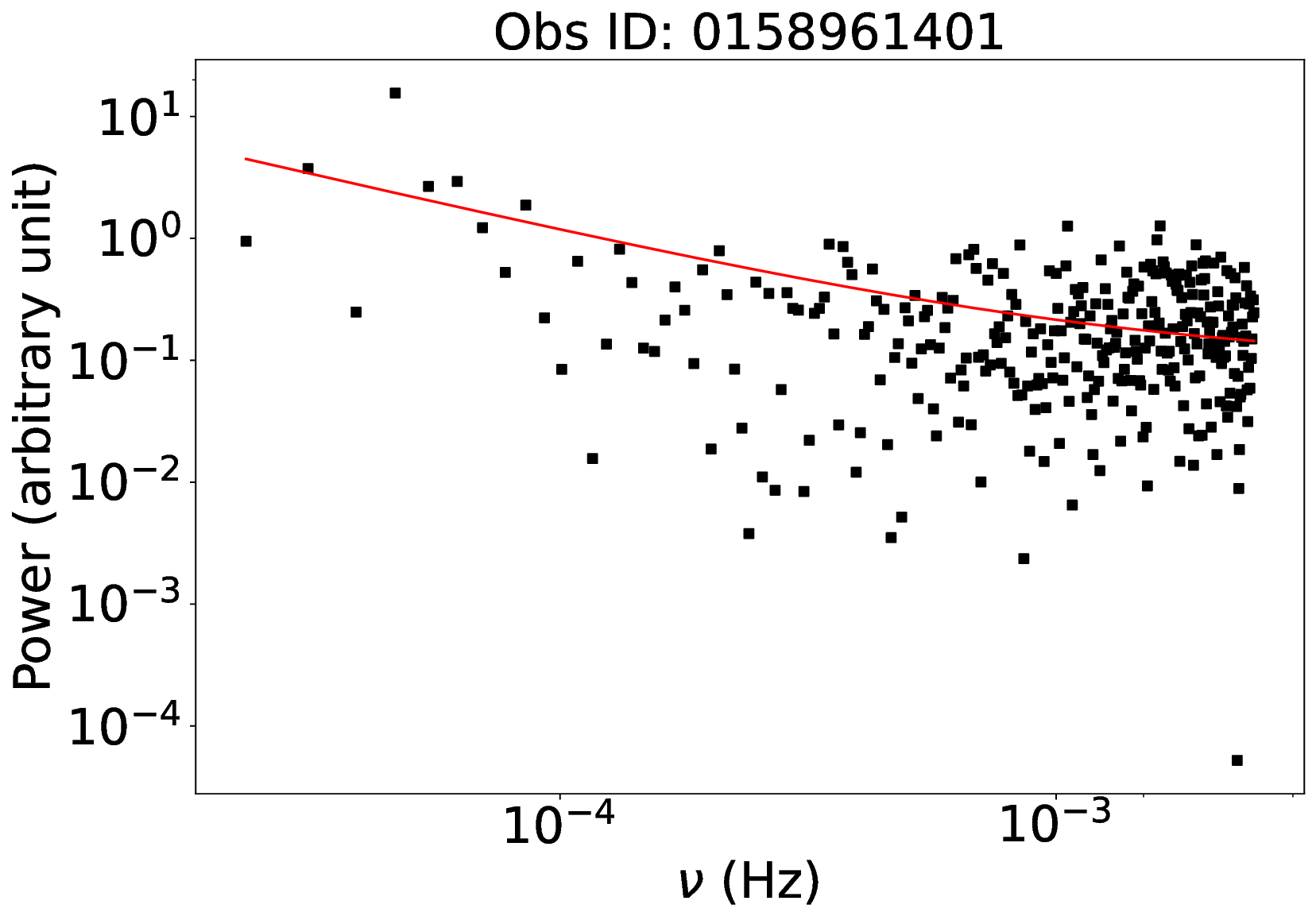}
    \includegraphics[width=5.8truecm]{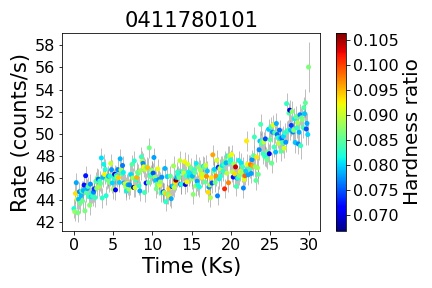}
    \includegraphics[width=5.8truecm]{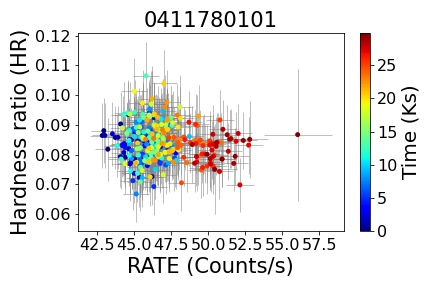}
    \includegraphics[width=5.8truecm]{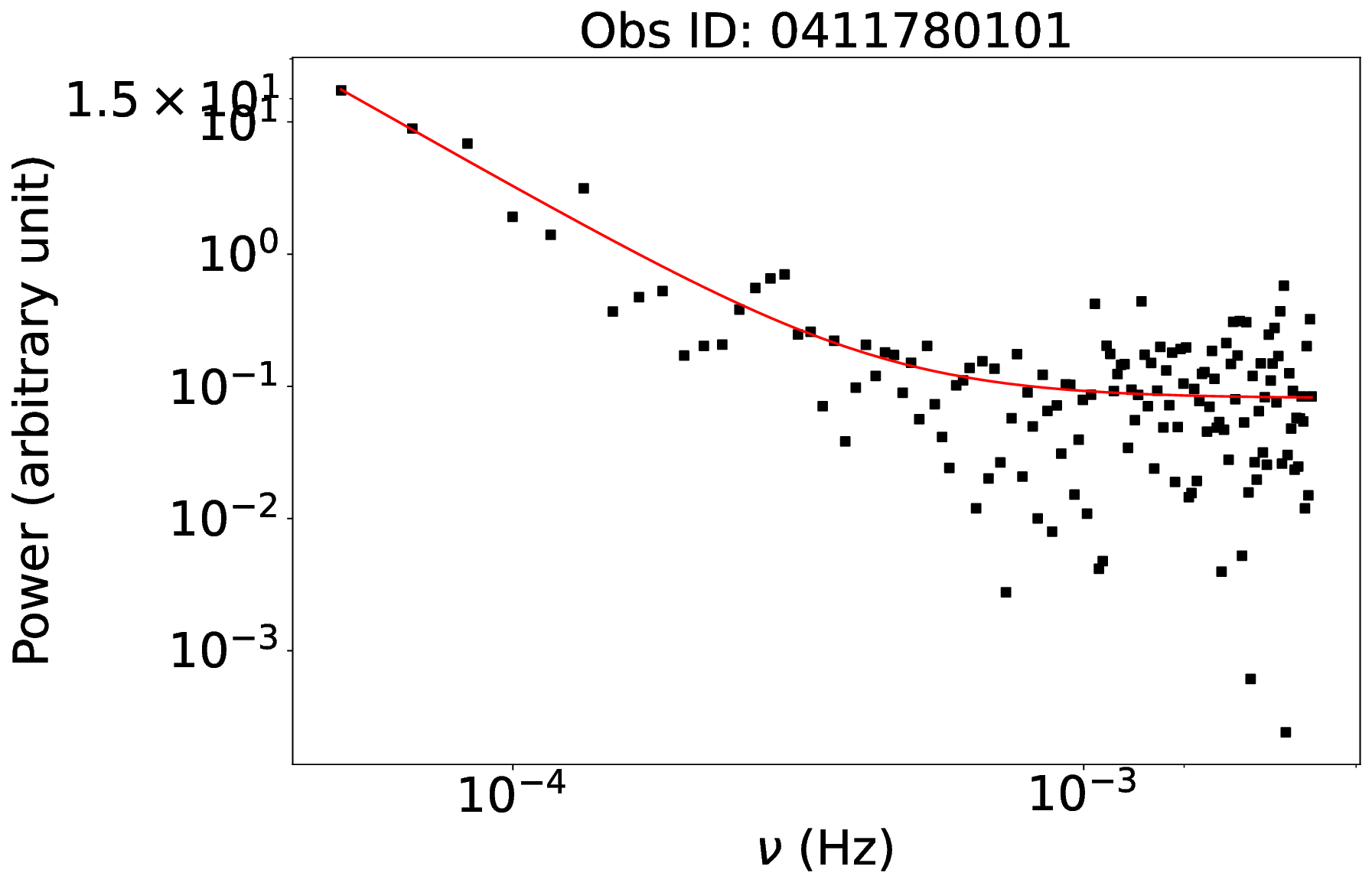}
    \includegraphics[width=5.8truecm]{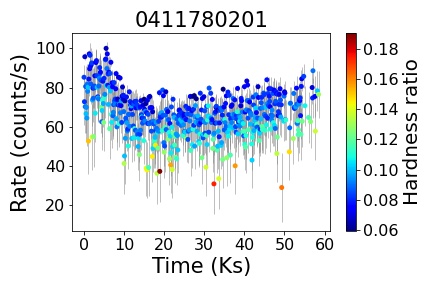}
    \includegraphics[width=5.8truecm]{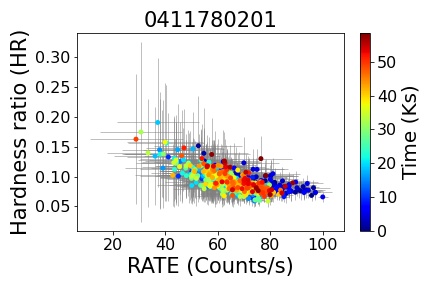}
    \includegraphics[width=5.8truecm]{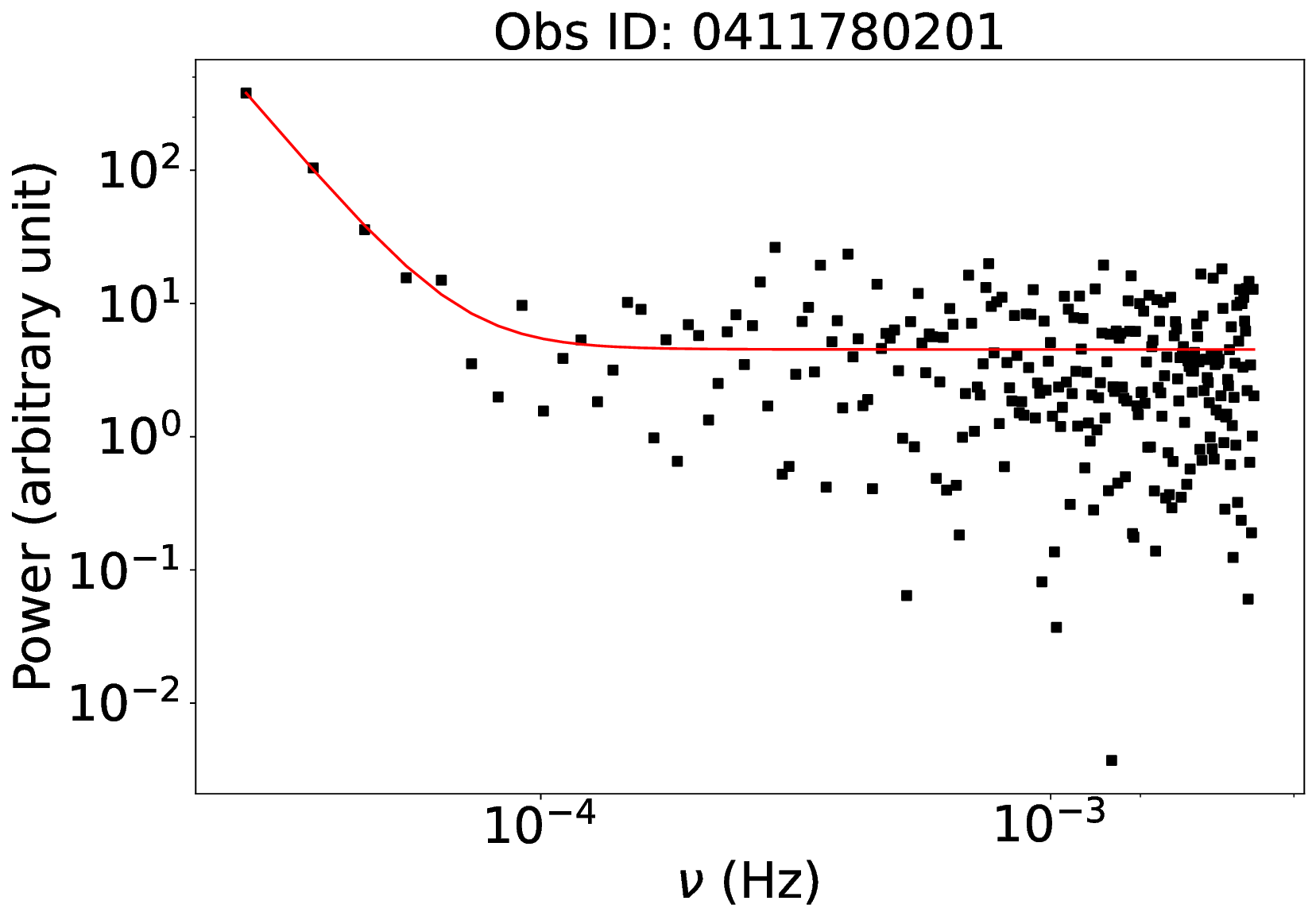}
    \includegraphics[width=5.8truecm]{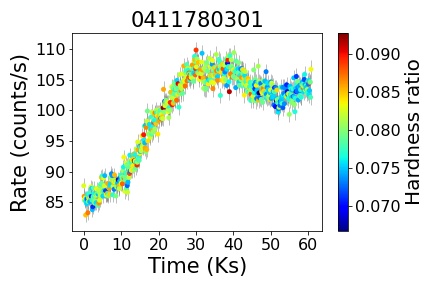}
    \includegraphics[width=5.8truecm]{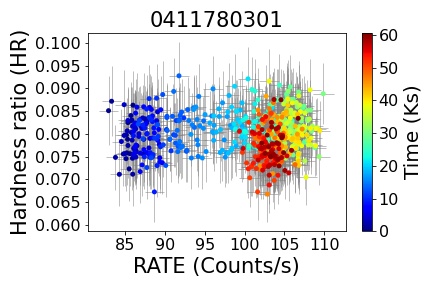}
    \includegraphics[width=5.8truecm]{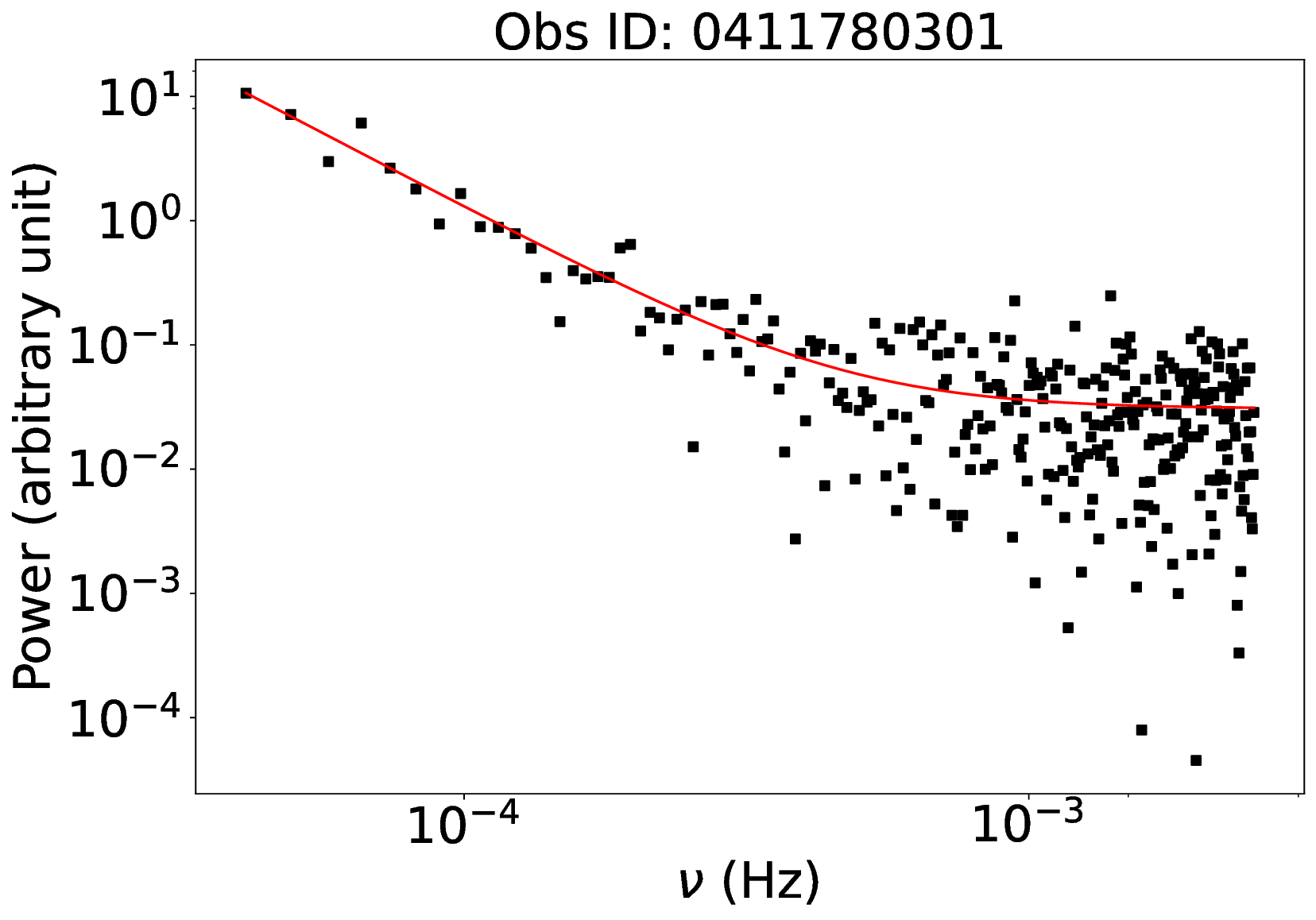}
    \includegraphics[width=5.8truecm]{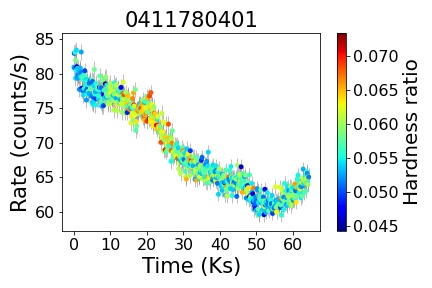}
    \includegraphics[width=5.8truecm]{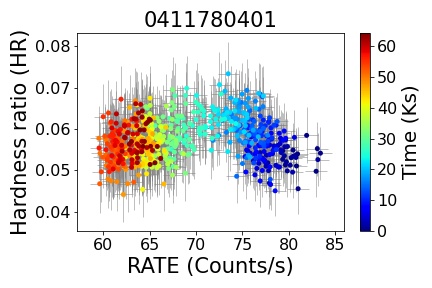}
    \includegraphics[width=5.8truecm]{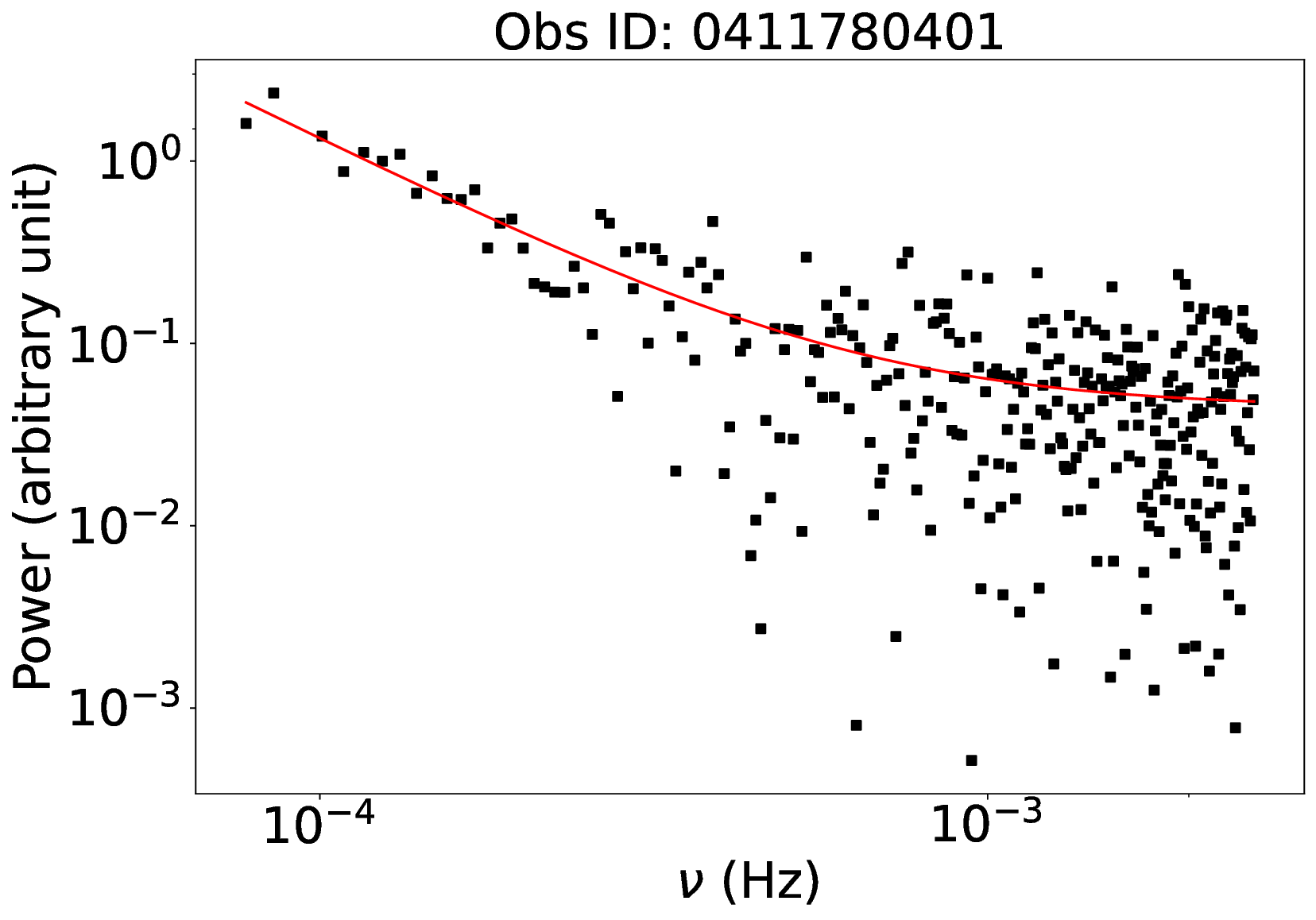}}
     \caption{LCs (Left), Hardness ratio (Middle) and PSD (Right) of Blazar PKS2155-304 from 2000 to 2014}
    \label{fig:2test}
\end{figure*}

\begin{figure*} 
    \centering{
    
    \includegraphics[width=5.8truecm]{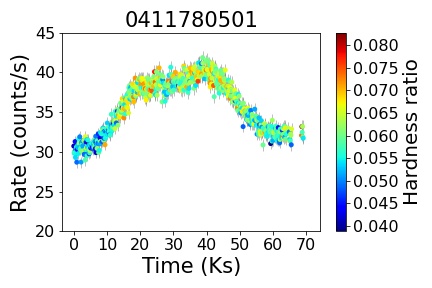}
    \includegraphics[width=5.8truecm]{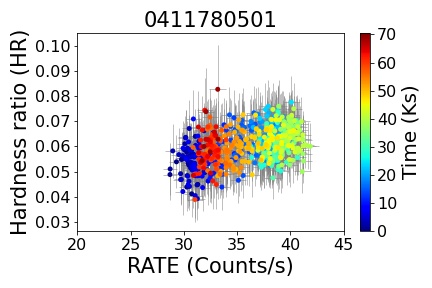}
    \includegraphics[width=5.8truecm]{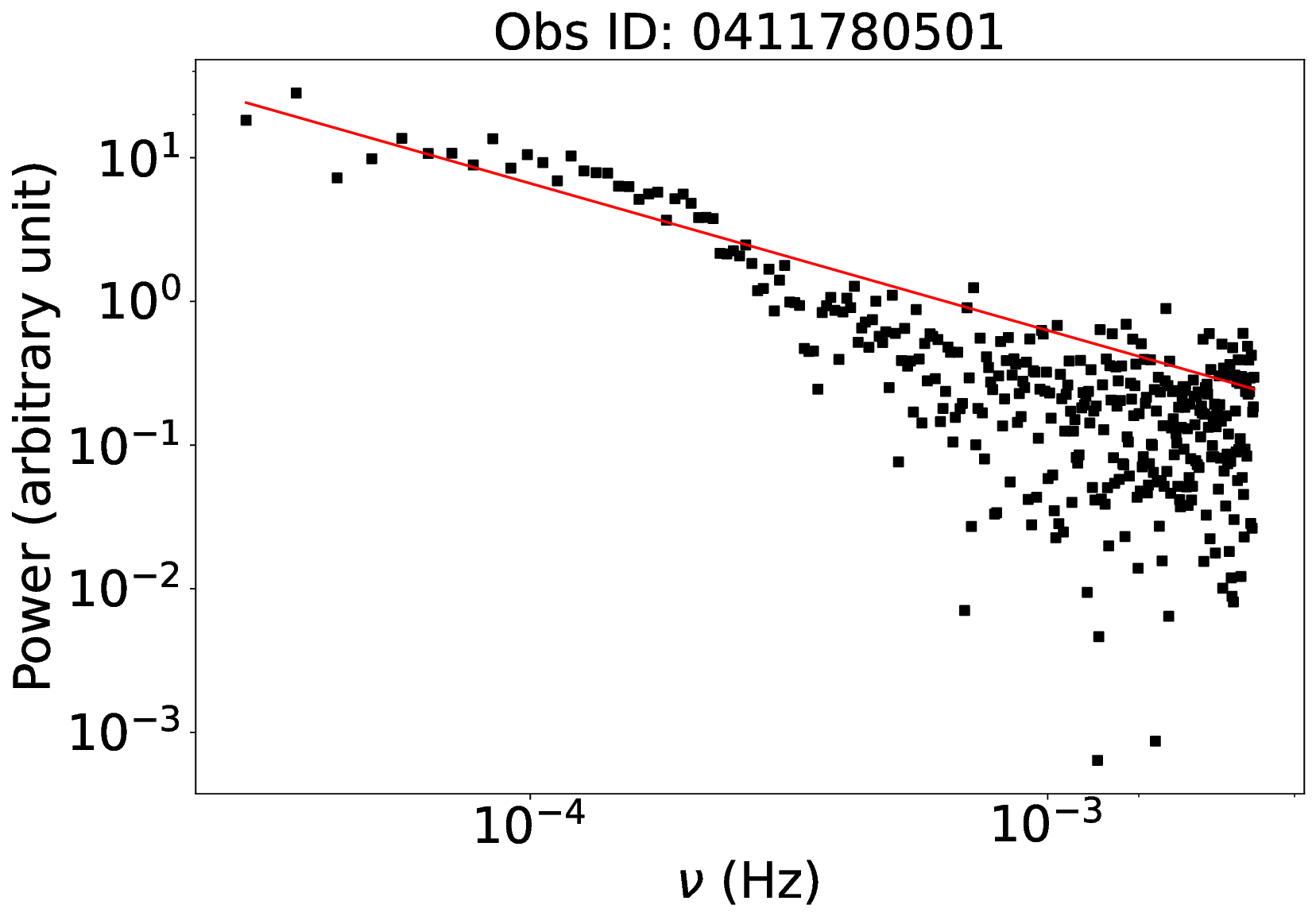}
    \includegraphics[width=5.8truecm]{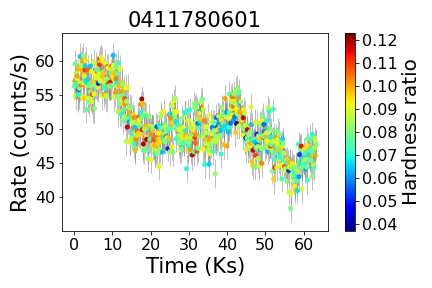}
    \includegraphics[width=5.8truecm]{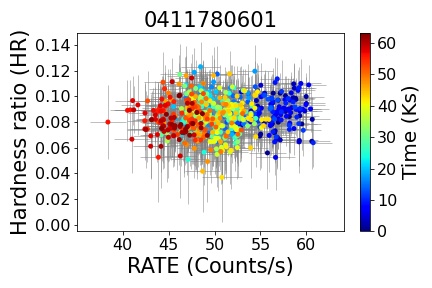}
    \includegraphics[width=5.8truecm]{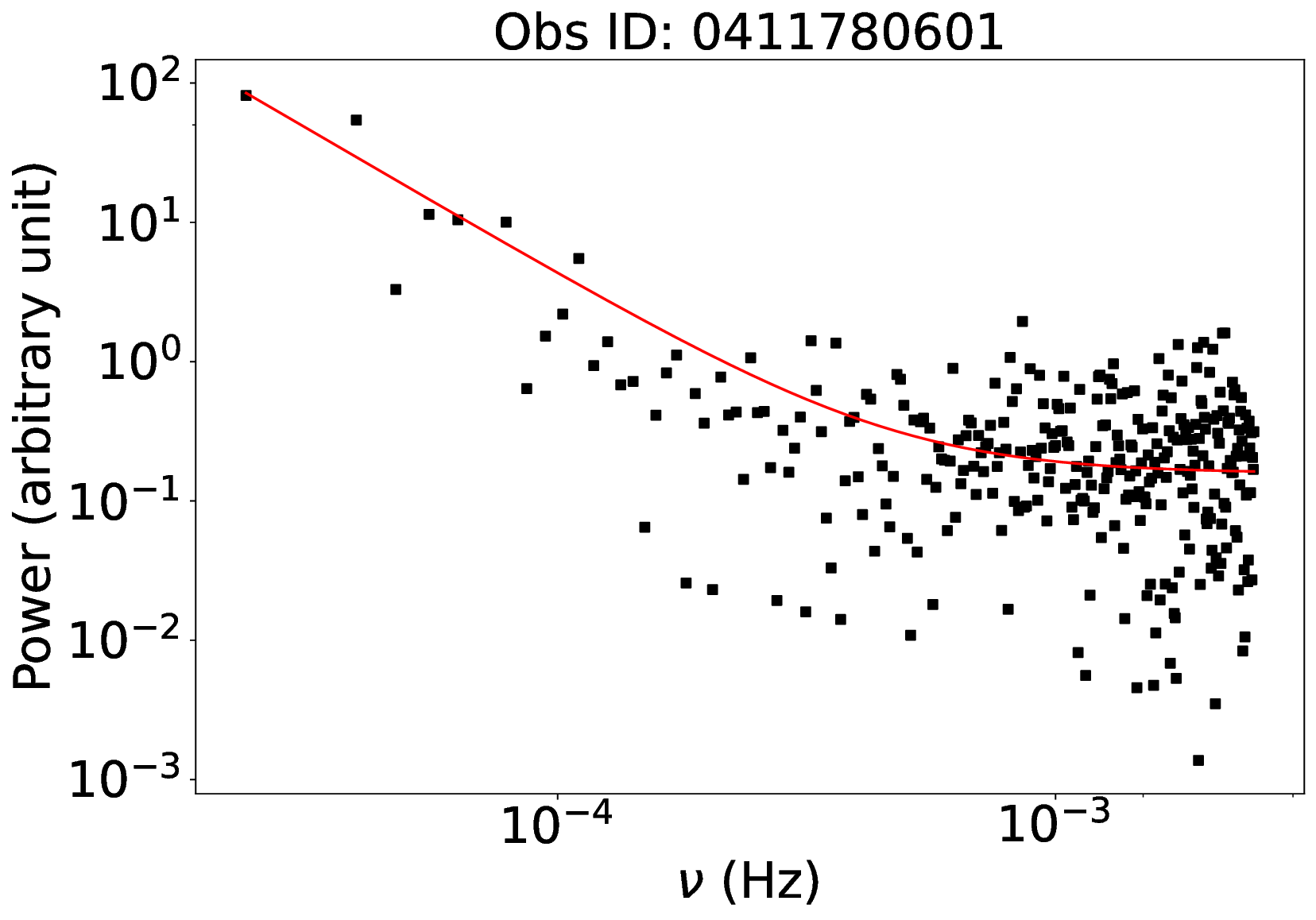}
     \includegraphics[width=5.8truecm]{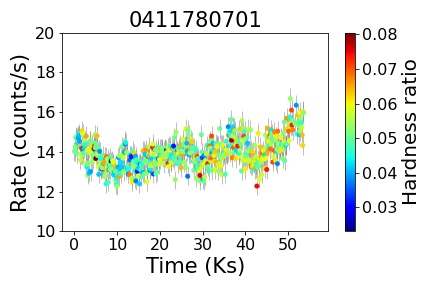}
    \includegraphics[width=5.8truecm]{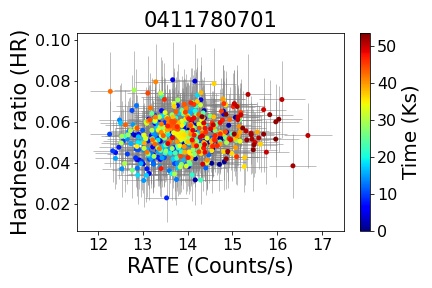}
    \includegraphics[width=5.8truecm]{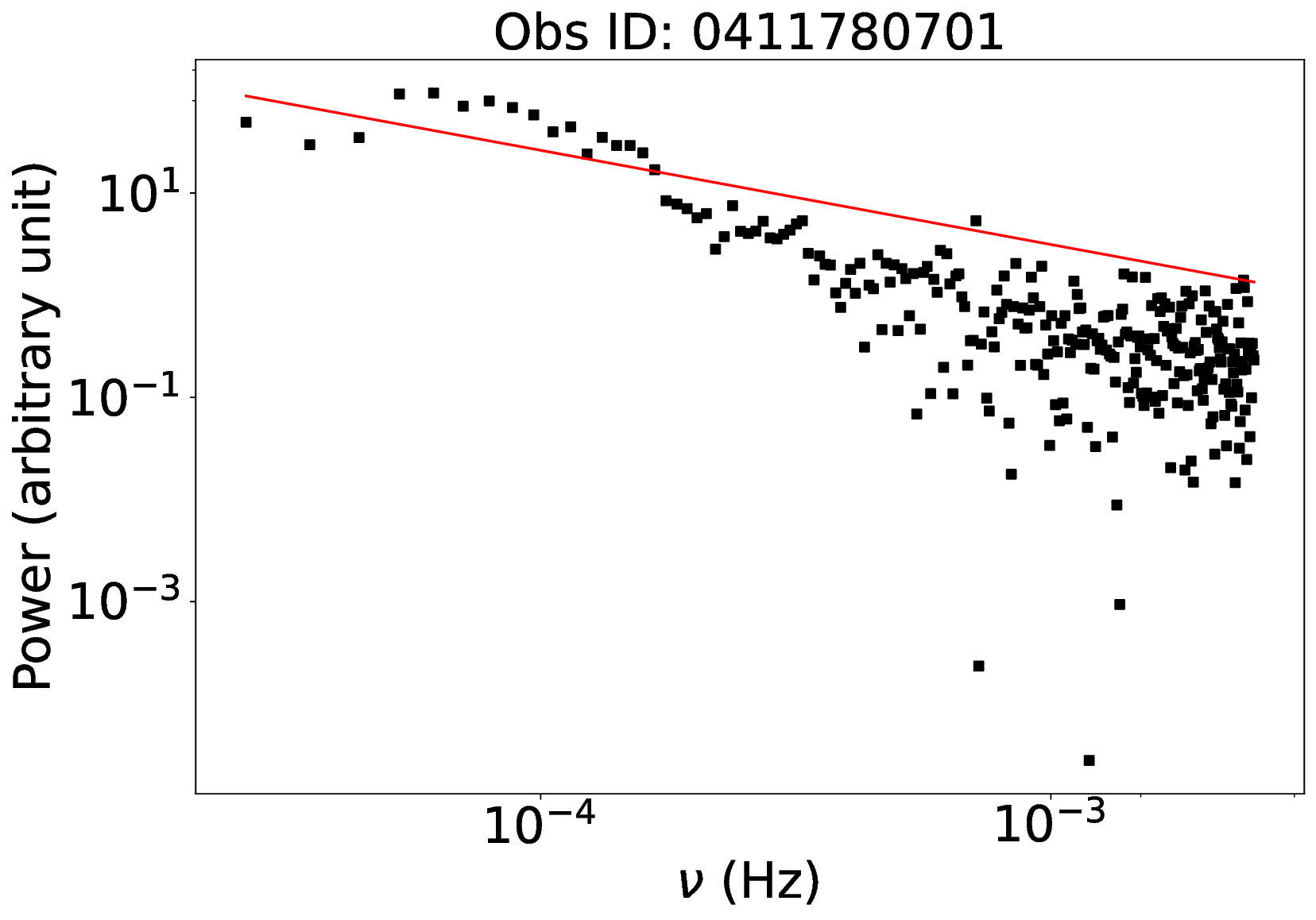}
    \includegraphics[width=5.8truecm]{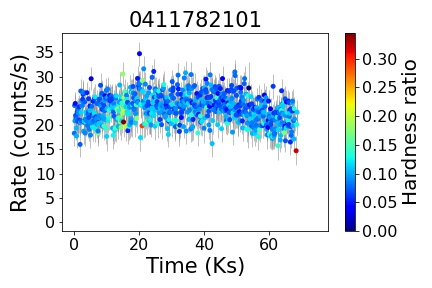}
    \includegraphics[width=5.8truecm]{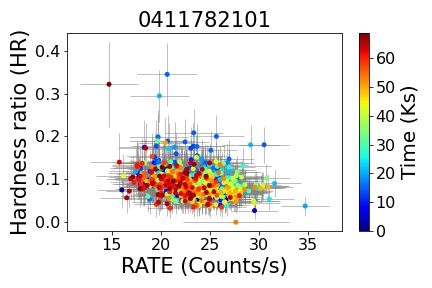}
    \includegraphics[width=5.8truecm]{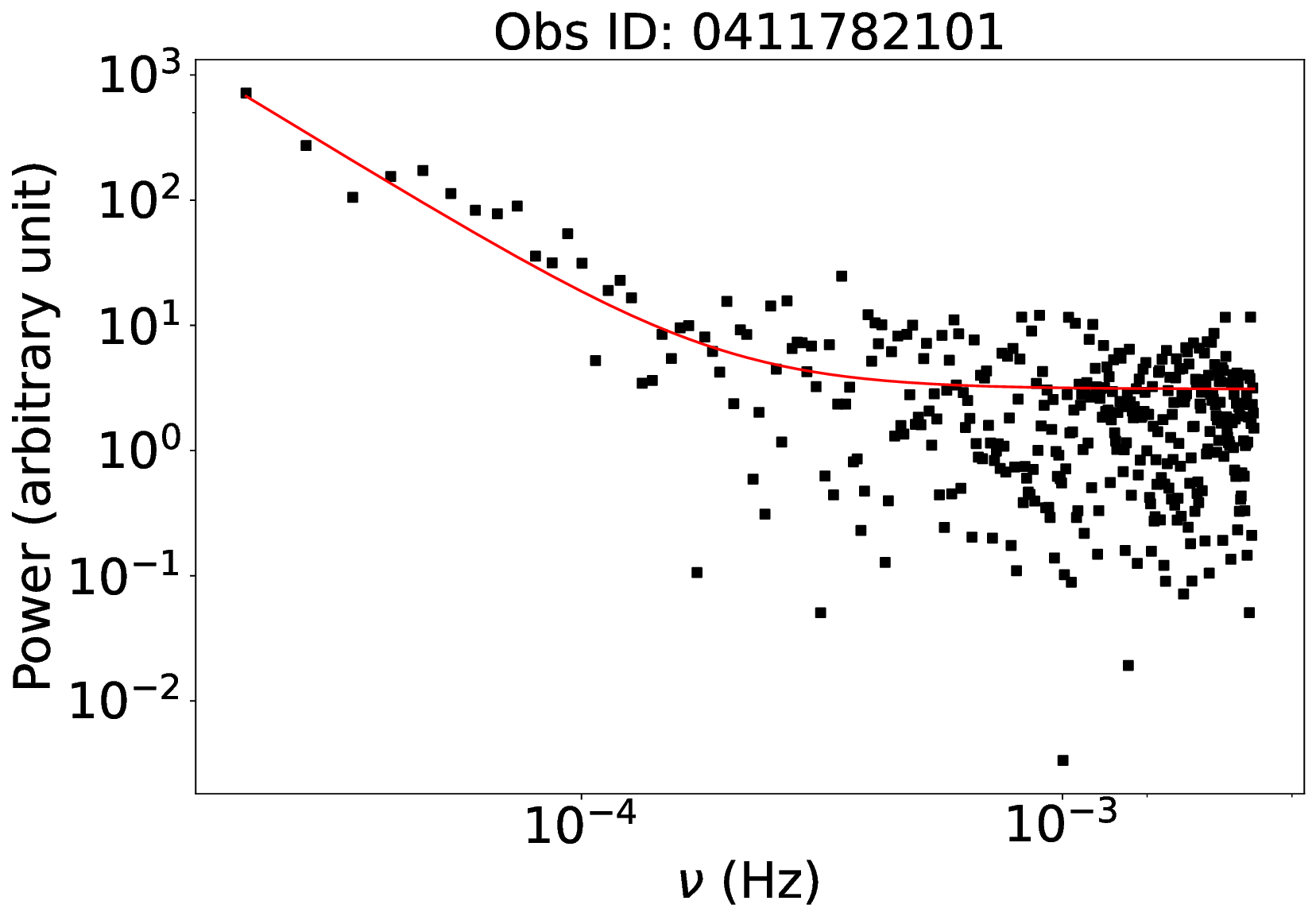}
    \includegraphics[width=5.8truecm]{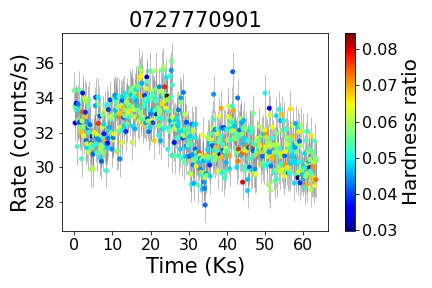}
    \includegraphics[width=5.8truecm]{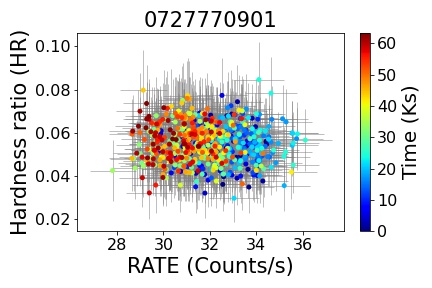}
    \includegraphics[width=5.8truecm]{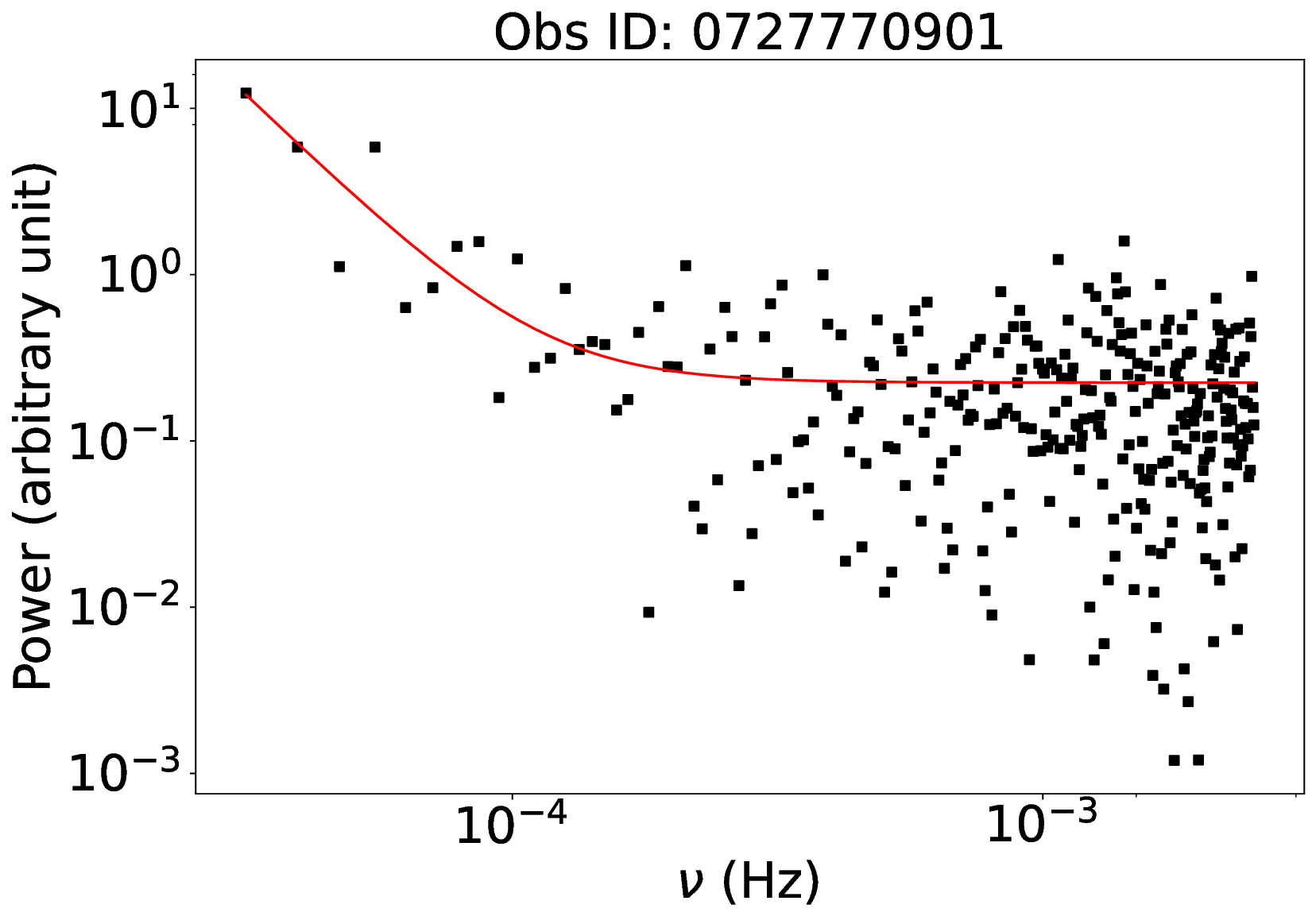}}
     \caption{Lightcurves of Blazar PKS2155-304 from 2000 to 2014}
    \label{fig:3test}
\end{figure*}

\begin{figure*}
	\centering
	\caption{The spectral ﬁtting of observations of PKS 2155–304 in 0.3–10 keV. Each spectra is ﬁtted using the powerlaw, BPL and log parabolic model and the data-to-model ratio is shown in the three subpanels for each spectra in olive, red and blue colors, respectively.}
	\begin{minipage}{.30\textwidth} 
		\centering
		\includegraphics[width=.990\linewidth]{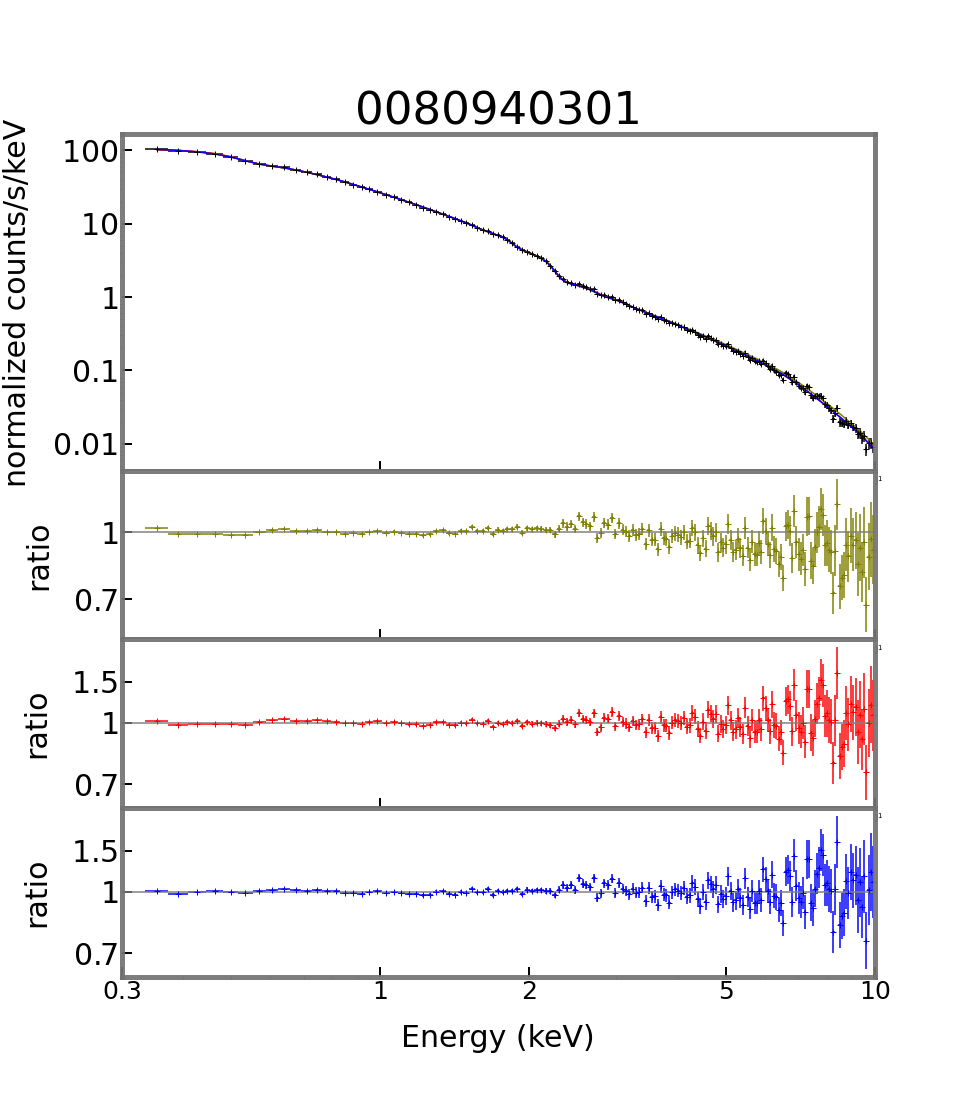}
	\end{minipage}
	\begin{minipage}{.30\textwidth} 
		\centering
		\includegraphics[width=.990\linewidth]{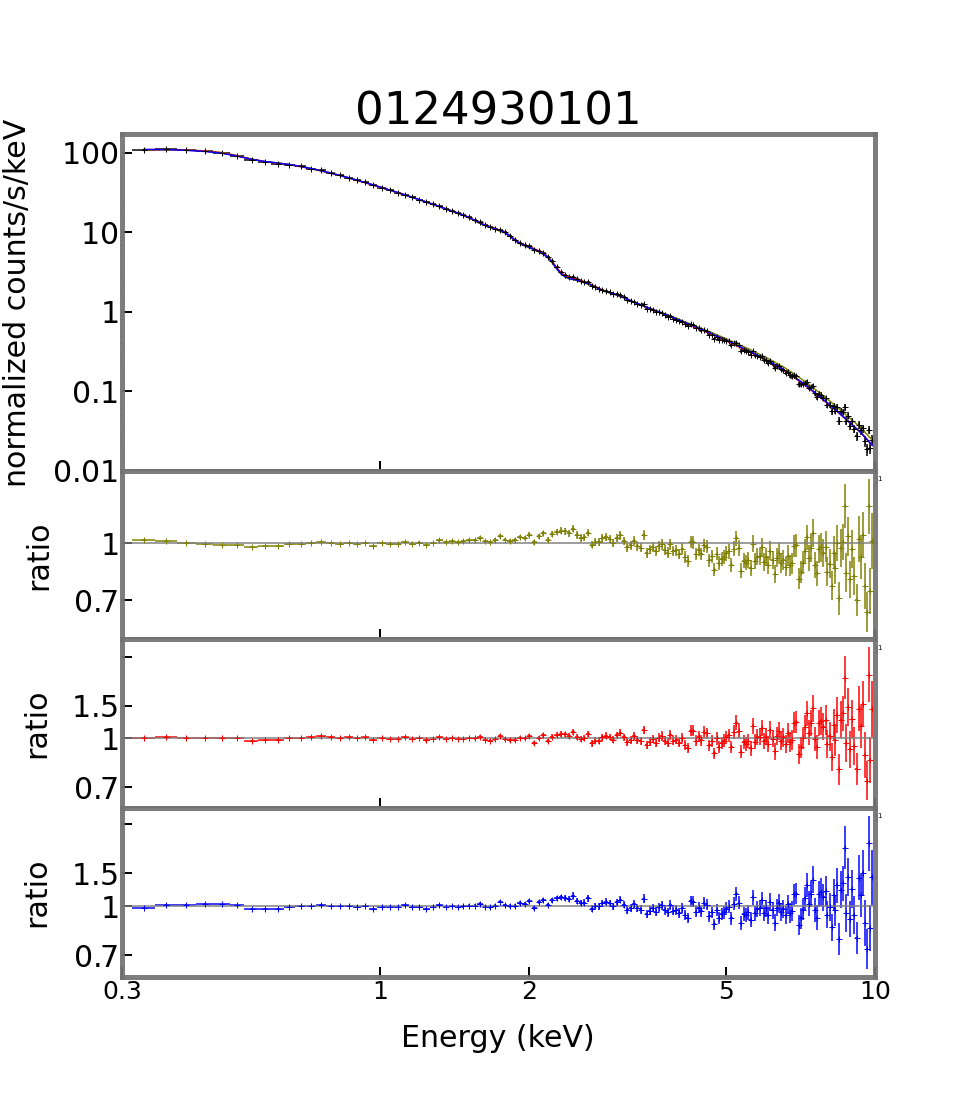}
	\end{minipage}
	\begin{minipage}{.30\textwidth} 
		\centering
		\includegraphics[width=.990\linewidth]{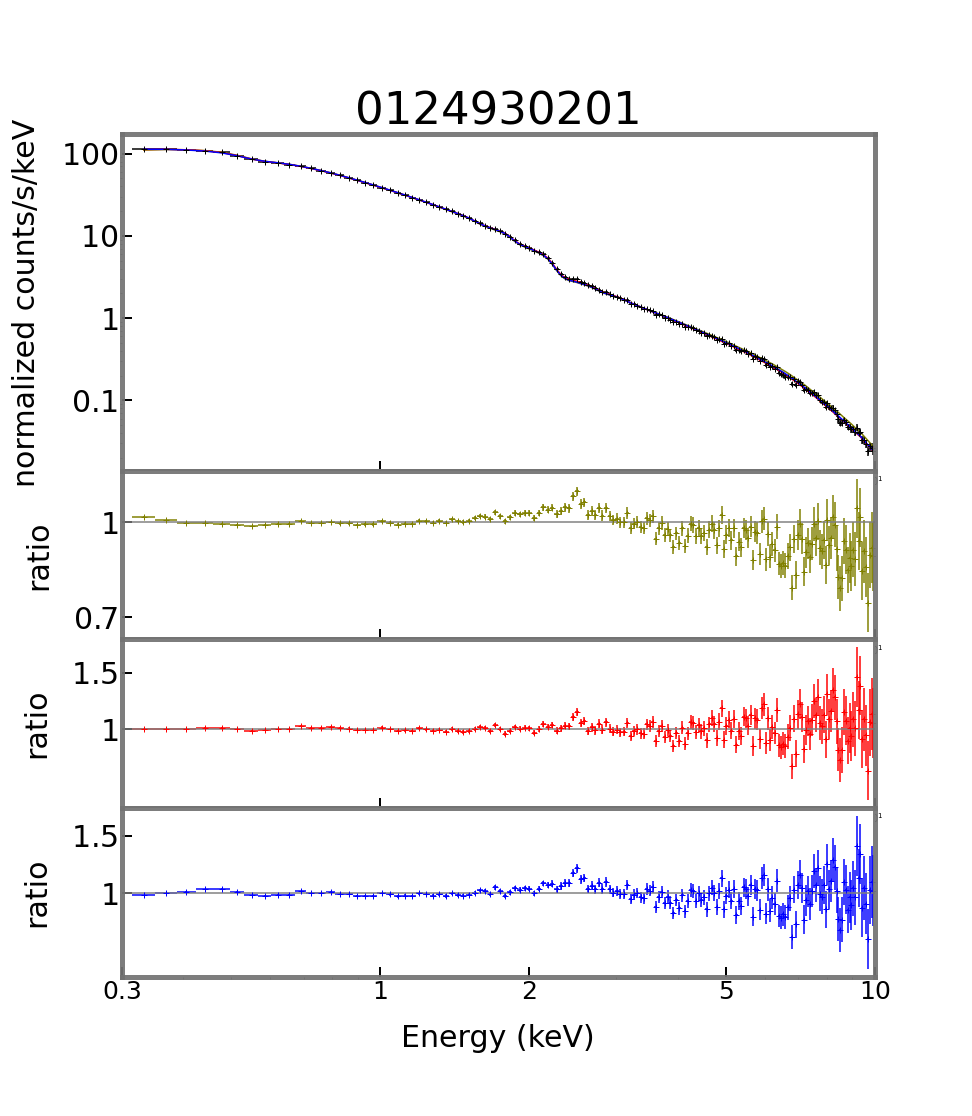}
	\end{minipage}
	\begin{minipage}{.30\textwidth} 
		\centering
		\includegraphics[width=.990\linewidth]{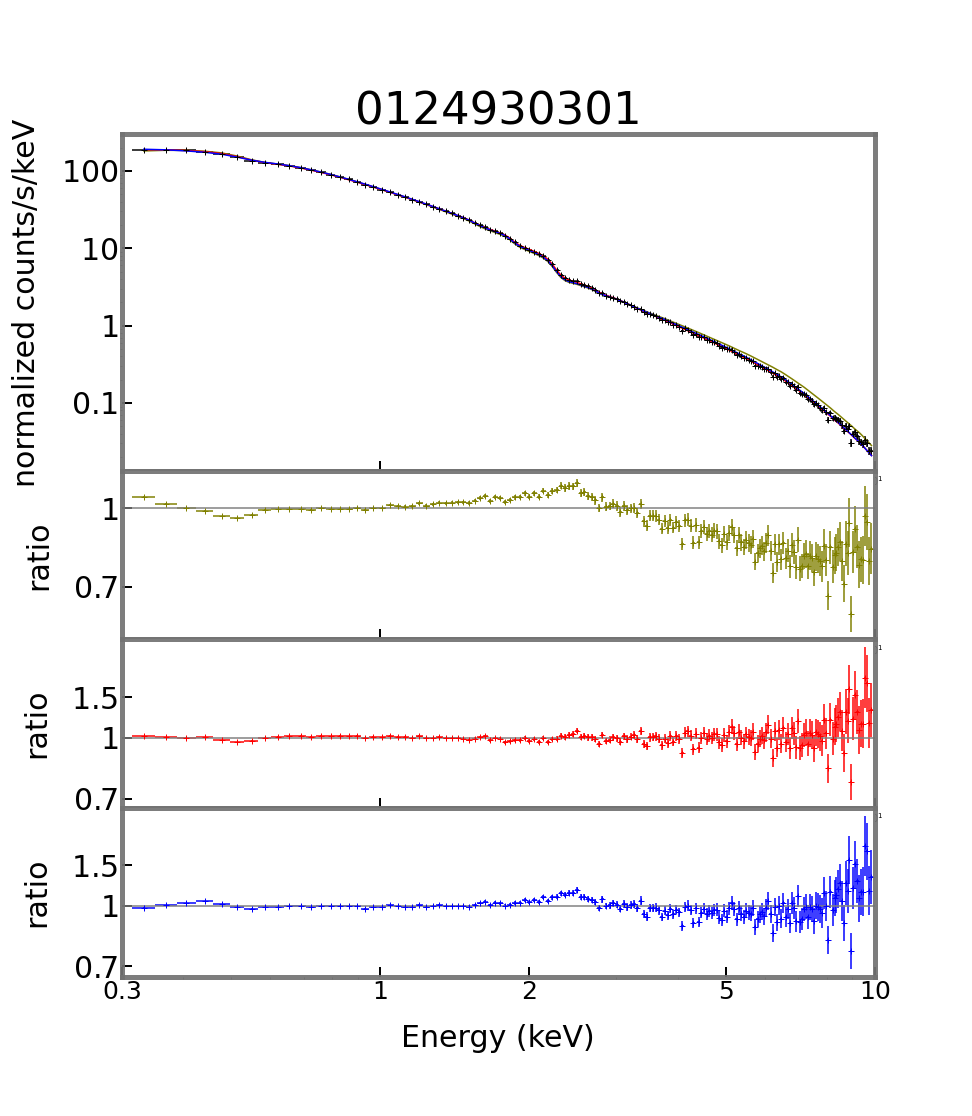}
	\end{minipage}
	\begin{minipage}{.30\textwidth} 
		\centering
		\includegraphics[width=.990\linewidth]{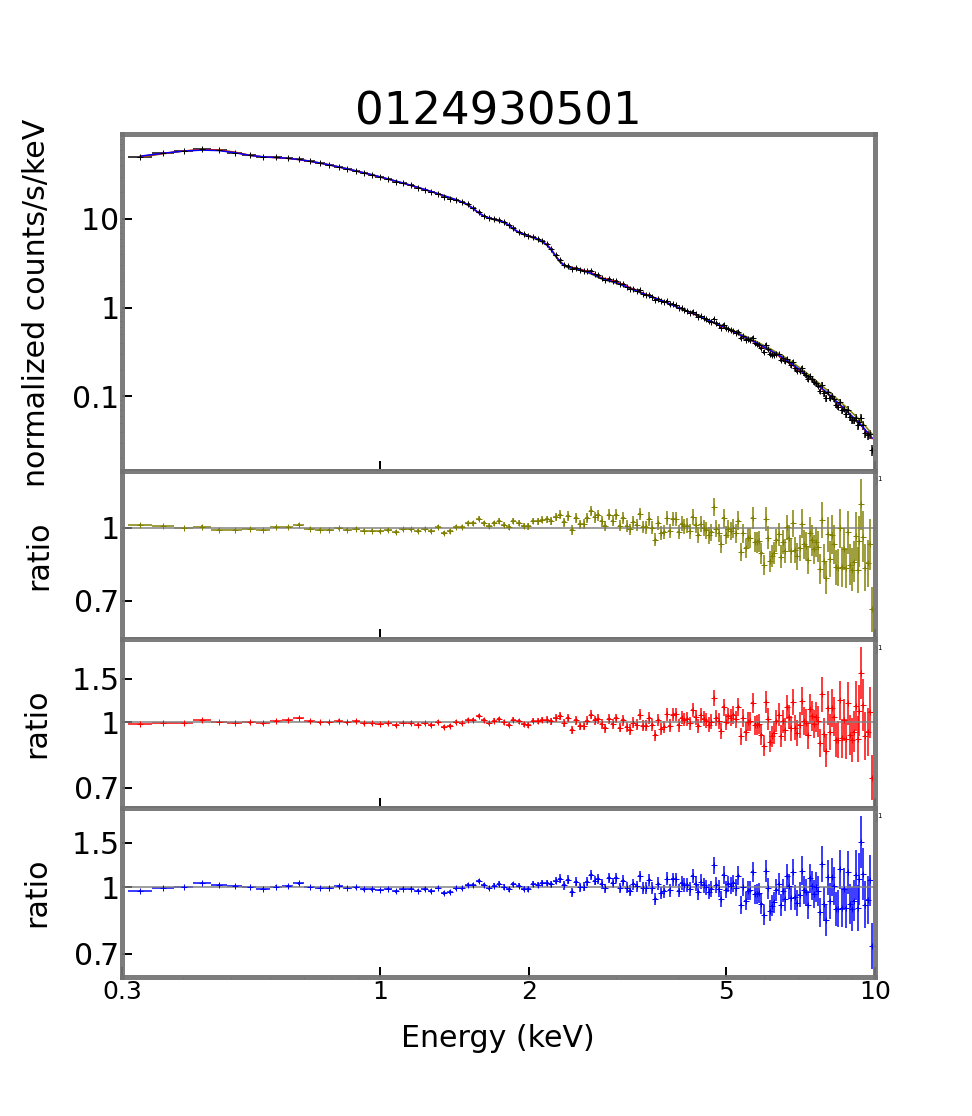}
	\end{minipage}
	\begin{minipage}{.30\textwidth} 
		\centering
		\includegraphics[width=.990\linewidth]{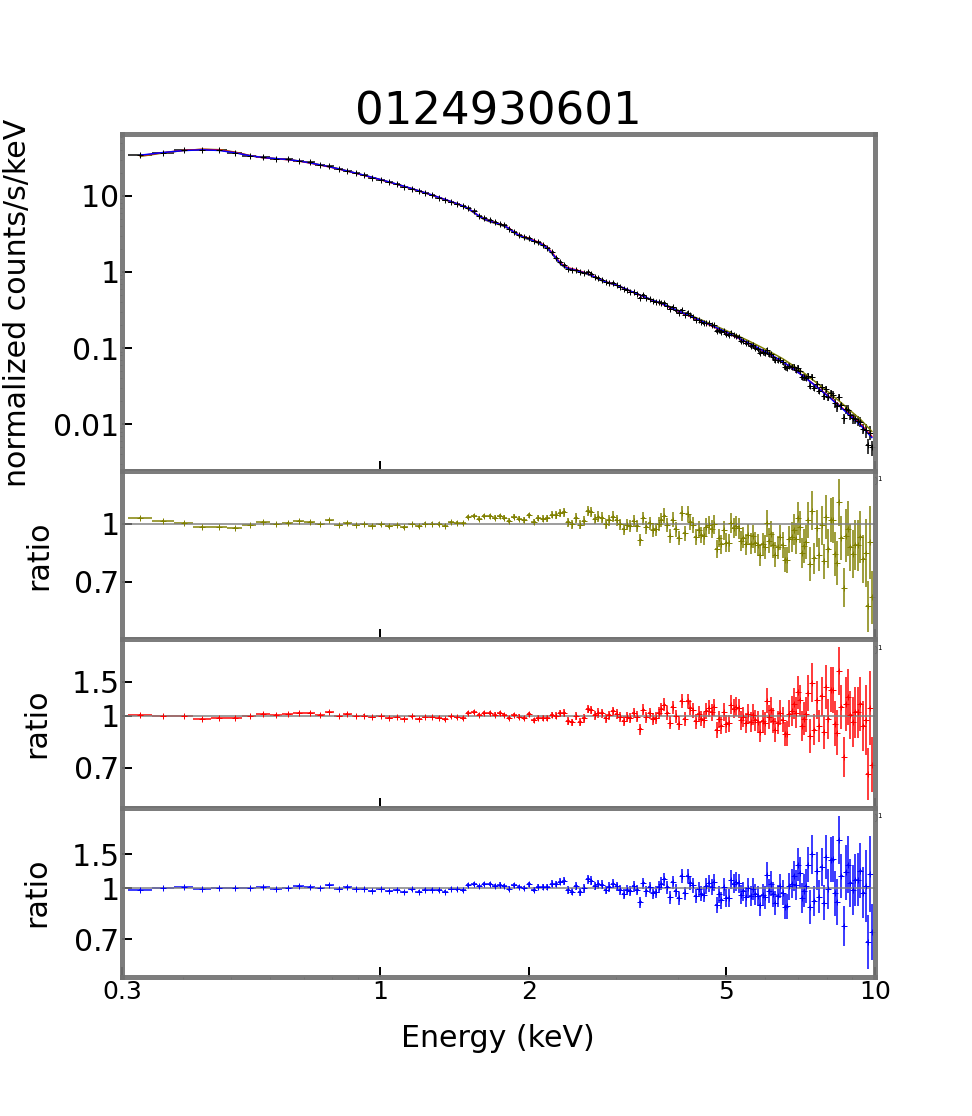}
	\end{minipage}
	\begin{minipage}{.30\textwidth} 
		\centering
		\includegraphics[width=.990\linewidth]{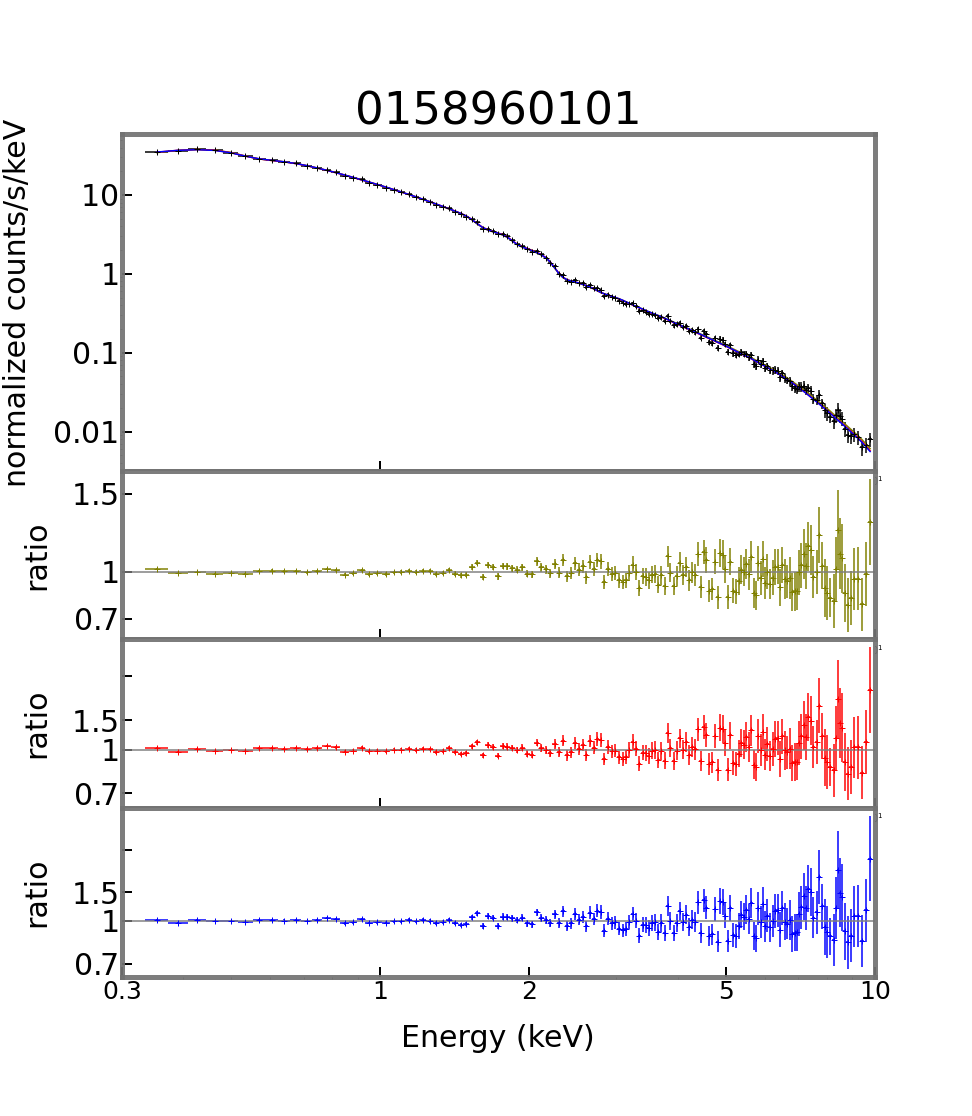}
	\end{minipage}
	\begin{minipage}{.30\textwidth} 
		\centering
		\includegraphics[width=.990\linewidth]{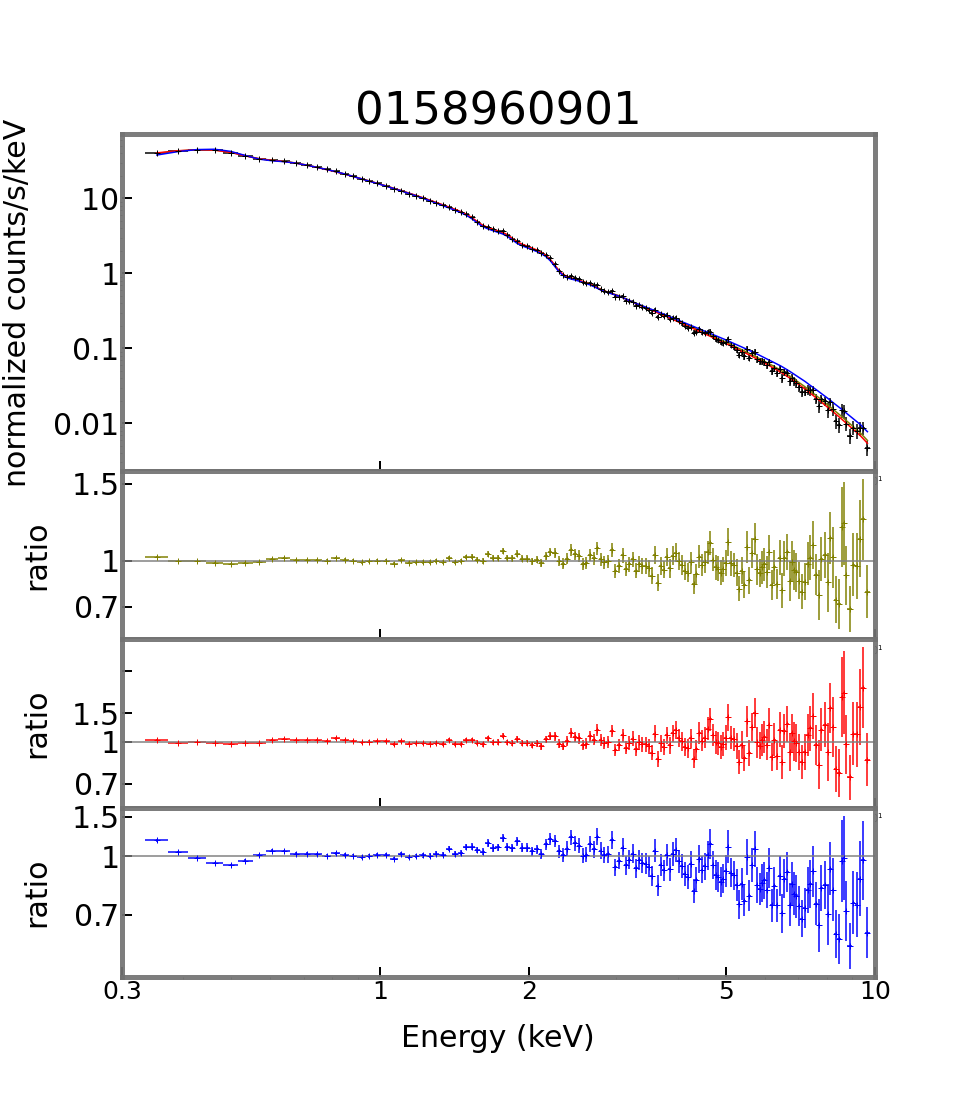}
	\end{minipage}
	\begin{minipage}{.30\textwidth} 
		\centering
		\includegraphics[width=.990\linewidth]{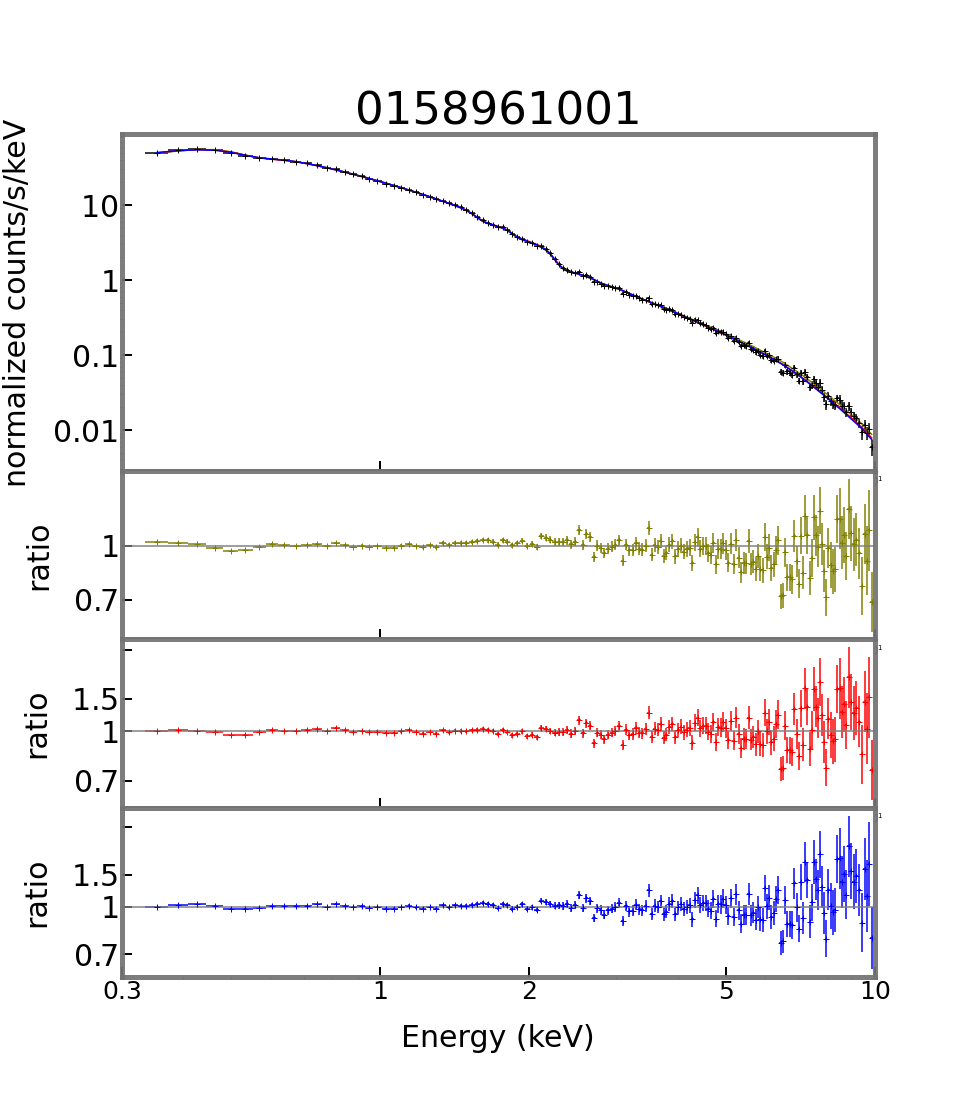}
	\end{minipage}
		\caption{Continue...}
	\end{figure*}

\begin{figure*}
	\centering
	\begin{minipage}{.30\textwidth} 
		\centering
		\includegraphics[width=.990\linewidth]{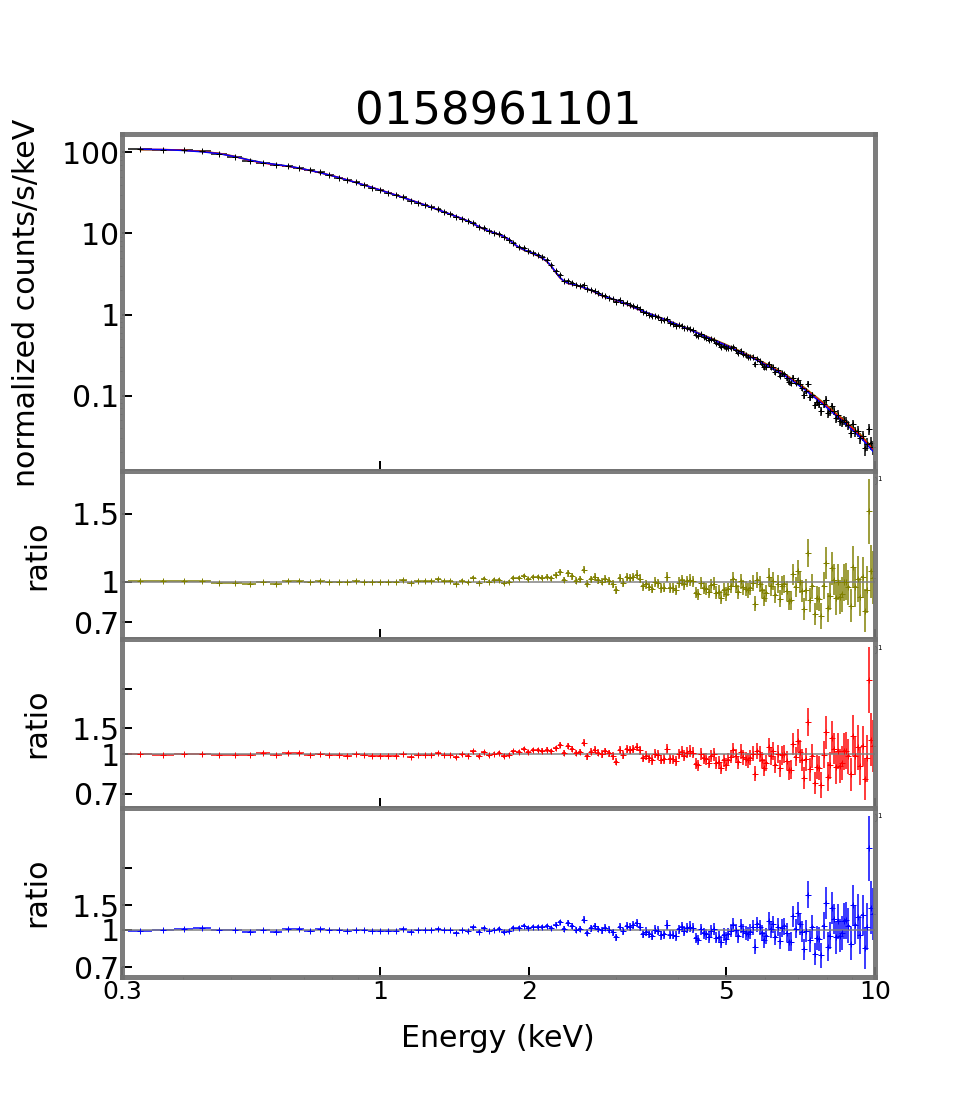}
	\end{minipage}
	\begin{minipage}{.30\textwidth} 
		\centering
		\includegraphics[width=.990\linewidth]{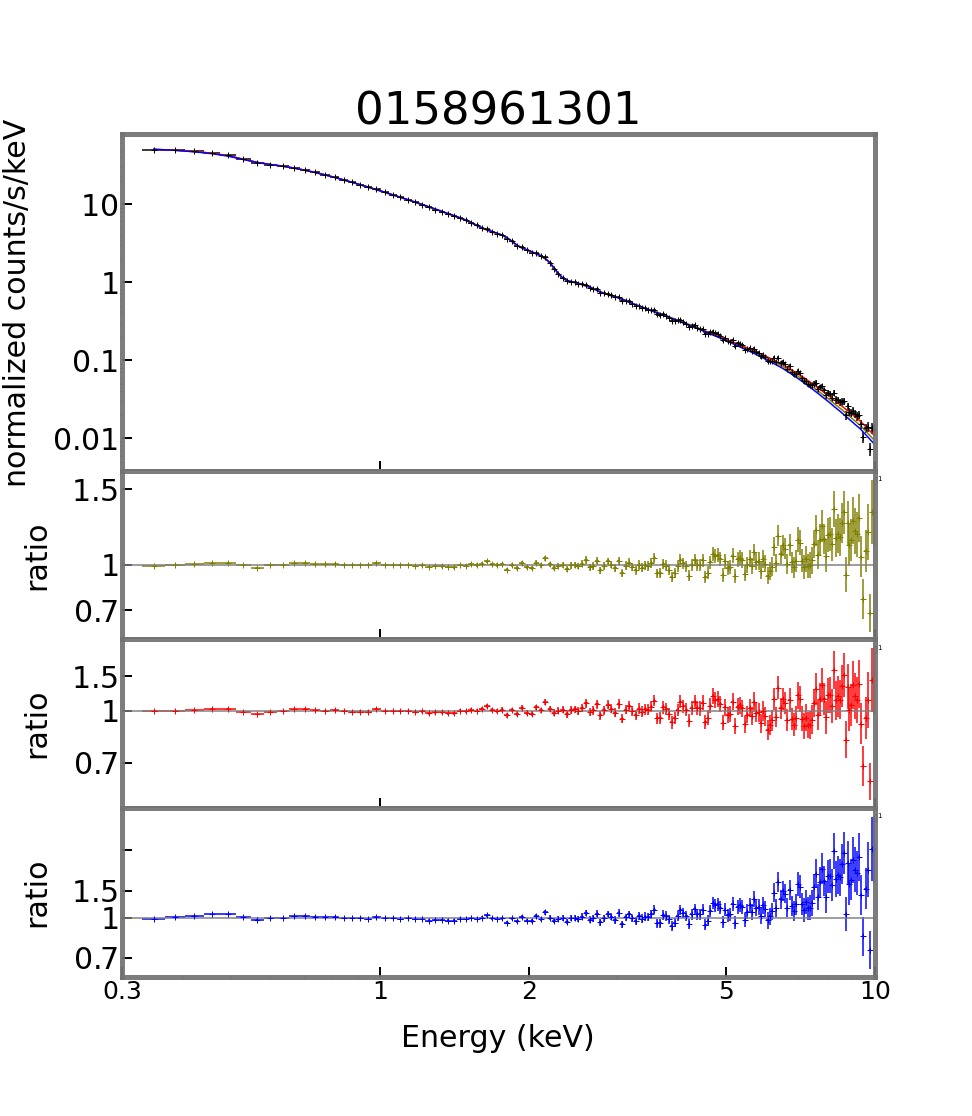}
	\end{minipage}
	\begin{minipage}{.30\textwidth} 
		\centering
		\includegraphics[width=.990\linewidth]{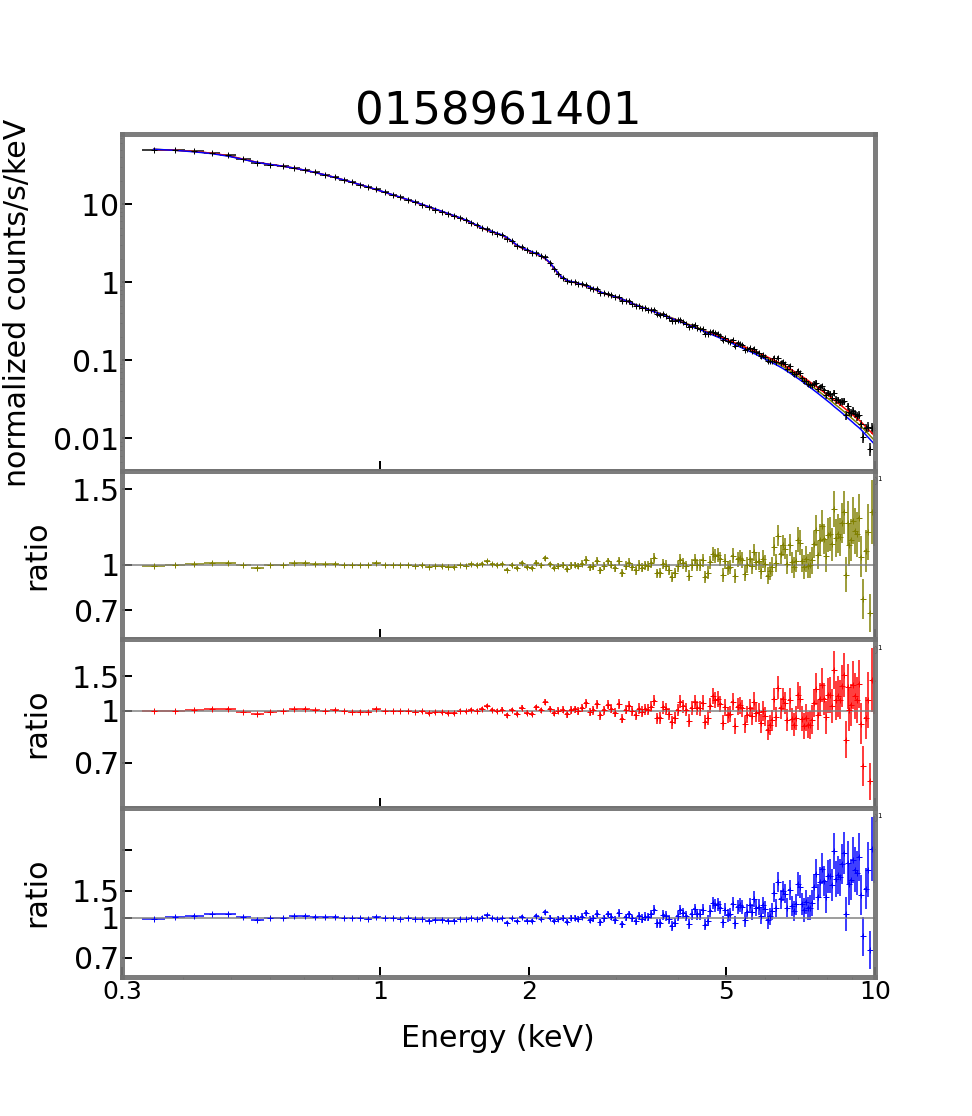}
	\end{minipage}
	\begin{minipage}{.30\textwidth} 
		\centering
		\includegraphics[width=.990\linewidth]{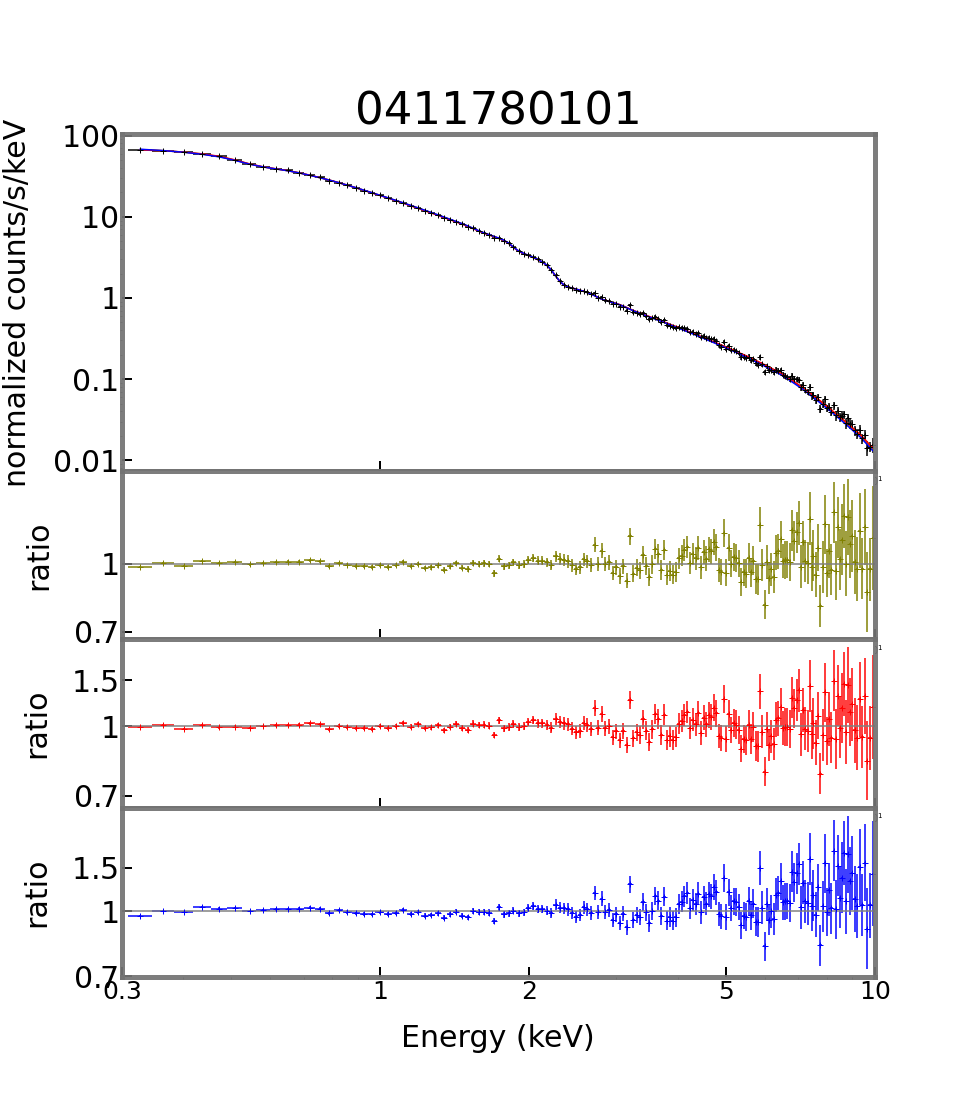}
	\end{minipage}
	\begin{minipage}{.30\textwidth} 
		\centering
		\includegraphics[width=.990\linewidth]{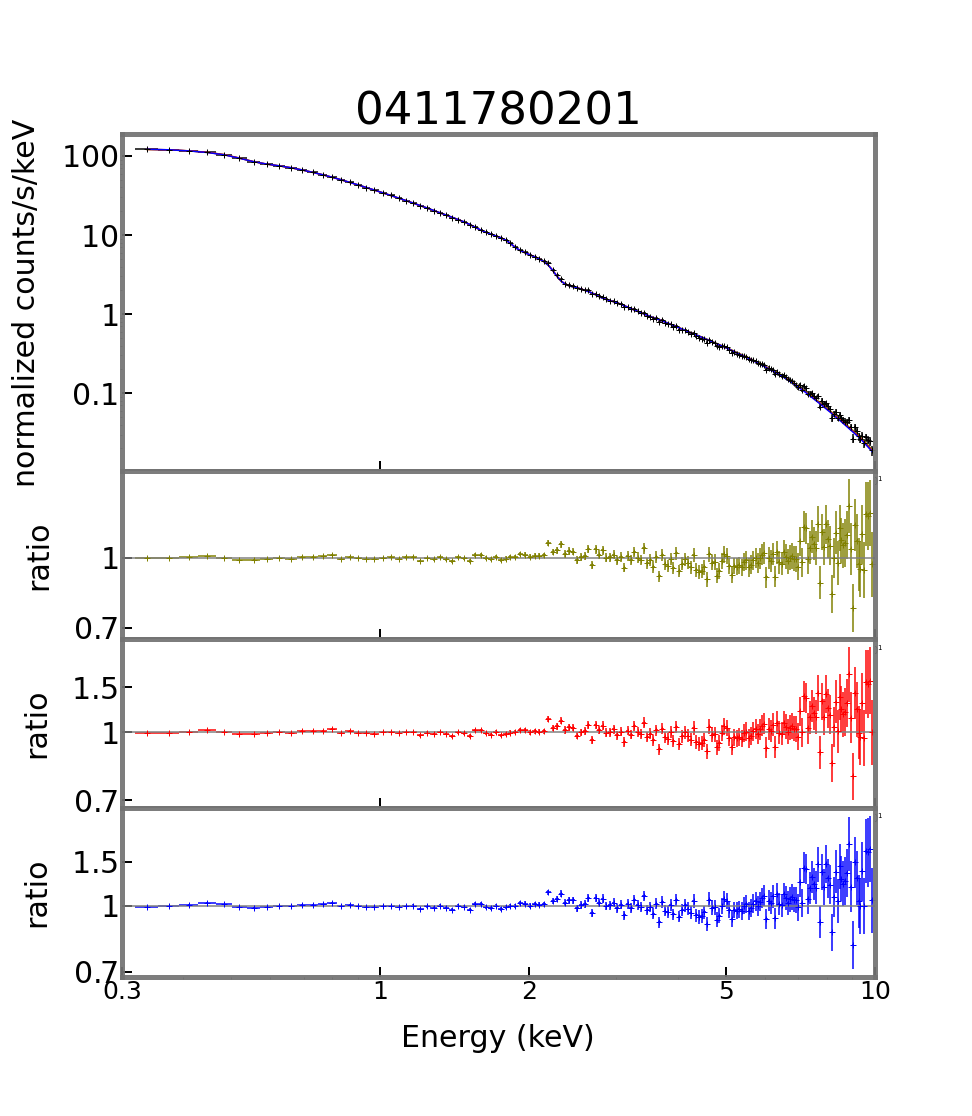}
	\end{minipage}
	\begin{minipage}{.30\textwidth} 
		\centering
		\includegraphics[width=.990\linewidth]{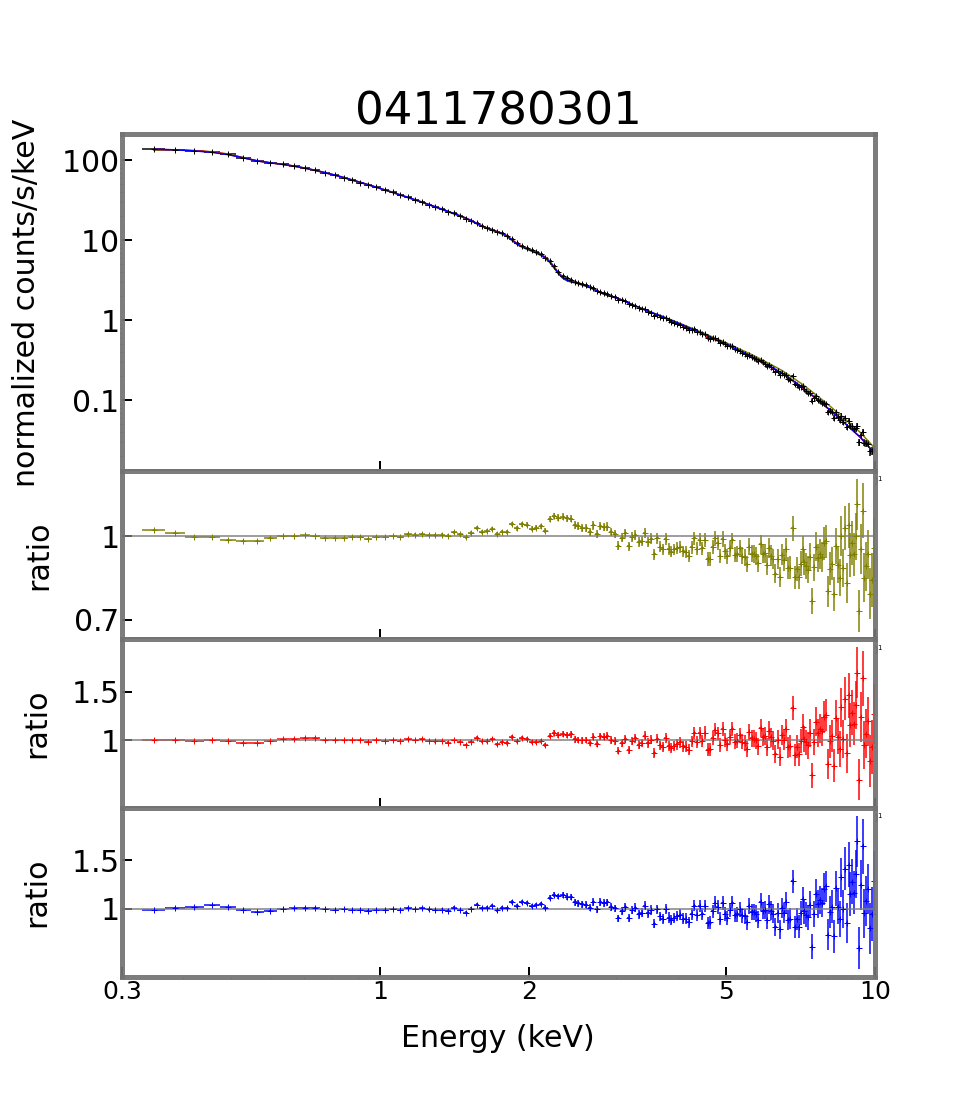}
	\end{minipage}
	\begin{minipage}{.30\textwidth} 
		\centering
		\includegraphics[width=.990\linewidth]{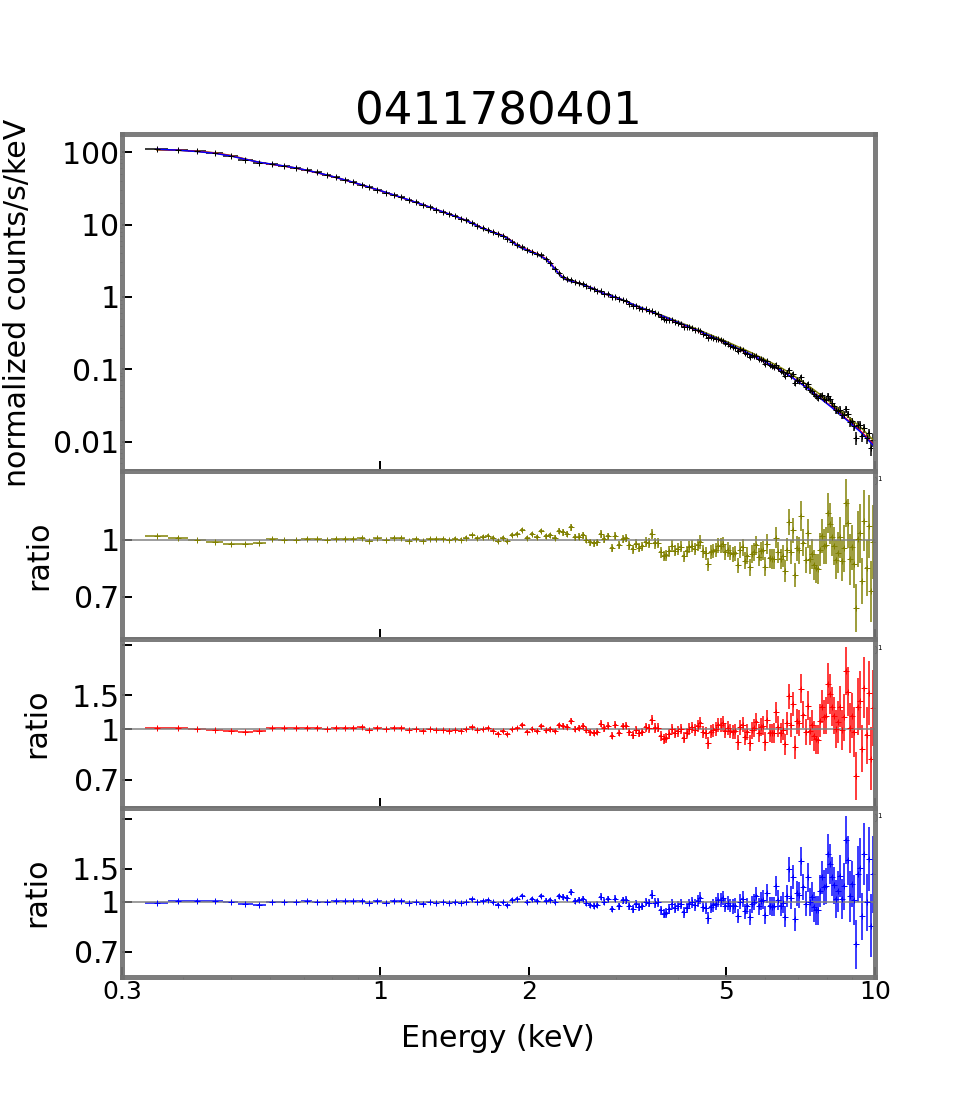}
	\end{minipage}
	\begin{minipage}{.30\textwidth} 
		\centering
		\includegraphics[width=.990\linewidth]{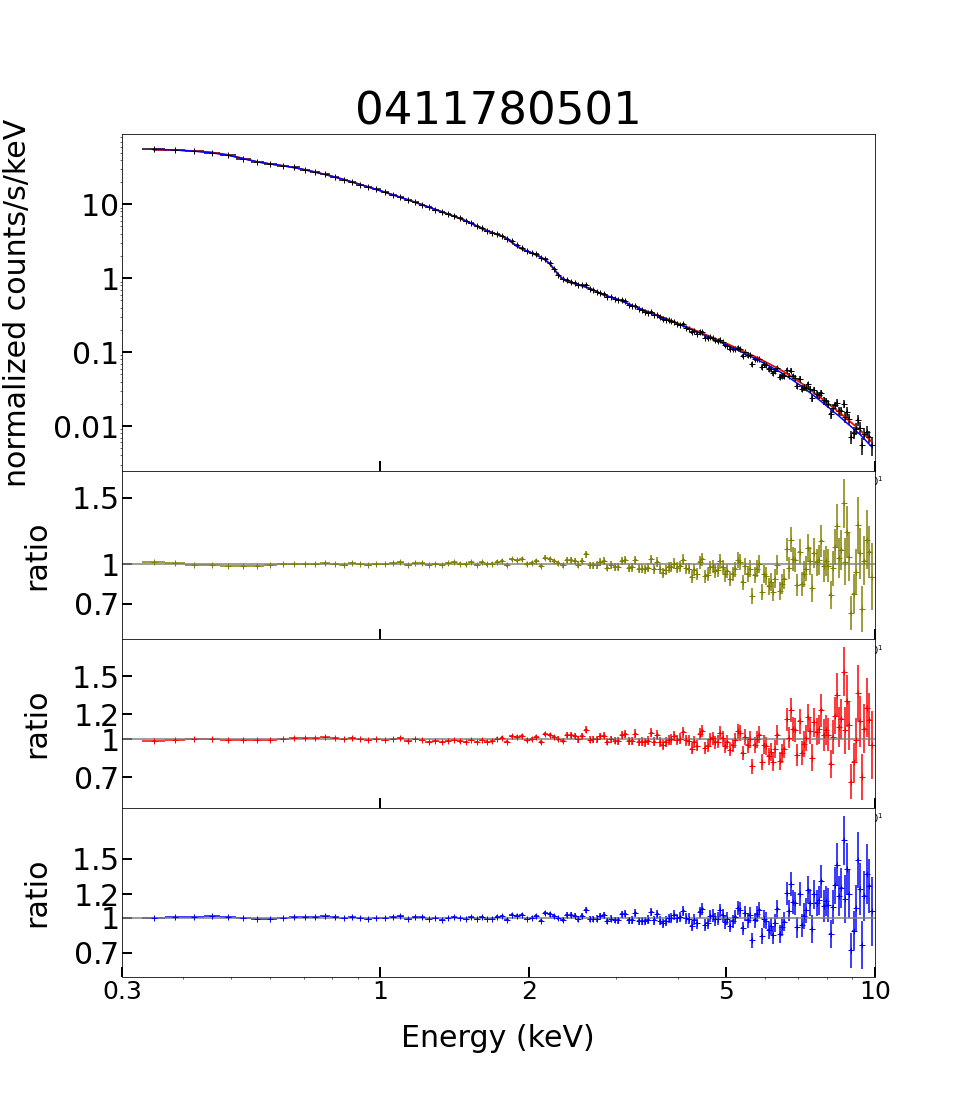}
	\end{minipage}
	\begin{minipage}{.30\textwidth} 
		\centering
		\includegraphics[width=.90\linewidth]{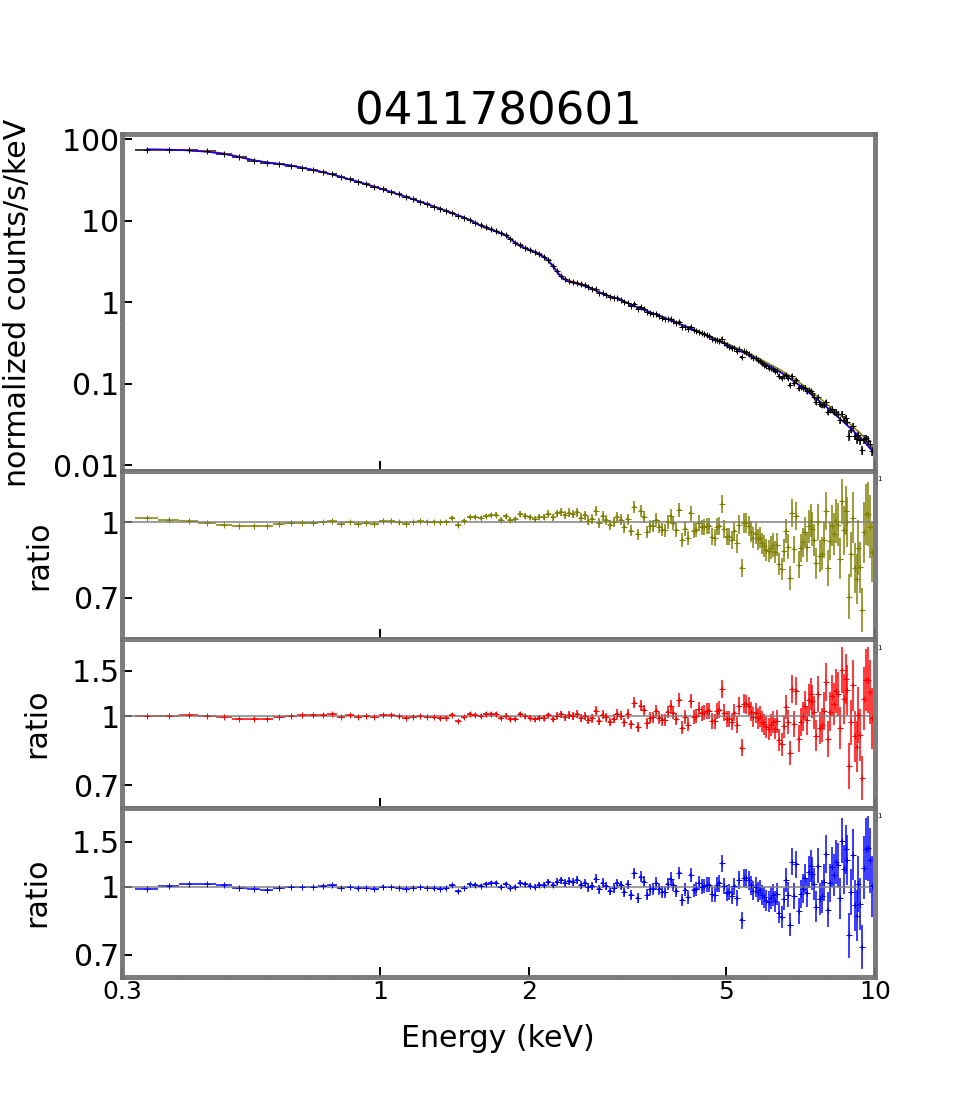}
	\end{minipage}
	\caption{Continue...}
\end{figure*}

\begin{figure*}
	\centering
	\begin{minipage}{.30\textwidth} 
		\centering
		\includegraphics[width=.990\linewidth]{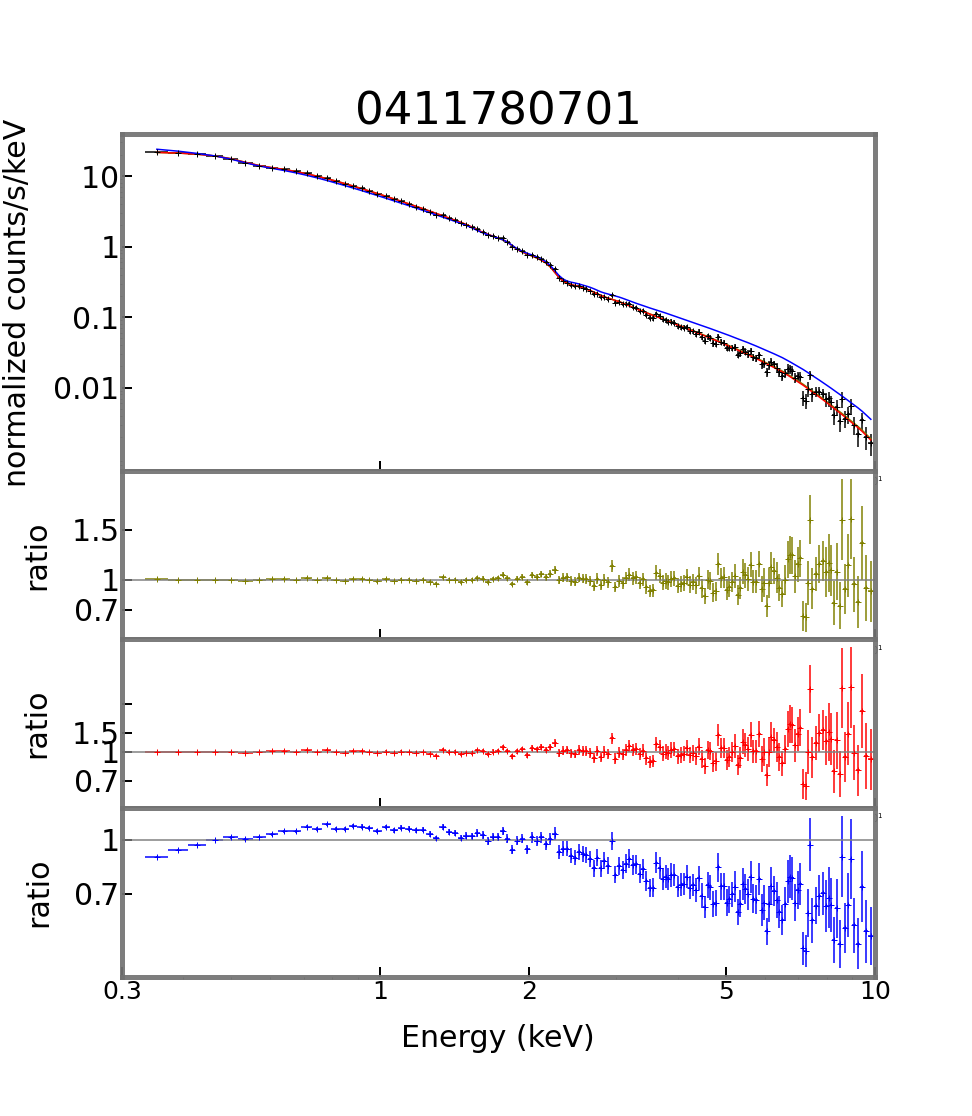}
	\end{minipage}
	\begin{minipage}{.30\textwidth} 
		\centering
		\includegraphics[width=.990\linewidth]{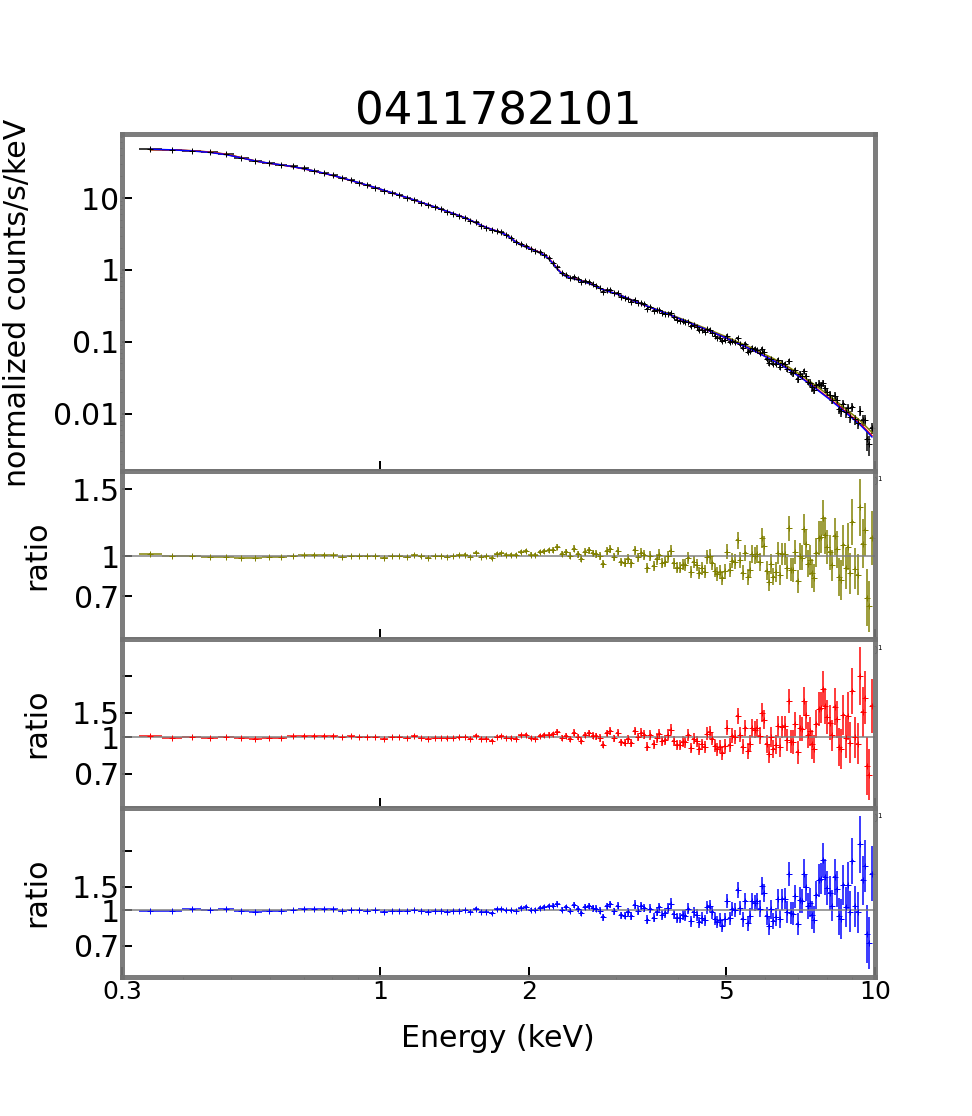}
	\end{minipage}
	\begin{minipage}{.30\textwidth} 
		\centering
		\includegraphics[width=.990\linewidth]{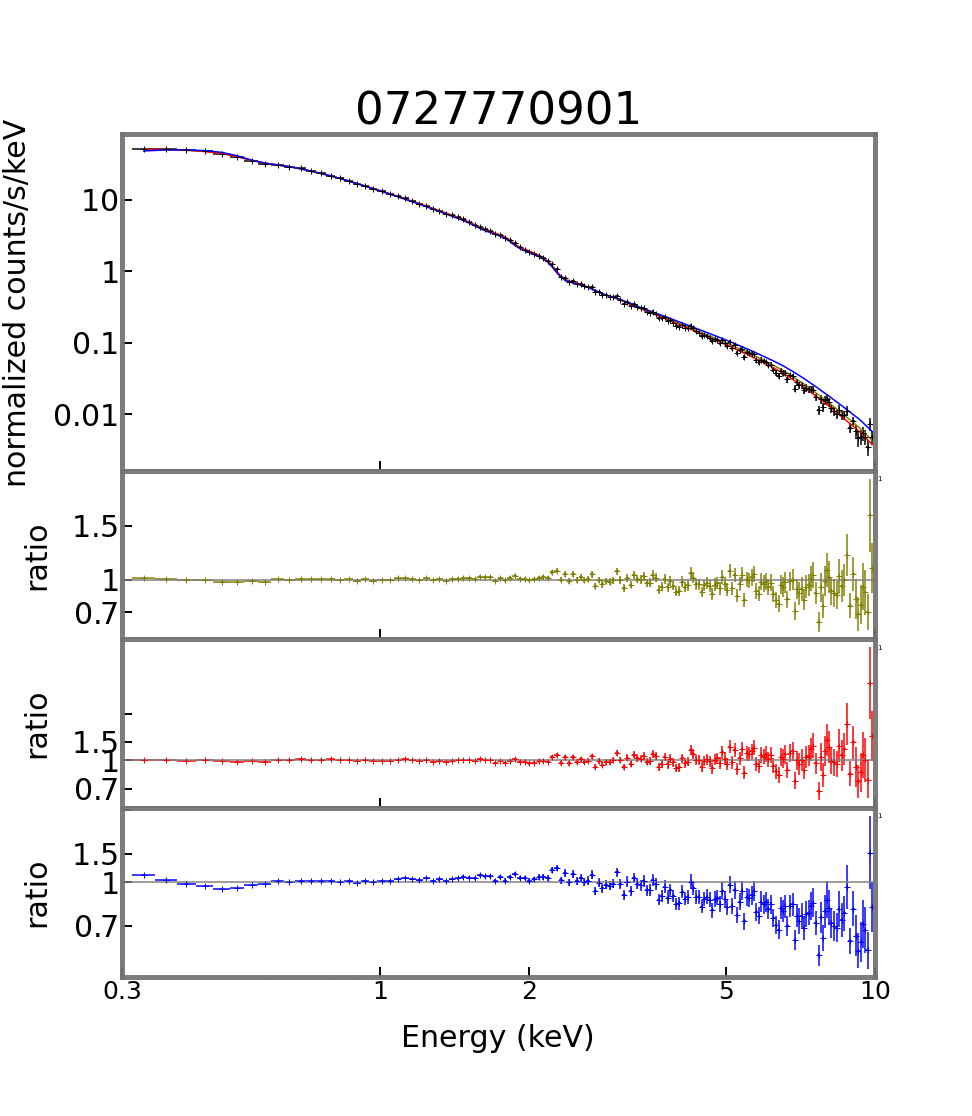}
	\end{minipage}
\end{figure*}

\begin{figure*}
    \centering
    \caption{Flux histograms for PKS 2155-304 observations.}
    \includegraphics[width=5.8truecm]{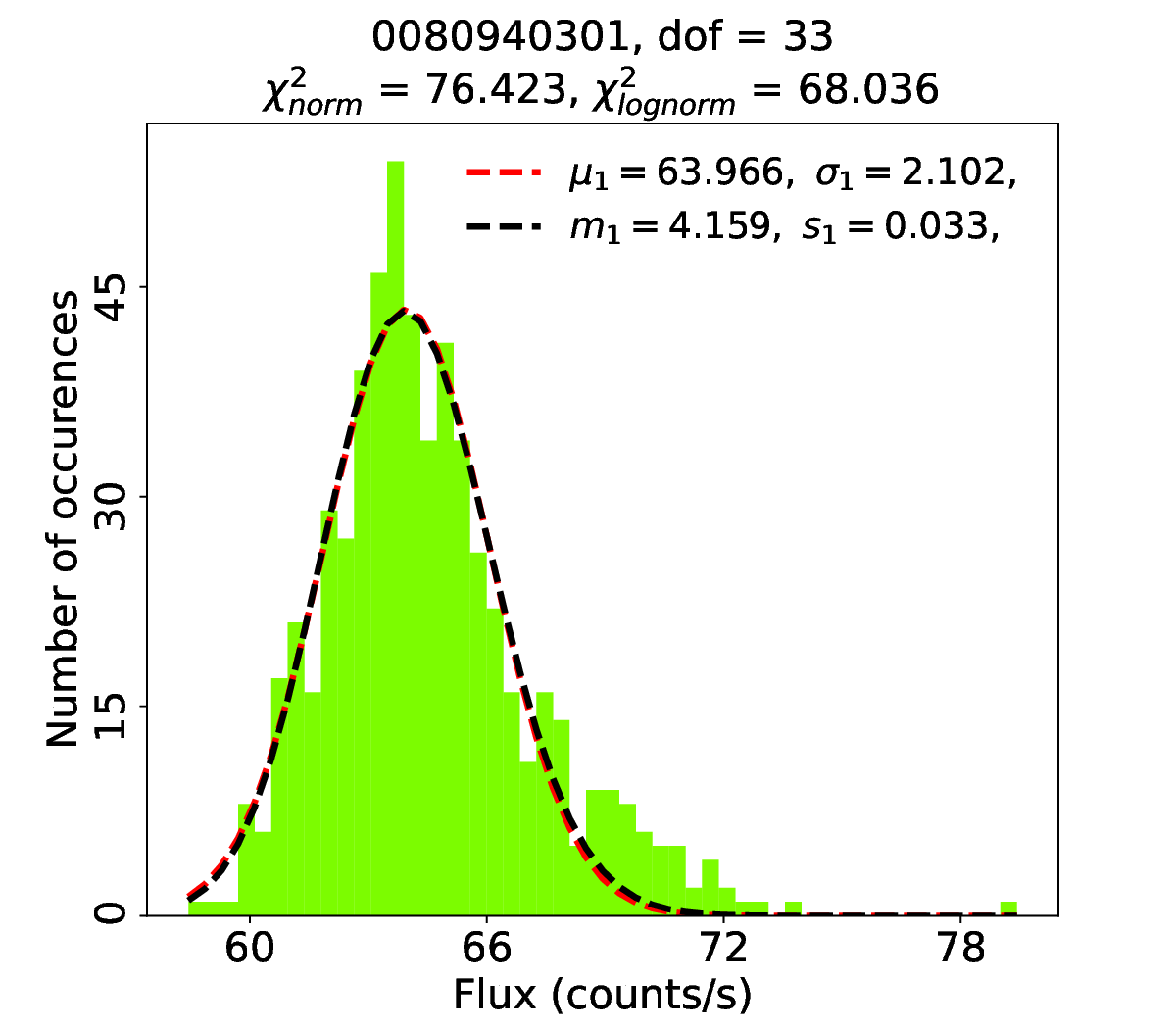}
    \includegraphics[width=5.8truecm]{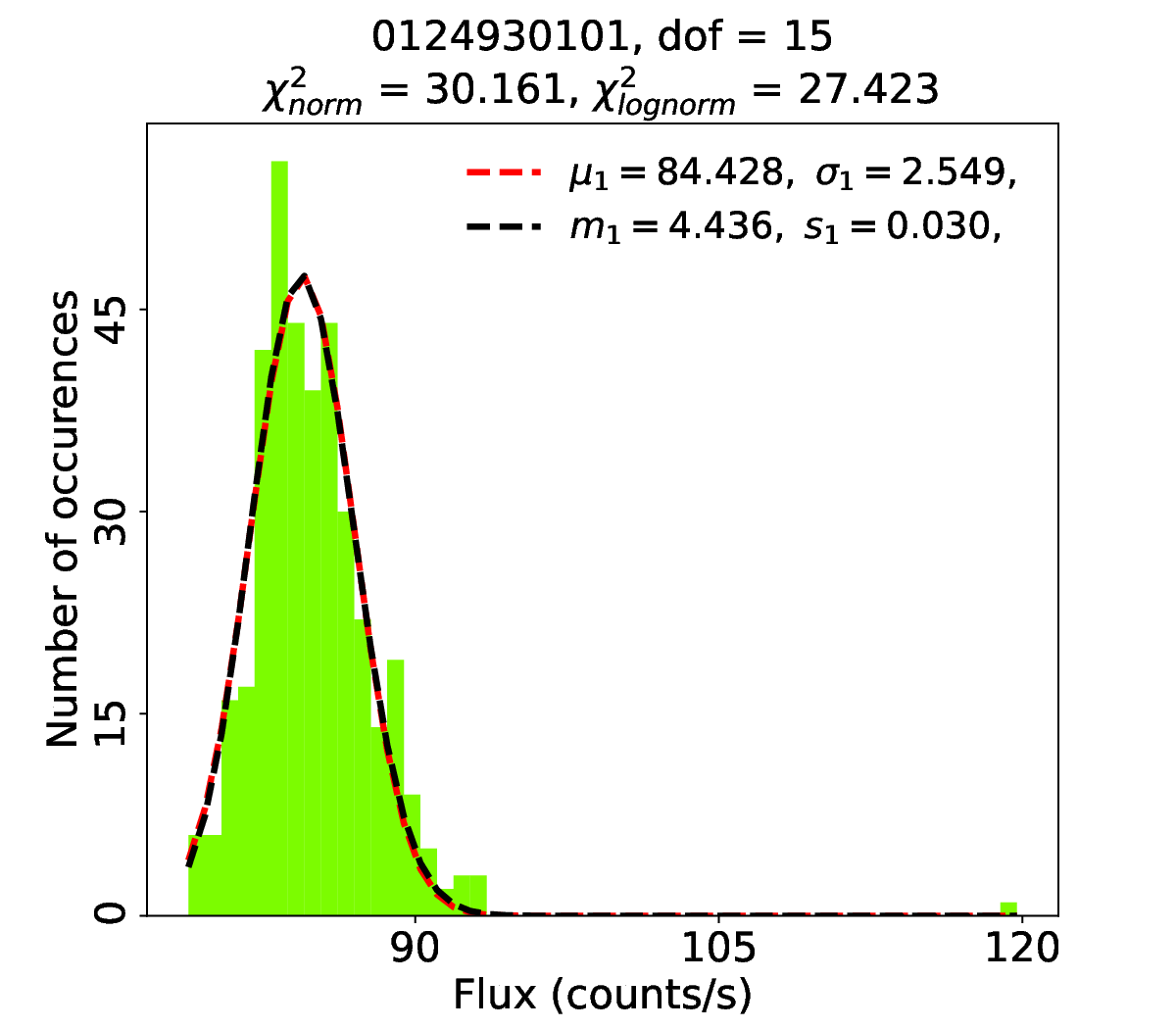}
    \includegraphics[width=5.8truecm]{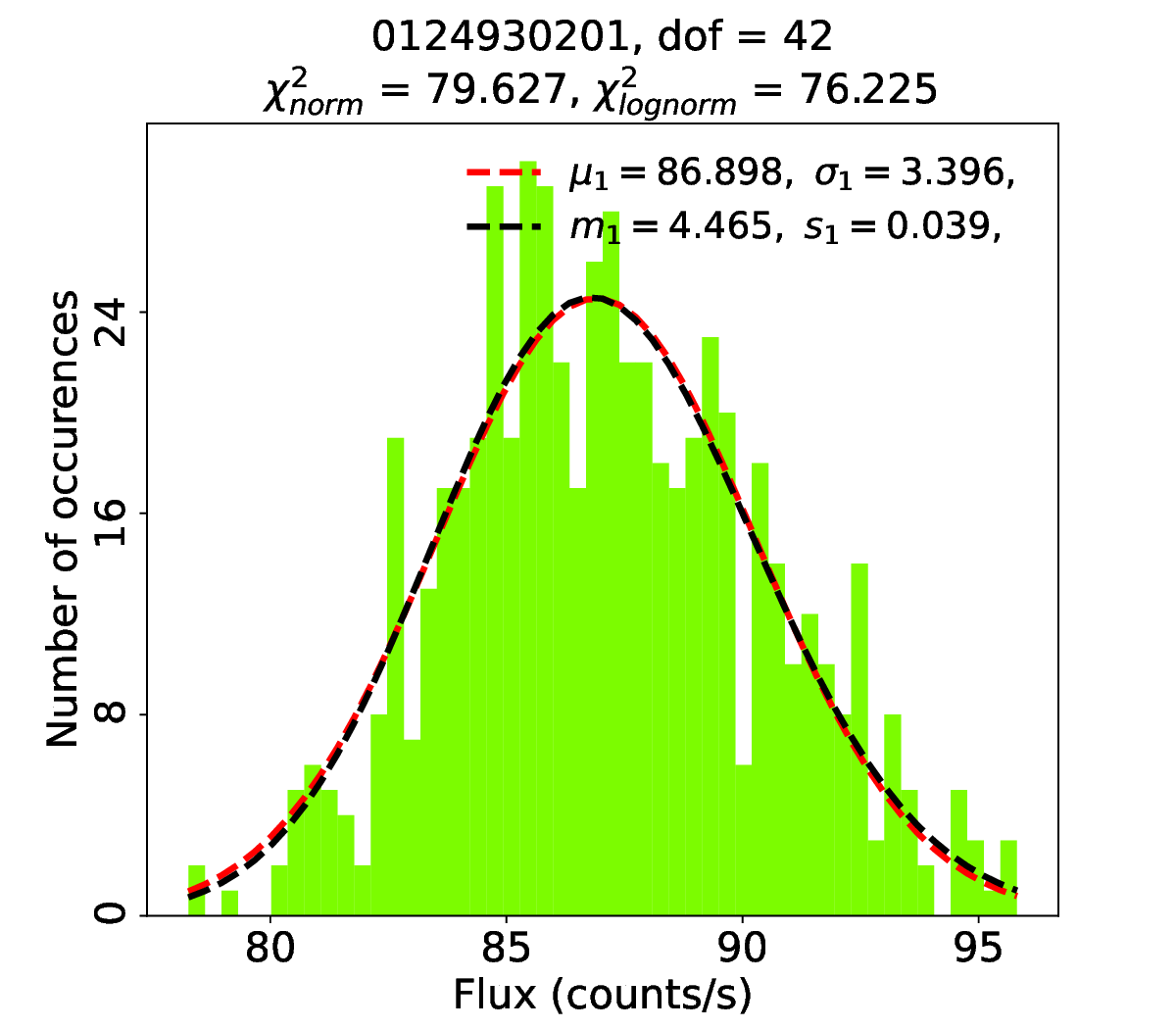}
    \includegraphics[width=5.8truecm]{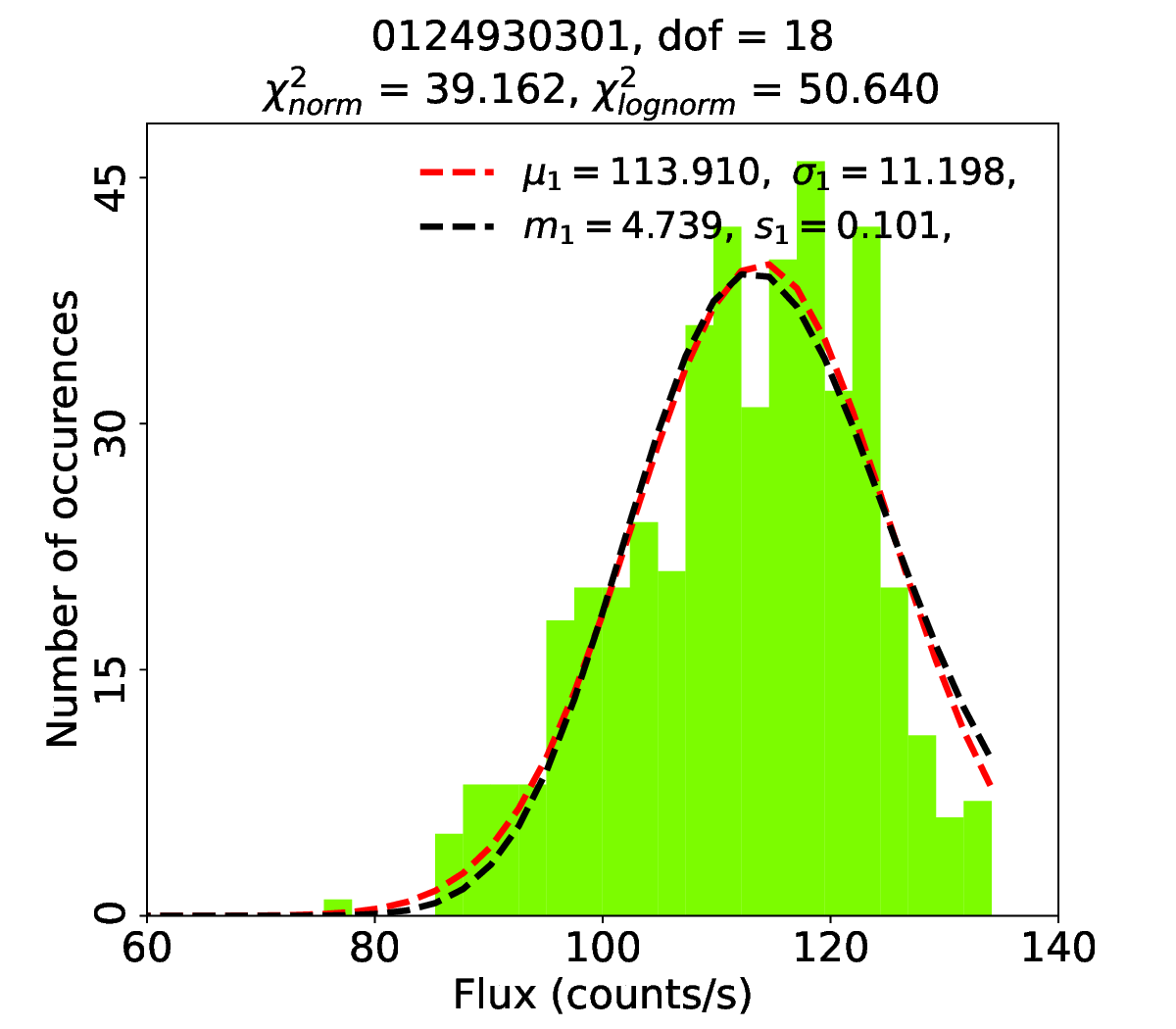}
    \includegraphics[width=5.8truecm]{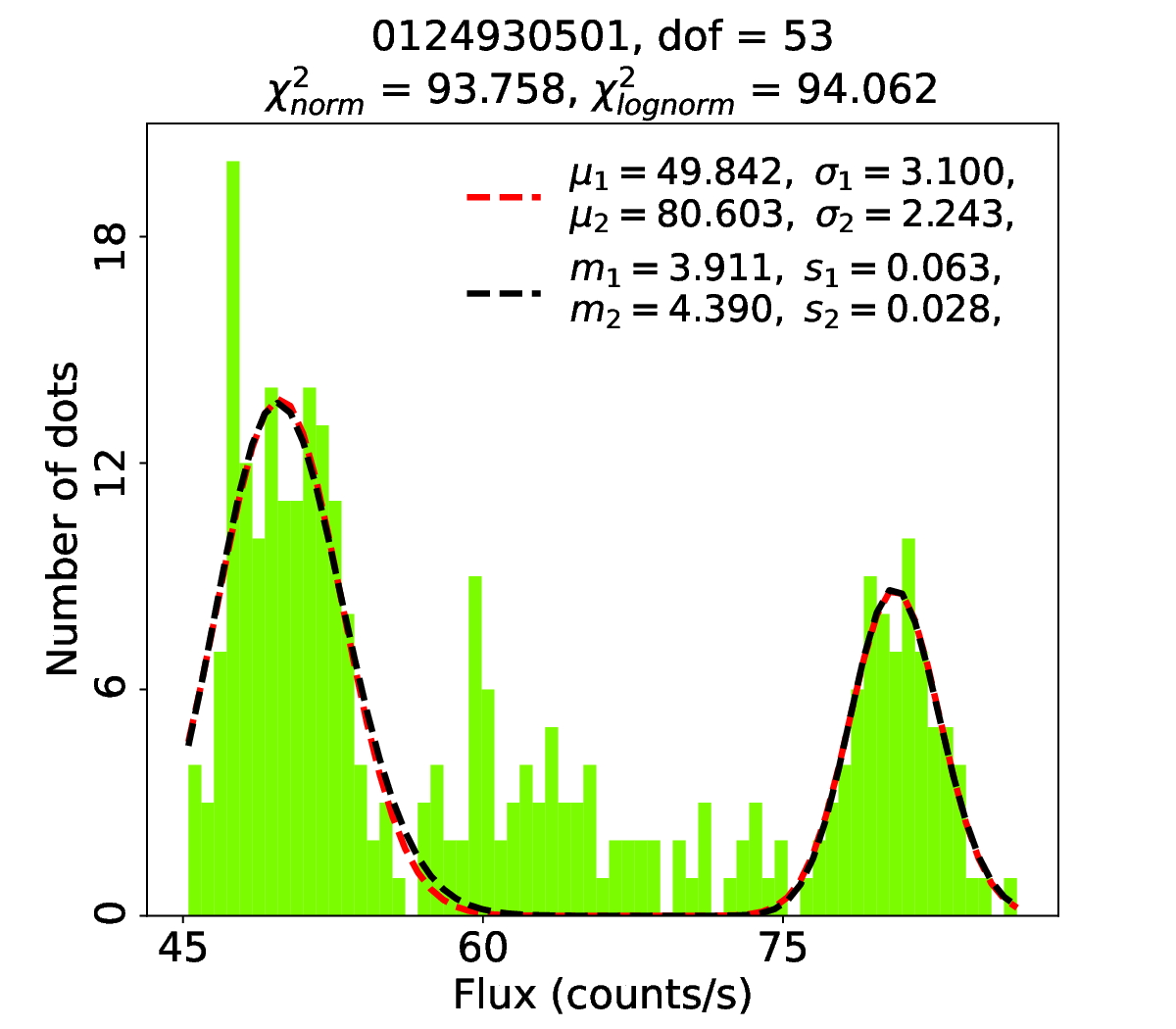}
    \includegraphics[width=5.8truecm]{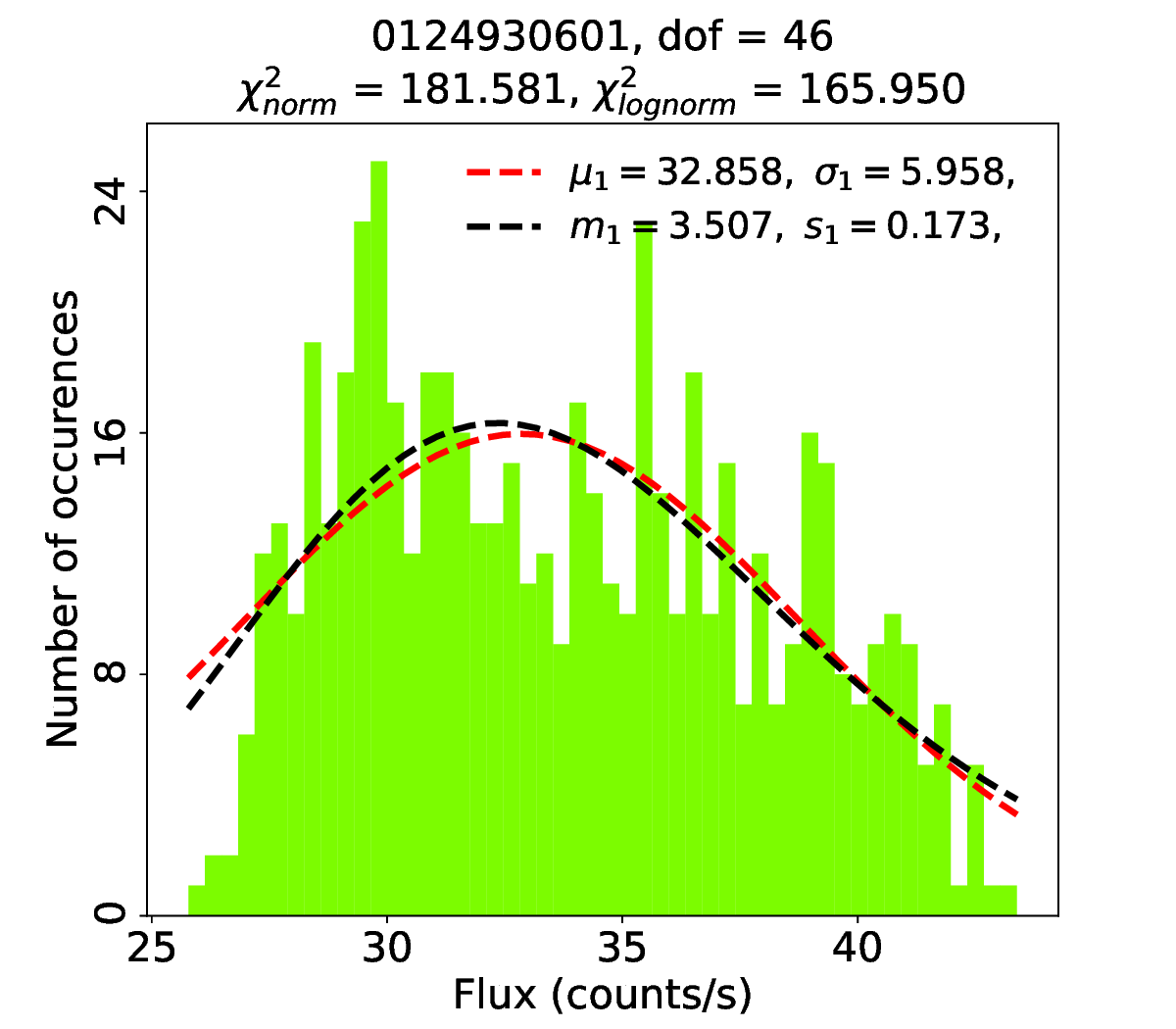}
    \includegraphics[width=5.8truecm]{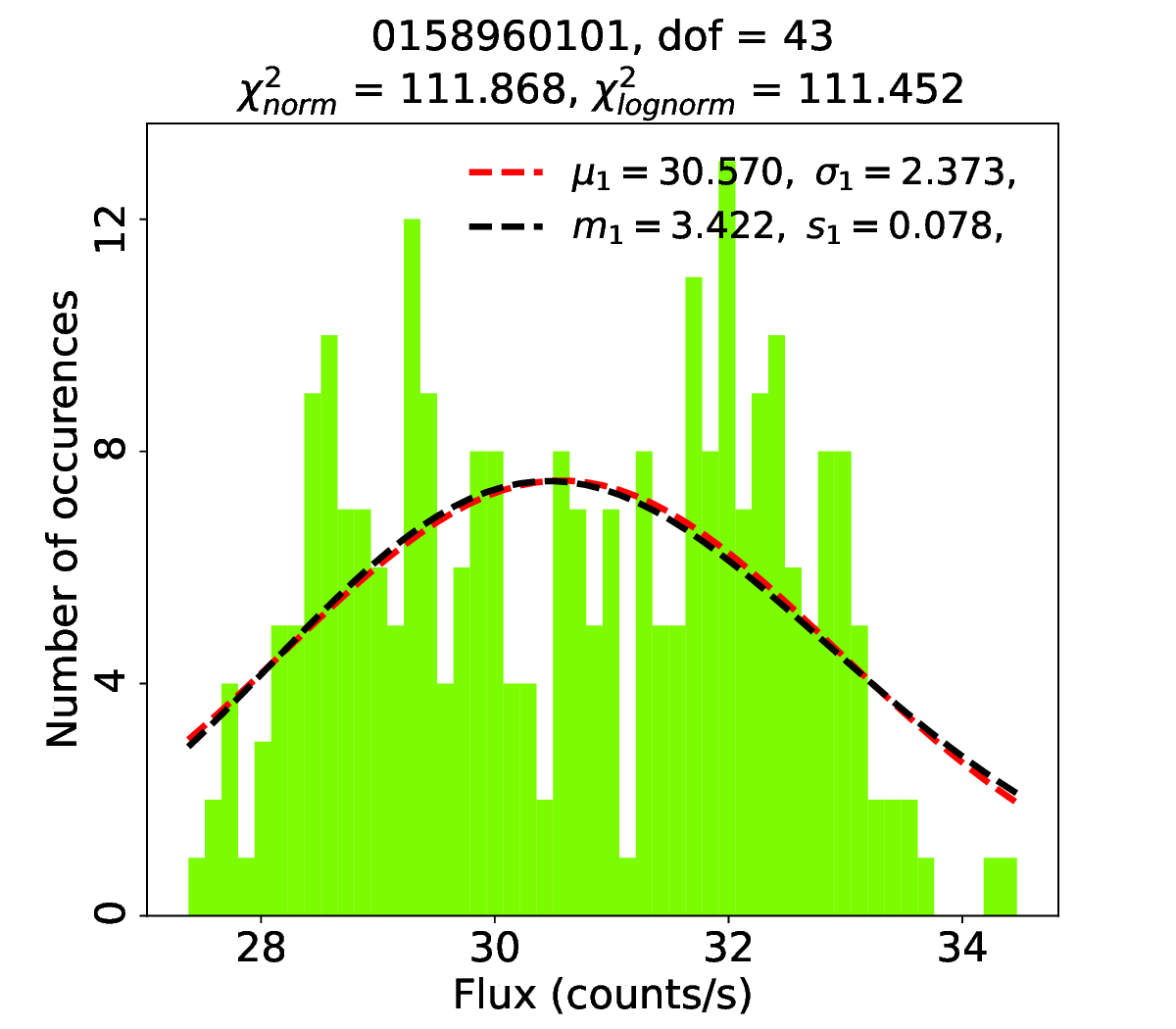}
    \includegraphics[width=5.8truecm]{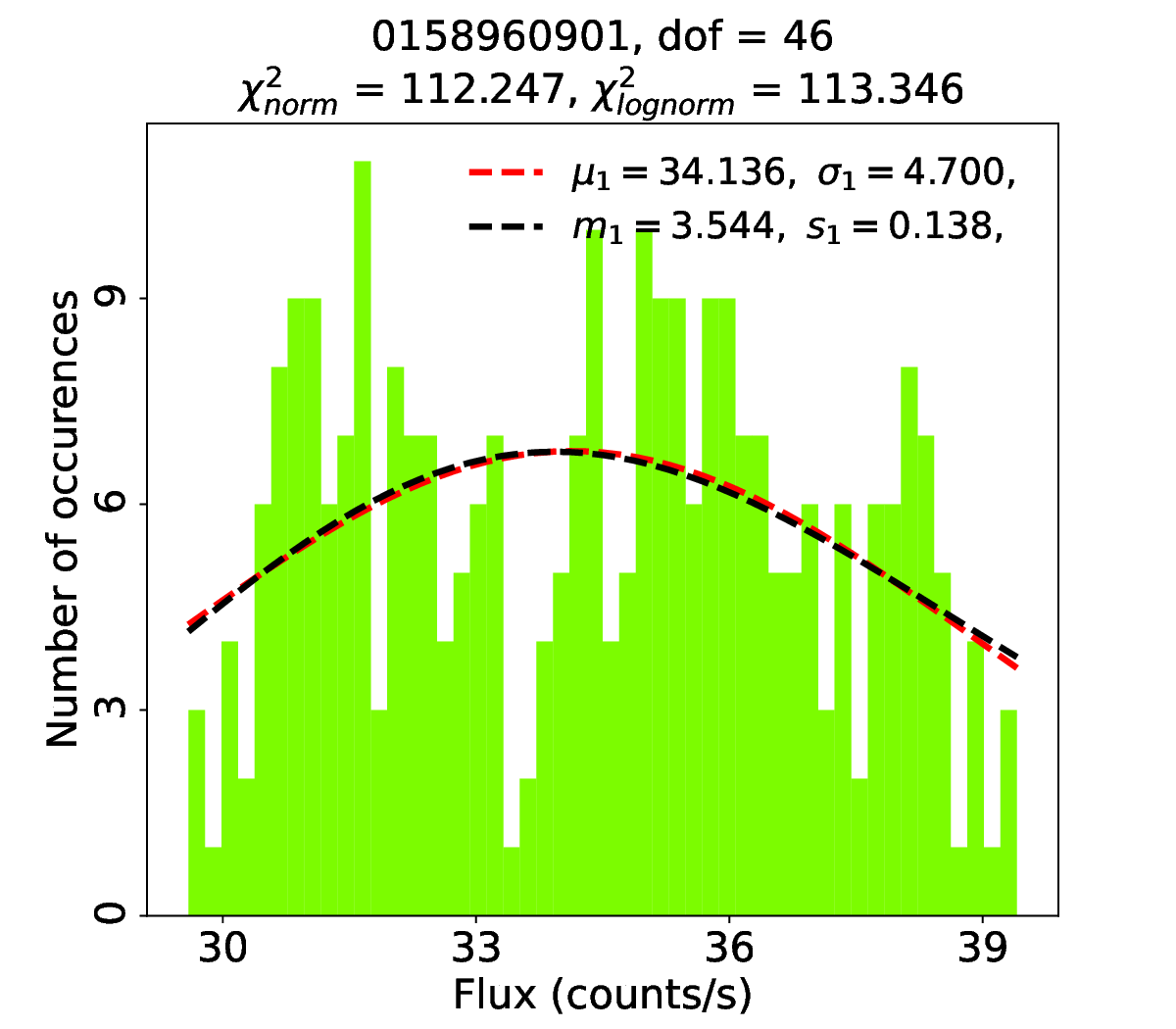}
    \includegraphics[width=5.8truecm]{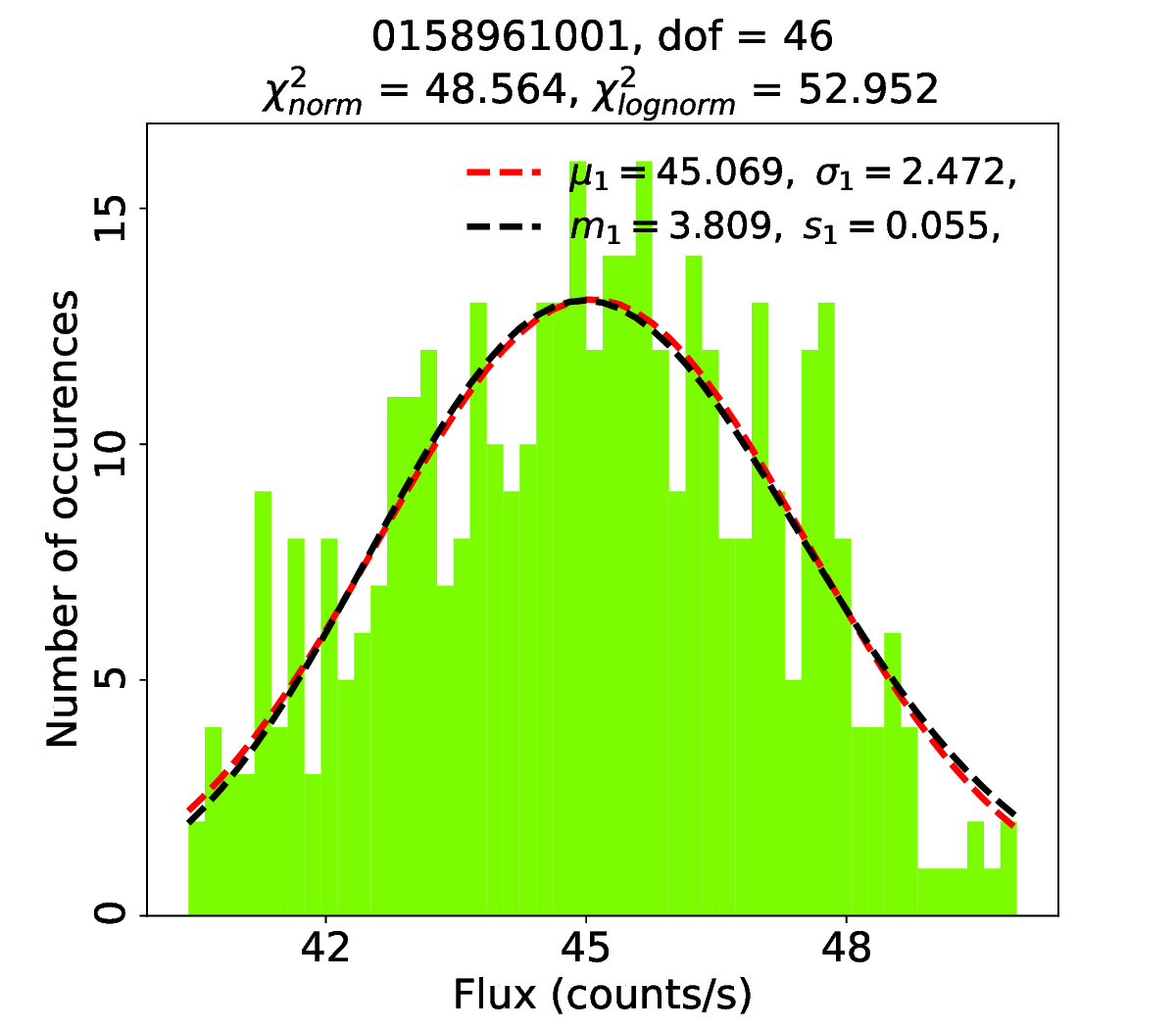}
    \includegraphics[width=5.8truecm]{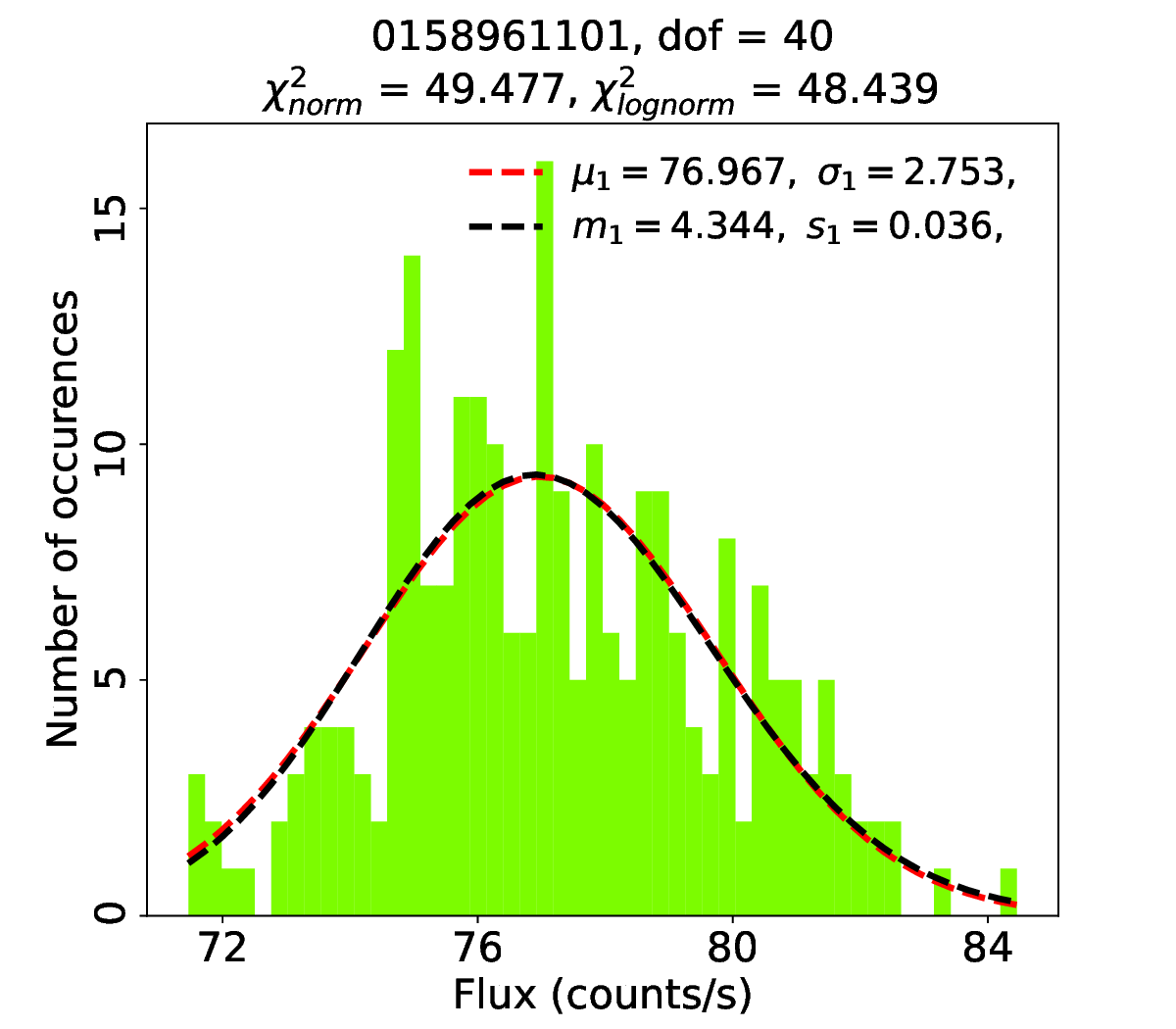}
    \includegraphics[width=5.8truecm]{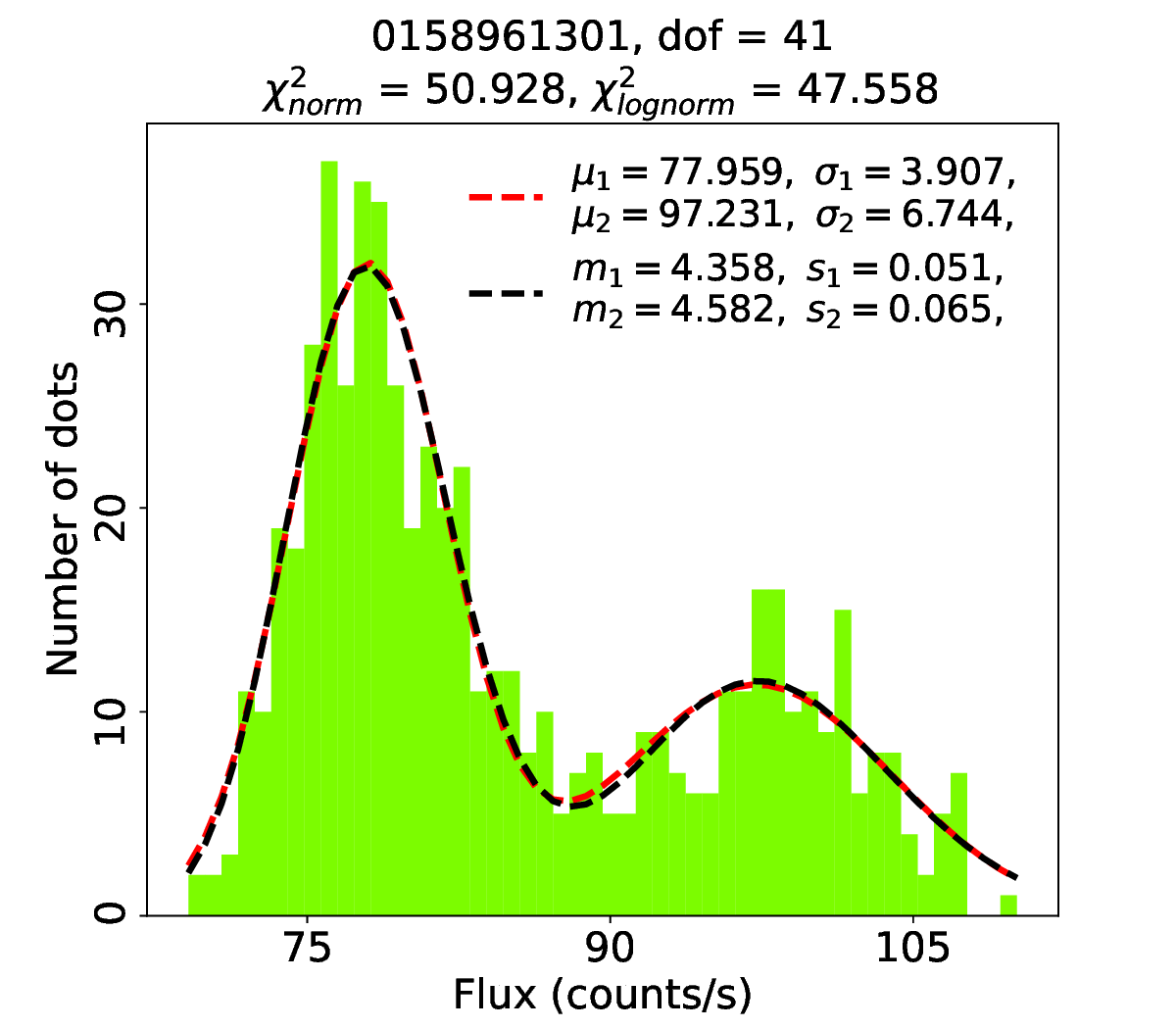}
    \includegraphics[width=5.8truecm]{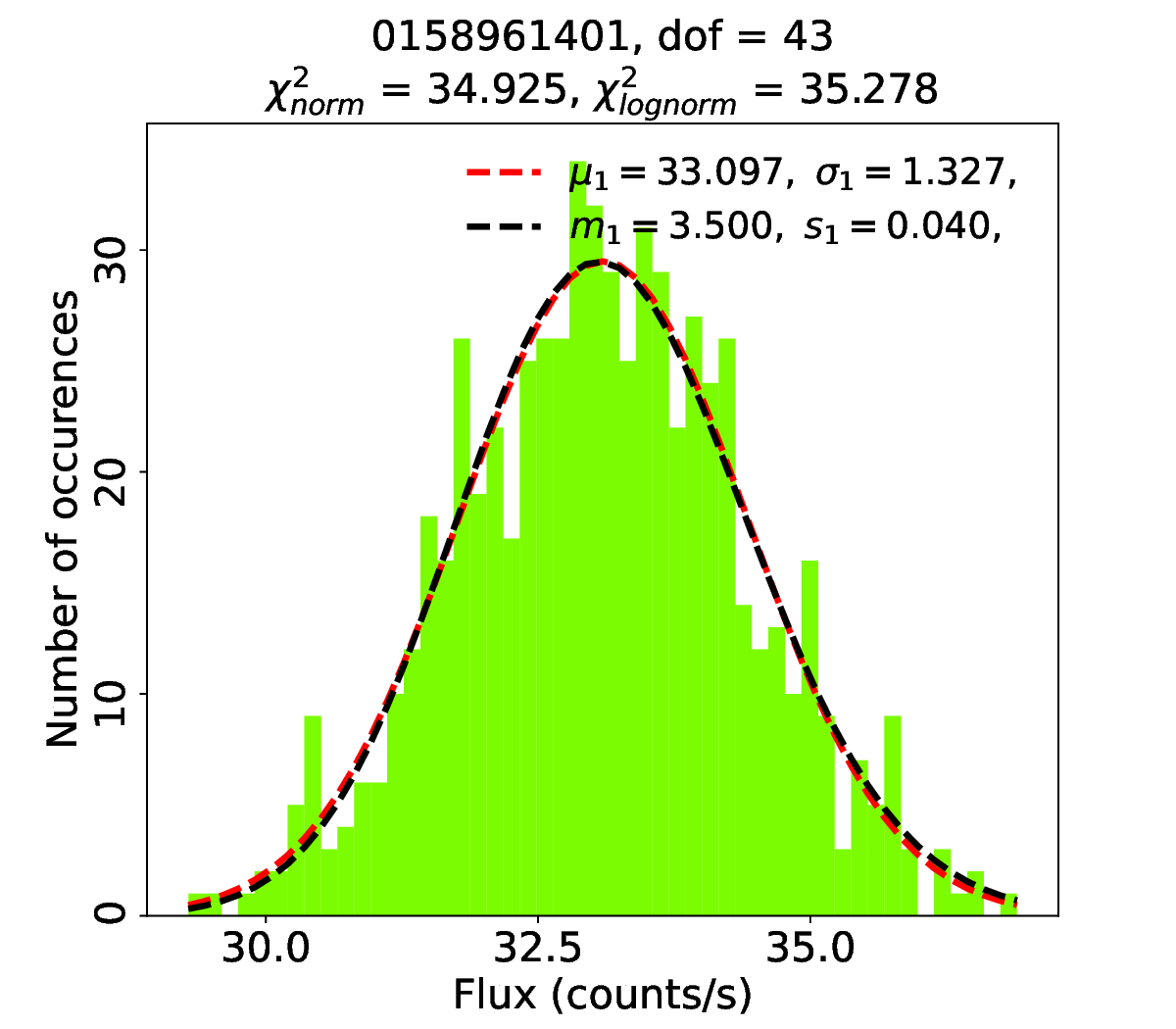}
    \caption{Continue...}
    \label{all_histo}
\end{figure*}

\begin{figure*}
    \centering
    \includegraphics[width=5.8truecm]{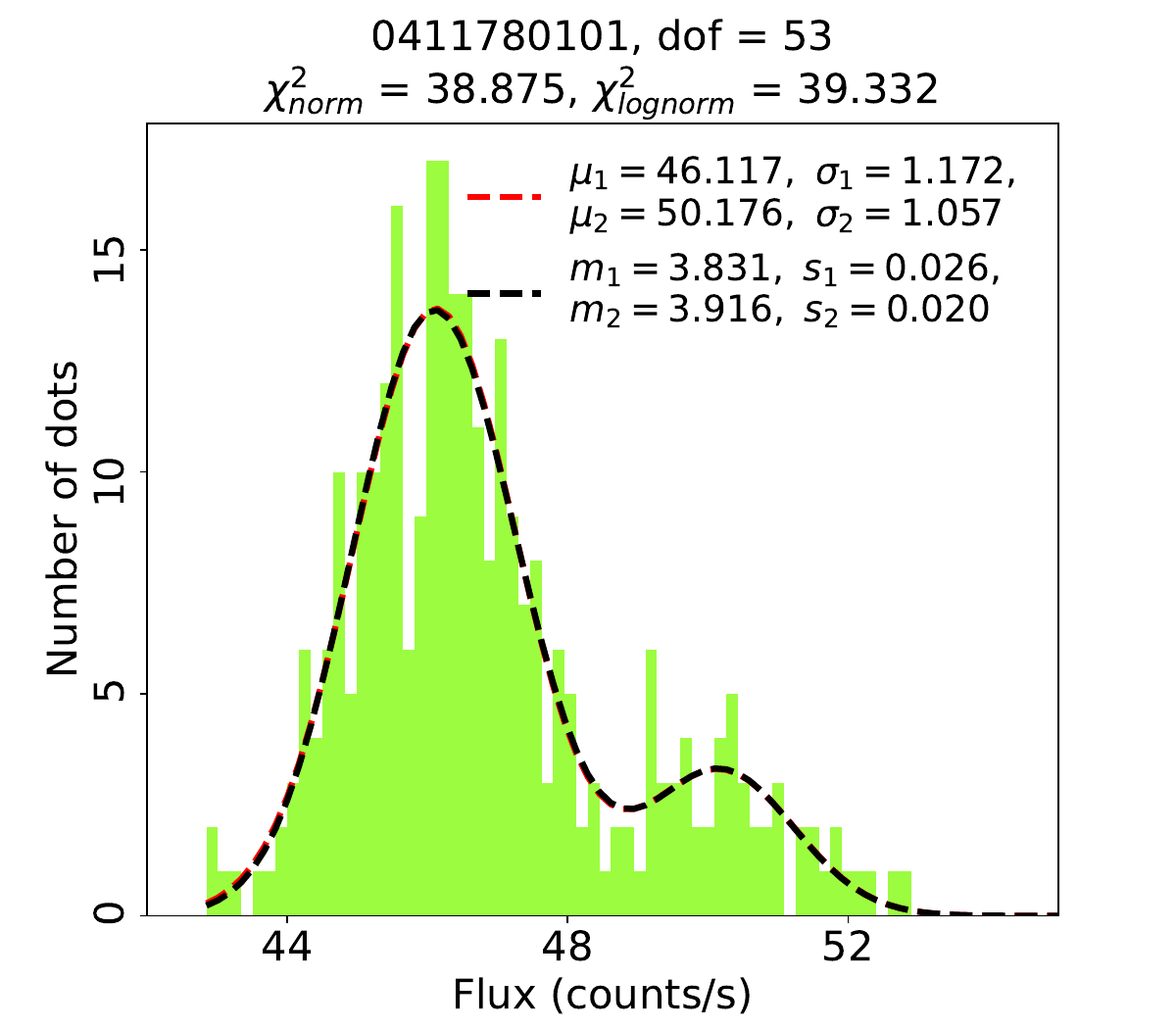}
    \includegraphics[width=5.8truecm]{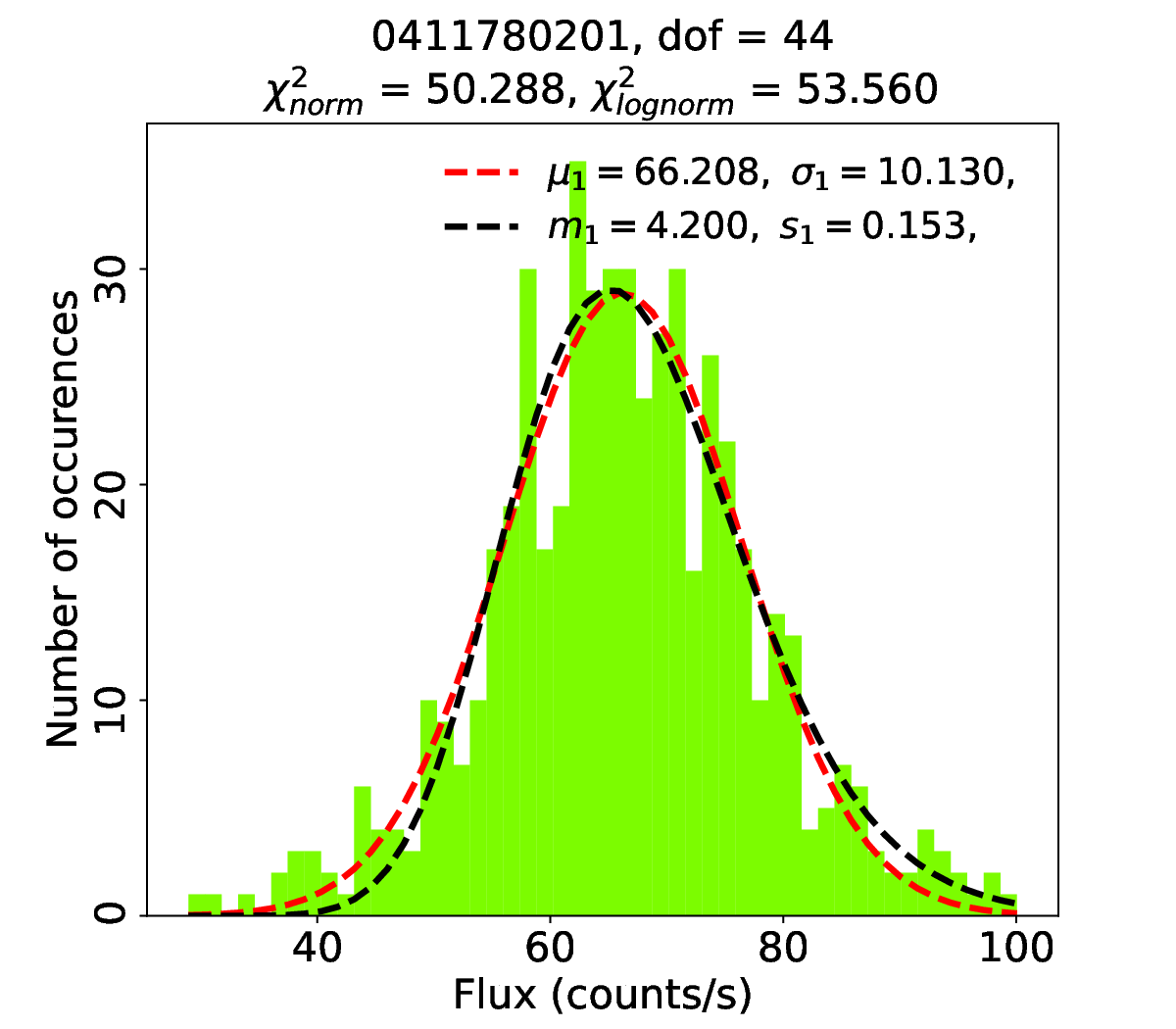}
    \includegraphics[width=5.8truecm]{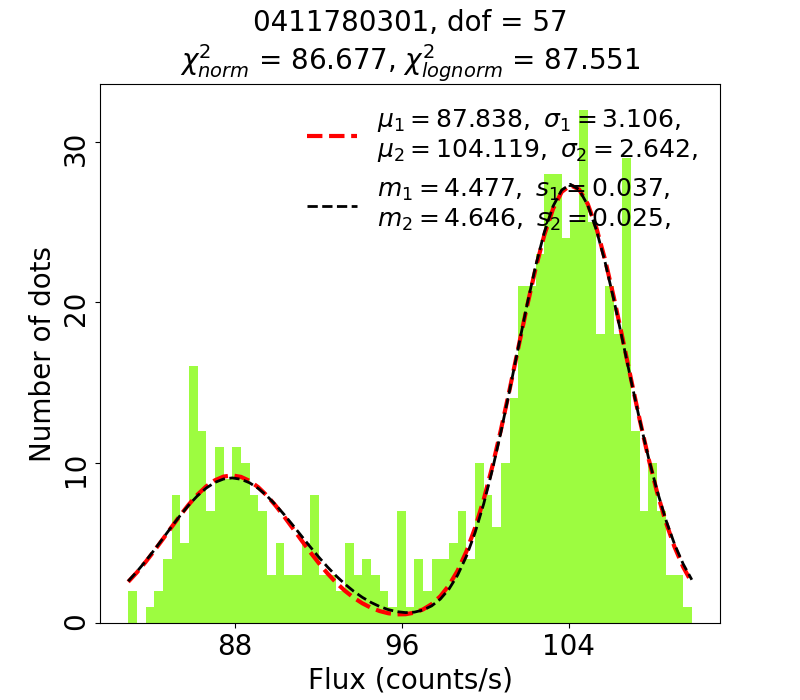}
    \includegraphics[width=5.8truecm]{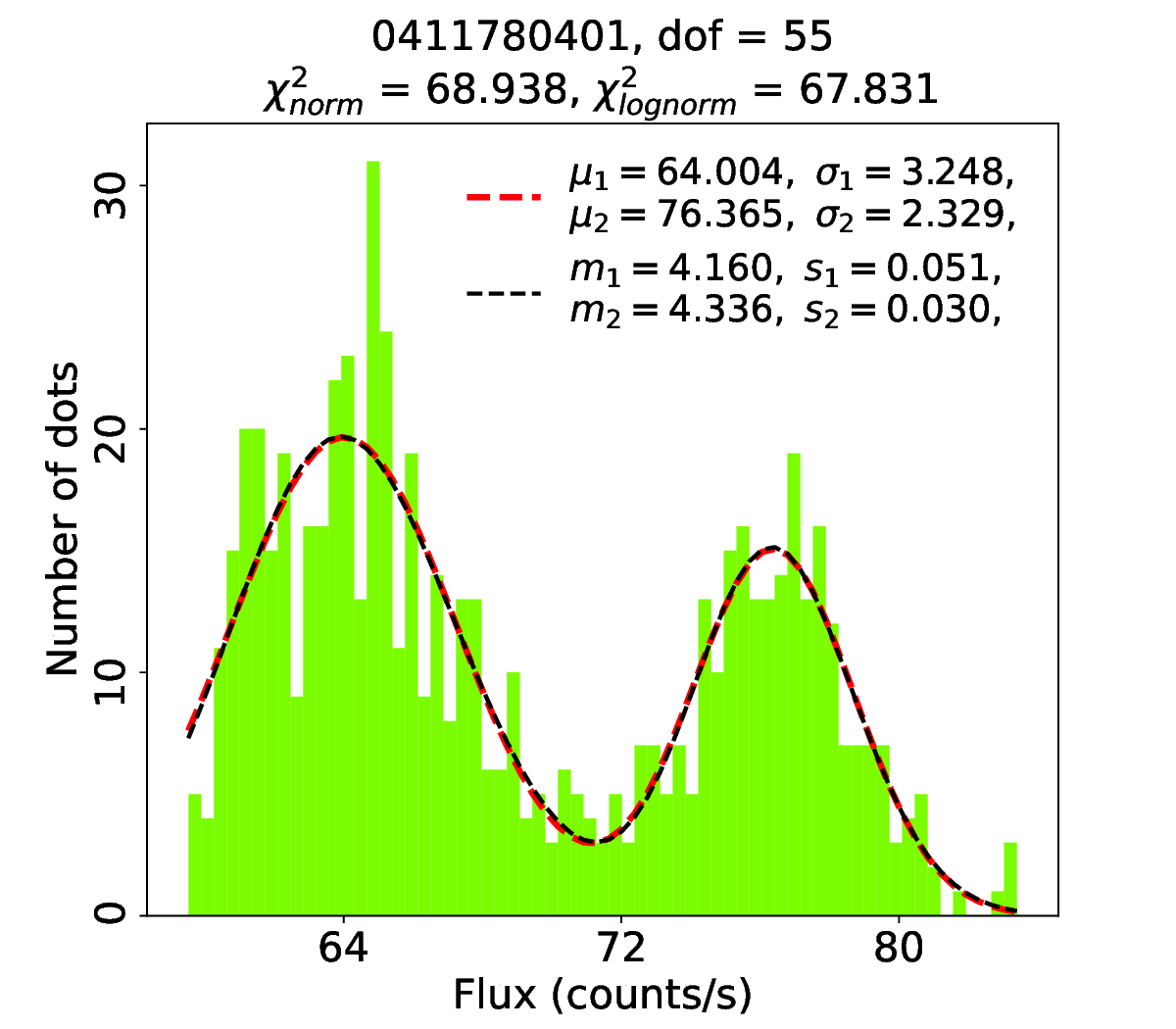}
    \includegraphics[width=5.8truecm]{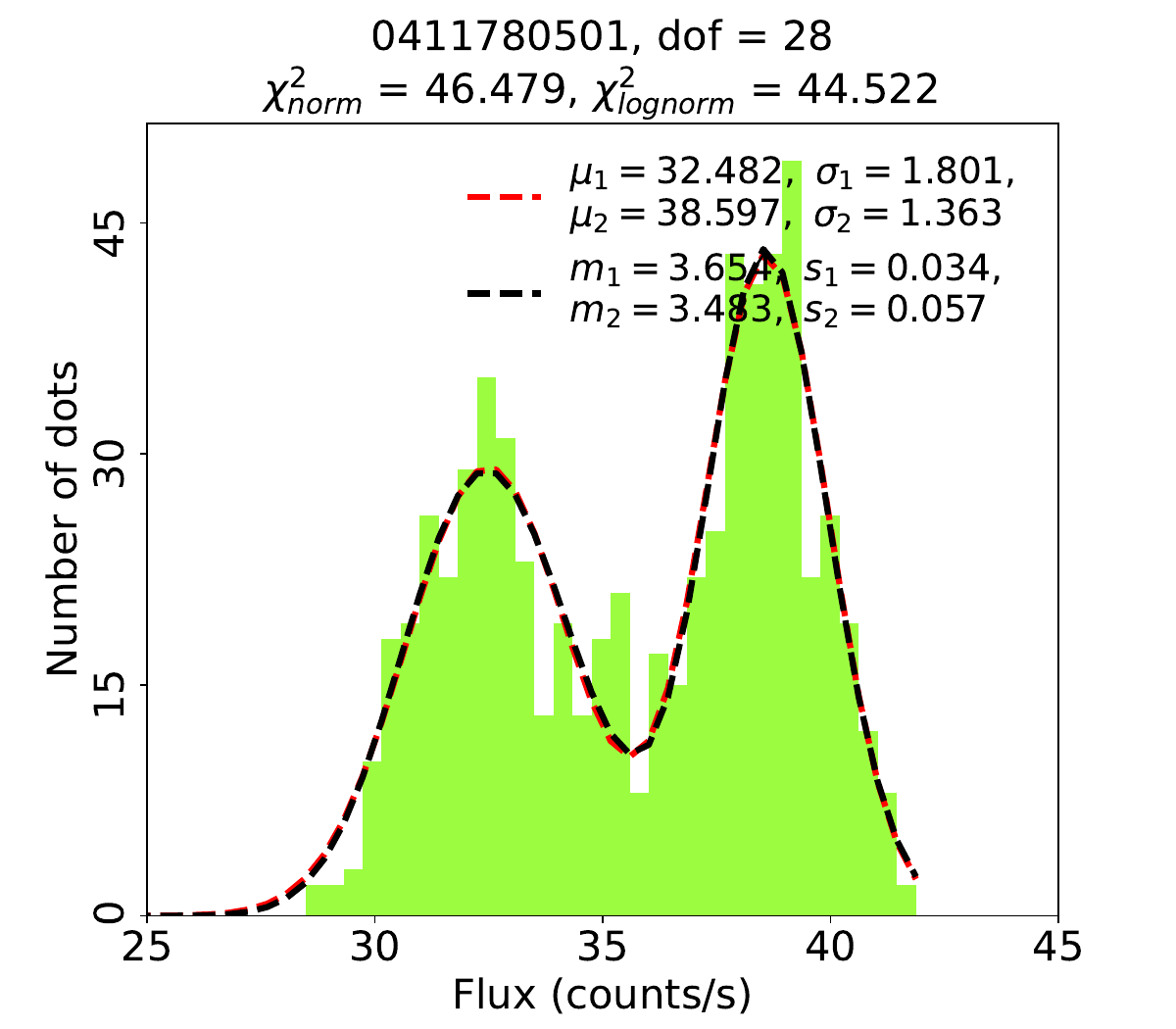}
    \includegraphics[width=5.8truecm]{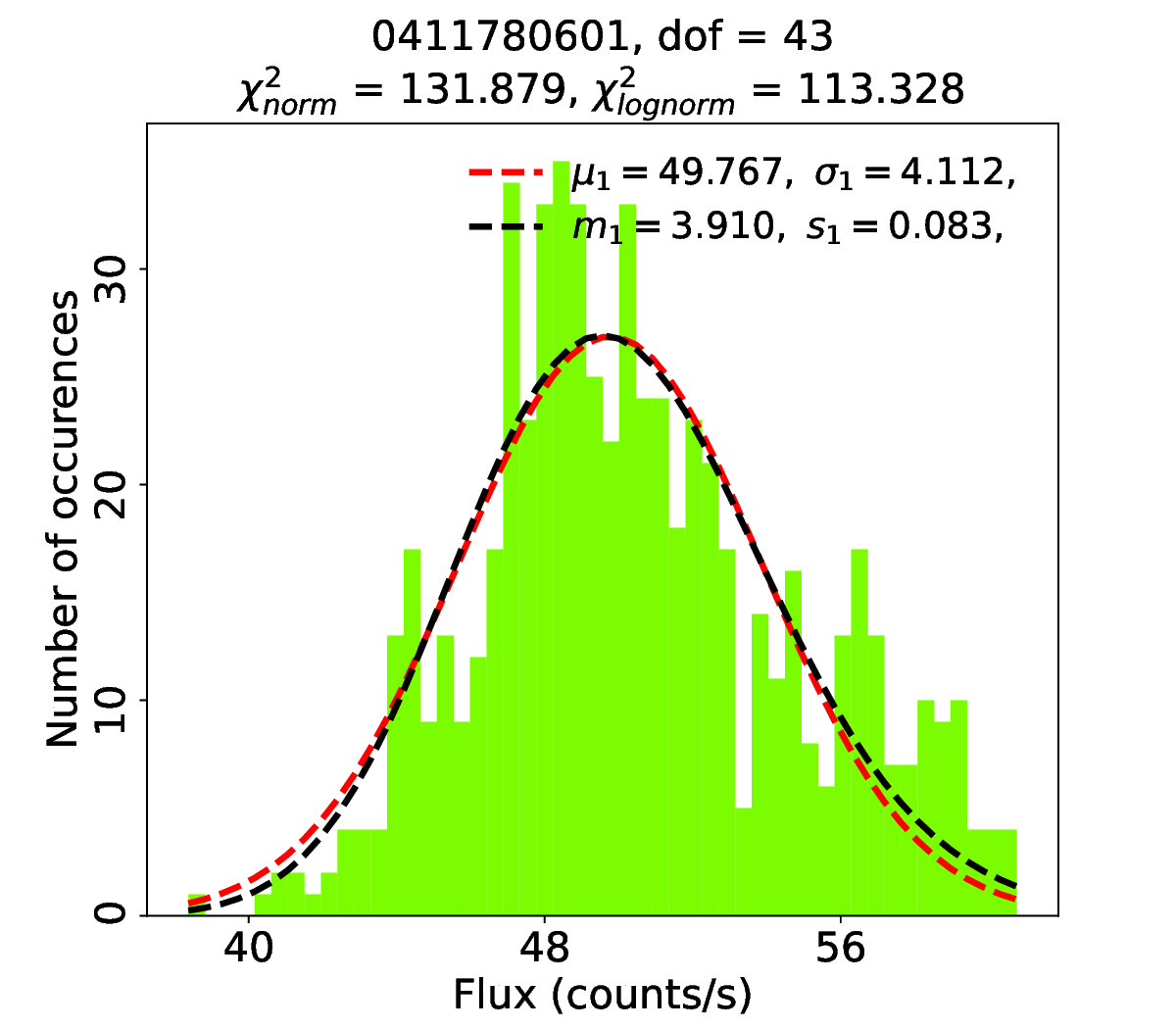}
    \includegraphics[width=5.8truecm]{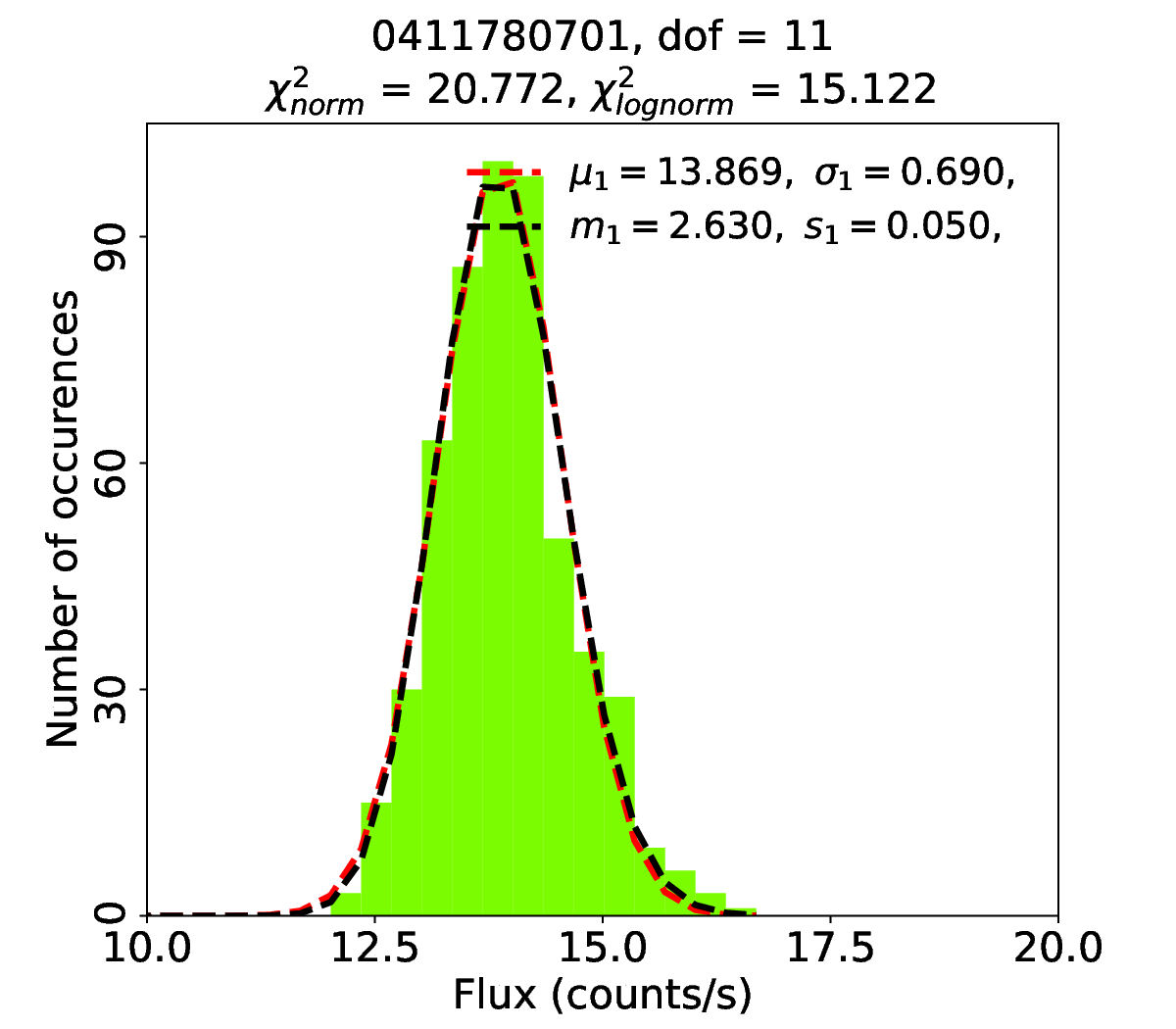}
    \includegraphics[width=5.8truecm]{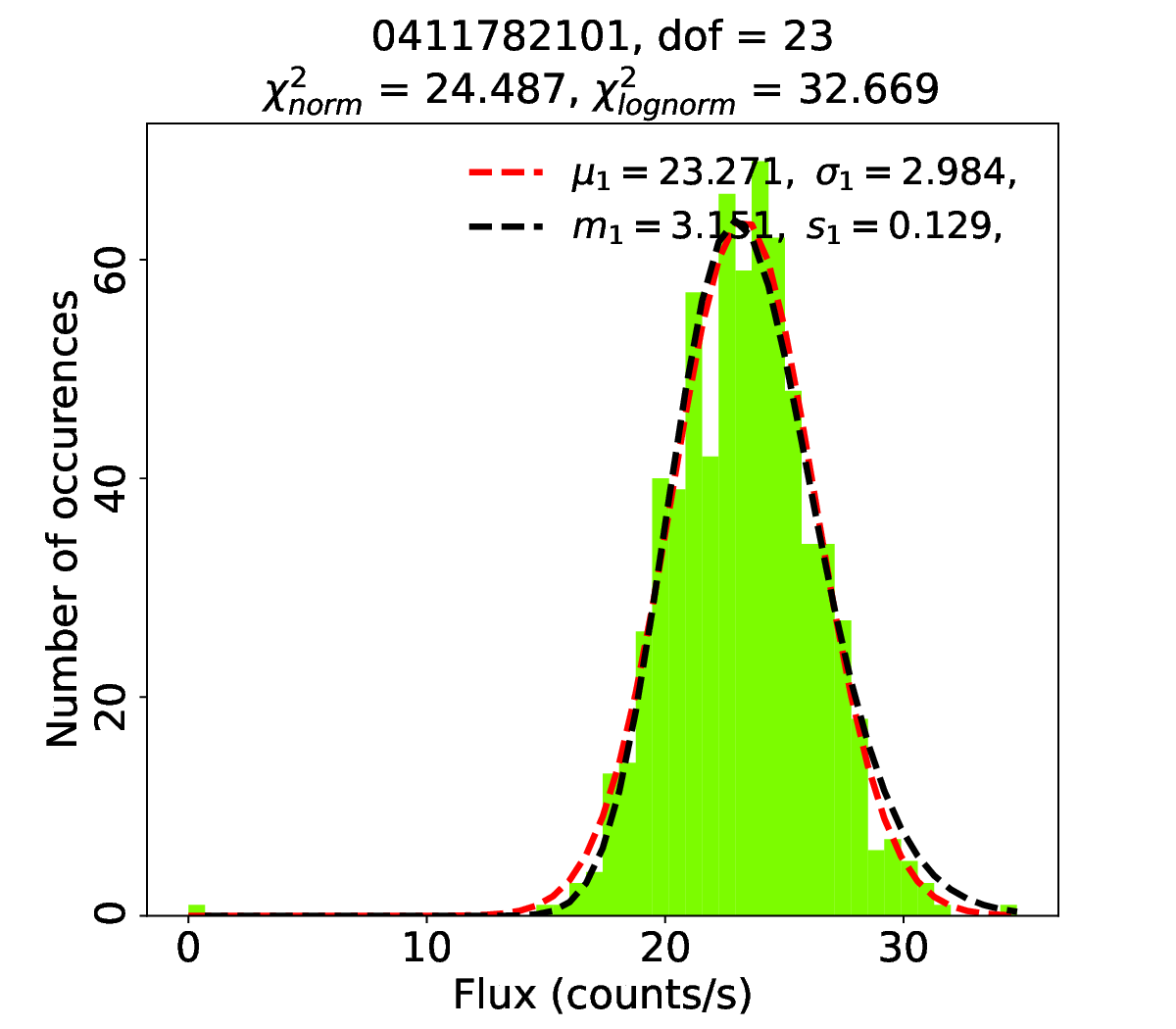}
    \includegraphics[width=5.8truecm]{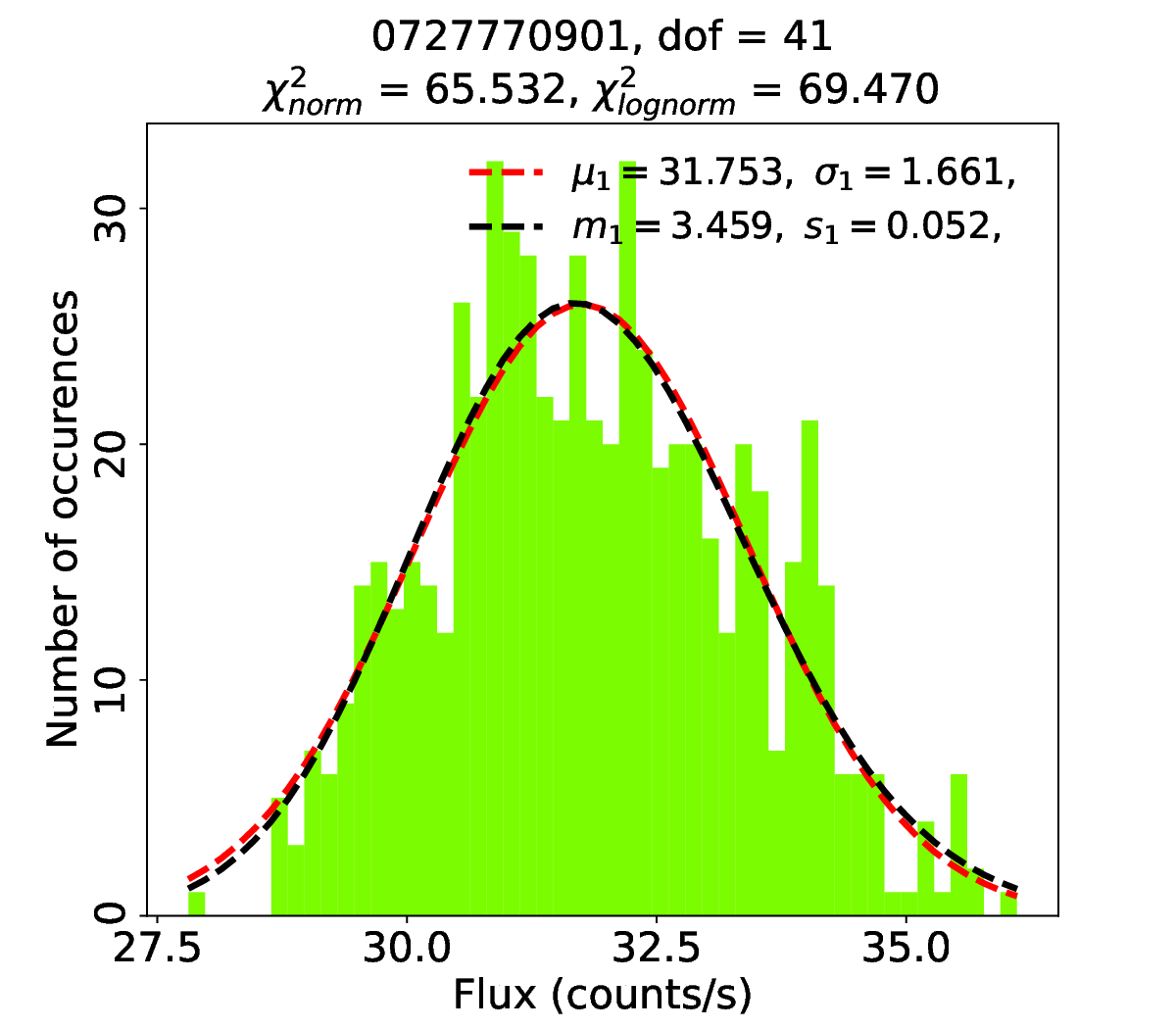}

    \label{fig:my_label}
\end{figure*}

\label{LastPage}

\end{document}